\documentclass[a4paper,UKenglish]{lipics}
\usepackage{microtype}
\newcommand\APPENDIX[1]{#1}
\newcommand\NOAPPENDIX[1]{}

\usepackage{amsopn}

\usepackage[T1]{fontenc}
\usepackage[utf8]{inputenc}

\usepackage{mathrsfs,amsfonts}
\usepackage{color}
\usepackage{ifthen}
\usepackage{tikz}
\usepgflibrary{arrows}
\usetikzlibrary{fit}
\usepackage{authblk}
\usepackage{mathdots}
\usepackage{aliascnt}
\usepackage{tikz}
\usepgflibrary{arrows}
\allowdisplaybreaks
\usepackage{mathtools}
\usepackage{stmaryrd}
\usepackage{dsfont}
\usepackage{savefnmark}
\usepackage{longtable}
\usepackage{booktabs}

\newcommand{\A}{\mathbb{A}}
\newcommand{\R}{\mathbb{R}}
\newcommand{\Q}{\mathbb{Q}}
\newcommand{\N}{\mathbb{N}}
\newcommand{\Ns}{\mathbb{N}^*}
\newcommand{\Z}{\mathbb{Z}}
\newcommand{\K}{\mathbb{K}}
\newcommand{\MAT}[3]{M_{#1\ifthenelse{\equal{#2}{}}{}{,#2}}\ifthenelse{\equal{#3}{}}{}{\left(#3\right)}}
\newcommand{\Rgen}{\R_{G}}
\newcommand{\Rpoly}{\R_{P}}
\newcommand{\Rp}{\R_{+}}
\newcommand{\Rs}{\R^{*}}
\newcommand{\Rps}{\R_{+}^{*}}
\newcommand{\intinterv}[2]{\llbracket#1,#2\rrbracket}
\newcommand{\powerset}[1]{\mathcal{P}(#1)}
\newcommand{\dom}{\operatorname{dom}}
\newcommand{\card}[1]{{\##1}}
\newcommand{\PTIME}{\ensuremath{\operatorname{P}}}
\newcommand{\EXPTIME}{\ensuremath{\operatorname{EXPTIME}}}
\newcommand{\NP}{\ensuremath{\operatorname{NP}}}
\newcommand{\NPTIME}{\NP}
\newcommand{\FP}{\ensuremath{\operatorname{FP}}}
\newcommand{\PSPACE}{\ensuremath{\operatorname{PSPACE}}}
\newcommand{\inorm}[2]{\left\lVert{#1}\right\rVert_{#2}}

\newcommand{\infnorm}[1]{\inorm{#1}{}}
\newcommand{\sigmap}[1]{{\Sigma{#1}}}
\newcommand{\degp}[1]{{\operatorname{deg}(#1)}}
\newcommand{\poly}{\operatorname{poly}}
\newcommand{\sgn}[1]{\operatorname{sgn}(#1)}
\newcommand{\intp}{\operatorname{int}}
\newcommand{\fracp}{\operatorname{frac}}
\newcommand{\fiter}[2]{#1^{[#2]}}
\newcommand{\frestrict}[2]{#1\restriction_{#2}}
\newcommand{\indicator}[1]{\mathds{1}_{#1}}
\newcommand{\idfun}{\operatorname{id}}
\newcommand{\ceil}[1]{\left\lceil#1\right\rceil}
\newcommand{\round}[1]{\left\lfloor#1\right\rceil}

\newcommand{\myclass}[1]{\operatorname{#1}}
\newcommand{\gval}[2][]{\ensuremath{\myclass{GVAL}}}
\newcommand{\gpval}[1][]{\ensuremath{\myclass{GPVAL}}}
\newcommand{\gc}[3][]{\ensuremath{\myclass{AC}(#2,#3)}}
\newcommand{\gpc}[1][]{\ensuremath{\myclass{AP}}}
\newcommand{\gexpc}[1][]{\ensuremath{\myclass{AEXP}}}
\newcommand{\gwc}[2]{\ensuremath{\myclass{AW}(#1,#2)}}
\newcommand{\gpwc}{\ensuremath{\myclass{AWP}}}
\newcommand{\grc}[3]{\ensuremath{\myclass{AR}(#1,#2,#3)}}
\newcommand{\gprc}{\ensuremath{\myclass{ARP}}}
\newcommand{\gsc}[3]{\ensuremath{\myclass{AS}(#1,#2,#3)}}
\newcommand{\gpsc}{\ensuremath{\myclass{ASP}}}
\newcommand{\goc}[3]{\ensuremath{\myclass{AO}(#1,#2,#3)}}
\newcommand{\gpoc}{\ensuremath{\myclass{AOP}}}
\newcommand{\guc}[4]{\ensuremath{\myclass{AX}(#1,#2,#3,#4)}}
\newcommand{\gpuc}{\ensuremath{\myclass{AXP}}}

\newcommand{\gplc}[1][]{\ensuremath{\myclass{ALP}}}
\newcommand{\Unaware}{Extreme}
\newcommand{\unaware}{extreme}
\newcommand{\unawarely}{extremely}

\newcommand{\argmin}{\operatornamewithlimits{argmin}}

\newcommand{\jacobian}[1]{J_{#1}}
\newcommand{\grad}[1]{\nabla{#1}}
\newcommand{\scalarprod}[2]{{#1}\cdot{#2}}
\newcommand{\bigO}[1]{\mathcal{O}\left(#1\right)}
\newcommand{\softO}[1]{\tilde{\mathcal{O}}\left(#1\right)}
\newcommand{\taylor}[3]{{T_{#1}^{#2}#3}}
\newcommand{\transpose}[1]{{#1}^T}
\newcommand{\pastsup}[2]{{\sup}_{#1}#2}
\newcommand{\mtt}[1]{\mathtt{#1}}
\newcommand{\ovl}[1]{\overline{#1}}

\newcommand{\PIVP}{{PIVP}}
\newcommand{\PIVPmap}{\operatorname{PIVP}}
\newcommand{\myop}[1]{\operatorname{#1}}

\newcommand{\rnd}{\myop{rnd}}

\newcommand{\lxh}{\myop{lxh}}
\newcommand{\hxl}{\myop{hxl}}
\newcommand{\plil}{\myop{plil}}
\newcommand{\reach}{\myop{reach}}

\newcommand{\norm}{\myop{norm}}
\newcommand{\mx}{\myop{mx}}
\newcommand{\sample}{\myop{sample}}

\newcommand{\crnd}{\myop{rnd}^*}

\newcommand{\glen}[1]{\myop{len}_{#1}}
\newcommand{\mix}[3]{\myop{mix}(#1,#2,#3)}
\newcommand{\lagrange}[1]{L_{#1}}
\newcommand{\lagreq}[2]{D_{#1=#2}}
\newcommand{\lagrneq}[2]{D_{#1\neq #2}}
\newcommand{\emptyword}{\lambda}

\newcommand{\machcfg}[1]{\mathcal{C}_{#1}}
\newcommand{\machstep}[1]{#1}
\newcommand{\realenc}[1]{\left\langle#1\right\rangle}
\newcommand{\idealrealstep}[1]{\realenc{#1}_\infty}
\newcommand{\realstep}[1]{\realenc{#1}}

\title{Polynomial Time Corresponds to Solutions of Polynomial Ordinary
  Differential Equations of Polynomial Length. \newline
{\normalsize  The General Purpose Analog Computer and Computable Analysis
are two efficiently equivalent models of computations}\footnote{Daniel Gra\c{c}a
was partially supported by \emph{Funda\c{c}\~{a}o para a Ci\^{e}ncia e a Tecnologia} and EU FEDER
POCTI/POCI via SQIG - Instituto de Telecomunica\c{c}\~{o}es through the FCT project UID/EEA/50008/2013.}}
\titlerunning{Polynomial Time  Corresponds to Solutions of Polynomial ODEs of Polynomial Length}

\date{}

\author[1]{Olivier Bournez
}
\author[,2,3]{Daniel S. Gra\c{c}a
}
\author[1,2]{Amaury Pouly}
\affil[1]{Ecole Polytechnique, LIX, 91128 Palaiseau Cedex, France}
\affil[2]{CEDMES/FCT, Universidade do Algarve, C. Gambelas, 8005-139 Faro, Portugal}
\affil[3]{SQIG/Instituto de Telecomunica\c{c}\~{o}es, Lisbon,
  Portugal}


\begin{document}

\maketitle

\begin{abstract}
The outcomes of this paper are twofold.
\bigskip

\noindent\textbf{Implicit complexity.} 
  We provide an implicit characterization of polynomial time computation in terms of
  ordinary differential equations: we characterize  the class $\PTIME$
  of languages computable in polynomial time in terms
  of differential equations with polynomial right-hand side.

  This result gives a purely continuous (time and space) elegant and
  simple characterization of $\PTIME$. We believe it is the first time such classes
  are characterized using only ordinary differential equations. Our
  characterization extends to functions computable in polynomial time
  over the reals in the sense of computable analysis. 

  Our results may provide a new perspective on classical complexity, by
  giving a way to define complexity classes, like $\PTIME$, in a very simple
  way, without any reference to a notion of (discrete) machine. This
  may also provide ways to state classical questions about computational complexity 
  via ordinary differential equations.

\bigskip

\noindent\textbf{Continuous-Time Models of Computation.} 
  Our results can also be interpreted in terms of analog computers or
  analog model of computation: As a side
  effect, we get that the 1941 General Purpose Analog Computer (GPAC) of
  Claude Shannon is provably equivalent to Turing machines both at the
  computability and complexity level, a fact that has never been
  established before. This result provides arguments in favour of a generalised form of
the Church-Turing Hypothesis, which states that any physically realistic (macroscopic) computer is equivalent to Turing machines both at a computability and at a computational complexity level.
\end{abstract}


\section{Introduction}

The outcomes of this paper are twofold, and are concerning a priori
not closely related topics.

\smallskip

\noindent \textbf{Implicit Complexity:}
Since the introduction of the $\PTIME$ and $\NPTIME$ complexity classes, much work
has been done to build a well-developed complexity theory based on Turing
Machines. In particular, classical computational complexity theory is based on
limiting resources used by Turing machines, like time and space.
Another approach is implicit computational complexity. The term ``implicit'' in
``implicit computational complexity'' can
sometimes be understood in various ways, but a common point of 
these characterizations is that they provide (Turing or equivalent)
machine-independent alternative definitions of classical complexity. 

Implicit characterization
theory has gained enormous interest in the last decade.  This has led
to many alternative characterizations of complexity classes using
recursive functions, function algebras, rewriting systems, neural
networks, lambda calculus and so on. 

However, most of --- if not all --- these models or
characterizations are essentially discrete: in particular they are
based on underlying discrete time models working on objects which are
essentially discrete such as words, terms, etc. that can be considered
as being defined in a discrete space.

Models of computation working on a continuous space have also been
considered: they include Blum Shub Smale machines \cite{BCSS98}, and
in some sense Computable Analysis \cite{Wei00}, or quantum computers \cite{Fey82}
which usually feature discrete-time and
continuous-space. Machine-independent characterizations of the
corresponding complexity classes have also been devised: see
e.g. \cite{JLC04,GM95}. However, the resulting characterizations are
still essentially discrete, since time is still considered to be
discrete.

In this paper, we provide a purely analog machine-independent
characterization of the $\PTIME$ class. Our characterization
relies only on a simple and natural class of
ordinary differential equations: $\PTIME$ is characterized using ordinary
differential equations (ODEs) with polynomial right-hand side. This shows
first that (classical) complexity theory can be presented in terms of
ordinary differential equations problems. This opens the way to state
classical questions, such as $\PTIME$ vs $\NPTIME$, as questions about
ordinary differential equations.

\smallskip

\noindent\textbf{Analog Computers:} 
Our results can also be interpreted in the context of
analog models of computation and actually originate as a side effect
from an attempt to understand continuous-time analog models of
computation, and if they could solve some problem more efficiently
than classical models.  Refer to \cite{LivreAnalogcomputing} for a very
instructive historical account of the history of Analog computers. See
also \cite{maclennan2009analog,CIEChapter2007} for other discussions.

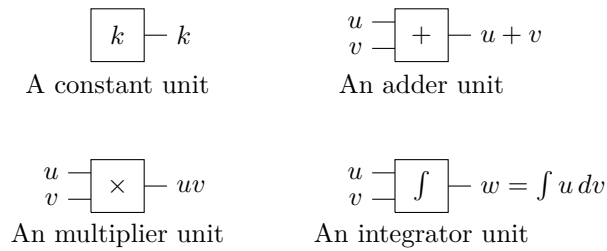
\begin{figure}[h]
\begin{center}
 \setlength{\unitlength}{1200sp}%
\begin{tikzpicture}
 \begin{scope}[shift={(0,0)},rotate=0]
  \draw (0,0) -- (0.7,0) -- (0.7,0.7) -- (0,0.7) -- (0,0);
  \node at (.35,.35) {$k$};
  \draw (.7,.35) -- (1,.35);
  \node[anchor=west] at (1,.35) {$k$};
  \node at (.35, -.3) {A constant unit};
 \end{scope}
 \begin{scope}[shift={(4,0)},rotate=0]
  \draw (0,0) -- (0.7,0) -- (0.7,0.7) -- (0,0.7) -- (0,0);
  \node at (.35,.35) {$+$};
  \draw (.7,.35) -- (1,.35); \draw (-.3,.175) -- (0,.175); \draw (-.3,.525) -- (0,.525);
  \node[anchor=west] at (1,.35) {$u+v$};
  \node at (.35, -.3) {An adder unit};
  \node[anchor=east] at (-.3,.525) {$u$};
  \node[anchor=east] at (-.3,.175) {$v$};
 \end{scope}
 \begin{scope}[shift={(0,-2)},rotate=0]
  \draw (0,0) -- (0.7,0) -- (0.7,0.7) -- (0,0.7) -- (0,0);
  \node at (.35,.35) {$\times$};
  \draw (.7,.35) -- (1,.35); \draw (-.3,.175) -- (0,.175); \draw (-.3,.525) -- (0,.525);
  \node[anchor=west] at (1,.35) {$uv$};
  \node at (.35, -.3) {An multiplier unit};
  \node[anchor=east] at (-.3,.525) {$u$};
  \node[anchor=east] at (-.3,.175) {$v$};
 \end{scope}
 \begin{scope}[shift={(4,-2)},rotate=0]
  \draw (0,0) -- (0.7,0) -- (0.7,0.7) -- (0,0.7) -- (0,0);
  \node at (.35,.35) {$\int$};
  \draw (.7,.35) -- (1,.35); \draw (-.3,.175) -- (0,.175); \draw (-.3,.525) -- (0,.525);
  \node[anchor=west] at (1,.35) {$w=\int u\thinspace dv$};
  \node at (.35, -.3) {An integrator unit};
  \node[anchor=east] at (-.3,.525) {$u$};
  \node[anchor=east] at (-.3,.175) {$v$};
 \end{scope}
\end{tikzpicture}
\end{center}
\caption{Circuit presentation of the GPAC: a circuit built from basic units}
\label{fig:gpac_circuit}
\end{figure}
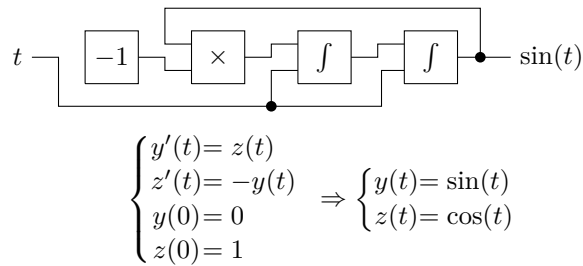
\begin{figure}[h]
\begin{center}
\begin{tikzpicture}
 \begin{scope}[shift={(-4.2,0)},rotate=0]
  \draw (0,0) -- (0.7,0) -- (0.7,0.7) -- (0,0.7) -- (0,0);
  \node at (.35,.35) {$-1$};
 \end{scope}
 \draw (-3.5,.35) -- (-3.15,.35) -- (-3.15,.175) -- (-2.8,.175);
 \begin{scope}[shift={(-2.8,0)},rotate=0]
  \draw (0,0) -- (0.7,0) -- (0.7,0.7) -- (0,0.7) -- (0,0);
  \node at (.35,.35) {$\times$};
 \end{scope}
 \draw (-2.1,.35) -- (-1.75,.35) -- (-1.75,.525) -- (-1.4,.525);
 \begin{scope}[shift={(-1.4,0)},rotate=0]
  \draw (0,0) -- (0.7,0) -- (0.7,0.7) -- (0,0.7) -- (0,0);
  \node at (.35,.35) {$\int$};
 \end{scope}
 \draw (-.7,.35) -- (-.35,.35) -- (-.35,.525) -- (0,.525);
 \begin{scope}[shift={(0,0)},rotate=0]
  \draw (0,0) -- (0.7,0) -- (0.7,0.7) -- (0,0.7) -- (0,0);
  \node at (.35,.35) {$\int$};
 \end{scope}
 \draw (.7,.35) -- (1.4,.35);
 \node[anchor=west] at (1.4,.35) {$\sin(t)$};
 \node[anchor=north] at (-1, -0.5) {$\left\lbrace
\begin{array}{@{}c@{}l}
y'(t)&=z(t)\\
z'(t)&=-y(t)\\
y(0)&=0\\
z(0)&=1
\end{array}
\right.\Rightarrow\left\lbrace\begin{array}{@{}c@{}l}
y(t)&=\sin(t)\\
z(t)&=\cos(t)
\end{array}\right.$};
 \draw (1,.35) -- (1,1) -- (-3.15,1) -- (-3.15,.525) -- (-2.8,.525);
 \fill (1,.35) circle[radius=.07];
 \draw (-4.9,.35) -- (-4.55,.35) -- (-4.55,-.3) -- (-.3,-.3) -- (-.3,.175) -- (0,.175);
 \draw (-1.75,-.3) -- (-1.75,.175) -- (-1.4,.175);
 \fill (-1.75,-.3) circle[radius=.07];
 \node[anchor=east] at (-4.9,.35) {$t$};
\end{tikzpicture}
\end{center}
\caption{Example of GPAC circuit: computing sine and cosine with two variables}
\label{fig:gpac_example_sin}
\end{figure}

Indeed, in 1941, Claude Shannon introduced in \cite{Sha41} the
General Purpose Analog Computer  (GPAC) model as a
model for the Differential Analyzer \cite{Bus31}, a mechanical programmable
machine, on which he worked
as an operator.
The GPAC model was later refined in \cite{Pou74}, \cite{GC03}.
Originally it was presented as a model based on circuits {(see Figure~\ref{fig:gpac_circuit})},
where several units performing basic operations (e.g.~sums,
integration) are interconnected {(see Figure~\ref{fig:gpac_example_sin}
)}.

However, Shannon himself realized that functions computed by a GPAC are nothing more
than solutions of a special class of polynomial differential
equations. In particular it can be shown that a function is computed
by Shannon's model if and only if it is a
(component of the) solution of an ordinary
differential equations (ODEs) with polynomial right-hand side \cite{Sha41}, \cite{GC03}.
In this paper, we consider the refined version presented in \cite{GC03}.

We note that the original model of the GPAC presented in \cite{Sha41}, \cite{GC03} is not equivalent to Turing machine based models. However, the original GPAC model performs computations in real-time: at time $t$ the output is $f(t)$, which different from the notion used by Turing machines. In \cite{Gra04} a new notion of computation for the GPAC, which uses ``converging computations'' as done by Turing machines was introduced and it was shown in \cite{BCGH06},\cite{BCGH07} that using this new notion of computation, the GPAC and computable analysis are two equivalent models of computation at a computability level.

In that sense, our paper extends this latter result and proves that the GPAC and computable analysis
are two equivalent models of computation, both at the computability
and at the complexity level. We also provide as a side effect a robust way to
measure time in the GPAC, or more generally in computations performed by
ordinary differential equations: basically, by considering the length
of the curve.

This paper is organized as follows. Section~\ref{sec:our_results} gives our
main definitions and results. Section~\ref{sec:discussion} discusses the related
work and consequences of our results. Section~\ref{sec:overwiew_proof} gives a very
high-level overview of the proof. It also contains more definitions and results
so that the reader can understand the big steps of the proof.

\section{Our Results}\label{sec:our_results}


We consider the following class of differential
equations:
\begin{equation}\label{eq:gpac}
y(0)=y_0\qquad y'(t)=p(y(t))
\end{equation}
where $y: I \to \R^d$ for some interval $I \subset \R$ and
where $p$ is a vector of polynomials.
Such systems are sometimes called \PIVP, for polynomial initial value
problems \cite{GBC09}. Observe that there is always a unique solution
 to the PIVP, which is analytic, defined on a maximum interval of life $I$ containing $y_0$, which we
refer to as ``the solution''. 




Our crucial and key idea is that, when using PIVPs to compute a function $f$,
the complexity should be measured as the length of the solution curve of the PIVP computing the function $f$.
We recall that the length of a curve $y\in C^1(I,\R^n)$ defined over
some interval $I=[a,b]$ is given by
$\glen{y}(a,b)=\int_I\infnorm{y'(t)}dt,$
where $\infnorm{y}$ refers to the infinite norm of $y$.

We assume the reader familiar with the notion of polynomial time
computable  function $f: [a,b] \to \R$ (see
\cite{Wei00} for an introduction to computable analysis). 
We take $\Rp=[0,+\infty[$ and denote by $\Rpoly$ the set of
polynomial time computable reals. 
For any vector $y$, $y_{i\ldots j}$ refers to the vector $(y_i, y_{i+1}, \ldots, y_j)$. For any sets $X$ and $Z$, $f:\subseteq X\rightarrow Z$
refers to any function $f:Y\rightarrow Z$ where $Y\subseteq X$ and $\dom{f}$ refers
to the domain of definition of $f$.

\begin{remark}[The space $\K$ of the coefficients]
In this paper, the coefficients of all considered polynomials will belong to $\K$.
Formally, $\K$ needs to a be \emph{generable field}, as introduced in \cite{TheseAmaury}.
However, without a significant loss of generality, the reader can
consider that $\K=\Rpoly$
which is the set of polynomial time computable real numbers. All the reader needs to
know about $\K$ is that it is a field and it is stable by generable functions
(introduced in Section~\ref{sec:turing}), meaning that if $\alpha\in\K$ and $f$
is generable then $f(\alpha)\in\K$.
It is shown in \cite{TheseAmaury} that there exists a small generable field $\Rgen$
lying somewhere between $\Q$ and $\Rpoly$, with expected strict inequality on both sides.
\end{remark}

Our main results (the class $\gpc$ is defined in Definition~\ref{def:glc},
and the notion of language recognized by a continuous system is given in Definition~\ref{def:discrete_rec})
are the following. Let us recall that $\PTIME(\R)$ is the class of polynomial time computable real
functions, as defined in \cite{Ko91}.

\begin{theorem}[An implicit characterization of $\PTIME(\R)$] \label{main:th}
Let $a,b\in\Rpoly$.
A function $f: [a,b] \to \R$ is computable in polynomial time iff its belongs
to the class $\gpc$.
\end{theorem}

\begin{theorem}[An implicit characterization of $\PTIME$]\label{th:p_gpac}
A decision problem (language) $\mathcal{L}$  belongs to class $\PTIME$
if and only if it is analog-recognizable.
\end{theorem}


\begin{definition}[Complexity Class $\gpc$]\label{def:glc}
We say that $f:\subseteq\R^n\rightarrow\R^m$ is in $\gpc$  if and only if there exists
a  vector $p$ of polynomials with $d\geqslant m$ variables and a
vector $q$ of polynomials with $n$ variables, both with coefficients
in $\K$, and a bivariate polynomial $\Omega$
such that for any $x\in\dom{f}$, there exists (a unique) $y:\Rp\rightarrow\R^d$ satisfying
for all $t\in\Rp$:
\begin{itemize}
\item $y(0)=q(x)$ and $y'(t)=p(y(t))$ \hfill$\blacktriangleright$ $y$ satisfies a PIVP
\item for all $\mu\in\Rp$, if $\glen{y}(0,t)\geqslant \Omega(\infnorm{x},\mu)$ then
$\infnorm{y_{1..m}(t)-f(x)}\leqslant e^{-\mu}$\hfill$\blacktriangleright$ $y_{1..m}$ converges\\
    \hphantom{}\hfill to $f(x)$
\item $\glen{y}(0,t)\geqslant t$
\hfill$\blacktriangleright$ technical condition: the length grows at least linearly with time\footnote{This is a
    technical condition required for the proof. This can be weakened,
    for example to $\infnorm{y'(t)}=\infnorm{p(y(t))}\geqslant\frac{1}{\poly(t)}$. The
    technical issue is that if the speed of the system
    becomes extremely small, it might take an exponential time to reach a
    polynomial length, and we want to avoid such ``unatural'' 
    cases. This is satisfied by all examples of computations we know \cite{LivreAnalogcomputing}.}
\end{itemize}
\end{definition}

Intuitively, a function f belongs to AP if there is a PIVP
that approximates f with a polynomial length to reach a given level of approximation. 

In definition~\ref{def:glc}, the PIVP was given its input $x$ as part of the initial condition:
this is very natural because $x$ was a real number. In the following, we will characterize languages with differential equations.
Since a language is made up of words, we need to discuss how to represent (encode) a word with a real number.
We fix a finite alphabet $\Gamma=\{0, .., k-2\}$ 
and define the encoding\footnote{Other encodings may be
  used, however, two crucial properties are necessary: (i) $\psi(w)$
  must provide a way to recover the length of the word, (ii)
  $\infnorm{\psi(w)}\approx\poly(|w|)$ in other words, the norm of the
  encoding is roughly the size of the word.}  $\psi(w)=\left(\sum_{i=1}^{|w|}w_ik^{-i},|w|\right)$ for a word $w=w_1w_2\dots w_{|w|}$. 

\begin{definition}[Analog recognizability]\label{def:discrete_rec}
A language $\mathcal{L}\subseteq\Gamma^*$  is called \emph{analog-recognizable} if there
exists a vector $q$ of bivariate polynomials and a vector $p$ of polynomials with $d$ variables,
both with coefficients in $\K$, and a polynomial $\Omega: \Rp \to \Rp$,
  such that for all
$w\in\Gamma^*$
there is a (unique) $y:\Rp\rightarrow\R^d$ such that for all $t\in\Rp$:
\begin{itemize}
\item $y(0)=q(\psi(w))$ and $y'(t)=p(y(t))$
\hfill$\blacktriangleright$ $y$ satisfies a differential equation
\item if $|y_1(t)|\geqslant1$ then 
$|y_1(u)|\geqslant1$ for all $u\geqslant t$
\hfill$\blacktriangleright$ the decision is stable
\item if $w\in\mathcal{L}$ (resp. $\notin\mathcal{L}$) and
  $\glen{y}(0,t)\geqslant \Omega(|w|)$
then $y_1(t)\geqslant1$ (resp. $\leqslant-1$)
\hfill$\blacktriangleright$ decision
\item $\glen{y}(0,t)\geqslant t$
\hfill$\blacktriangleright$ technical condition
\end{itemize}
\end{definition}

Intuitively this definition says that a language is analog-recognizable if
there is a PIVP such that, if the initial condition is set to be (the
encoding of) some word $w\in\Gamma^*$, then by using a portion of
\emph{polynomial length} of the curve, we are able to tell if this word should be accepted or rejected,
by watching to which region of the space the trajectory will go: the value of $y_1$ determines if the word has been accepted or not, or if the computation is still in progress.

\section{Discussion}\label{sec:discussion}

\newcommand\myparagraph[1]{\noindent \textbf{#1}:}
\newcommand\myparagraphd[1]{\noindent \textbf{#1}}

\myparagraph{Extensions} Our characterizations of the polynomial time can easily be extended to
characterizations of deterministic complexity classes above polynomial
time. For example, $\EXPTIME$ can be shown to correspond to the case where polynomial
$\Omega$ is replaced by some exponential function\APPENDIX{(see
Appendix~\ref{app:exptime})}. 
\smallskip

\begin{theorem} 
Let $a$ and $b$ in $\Rpoly$.
A function $f: [a,b] \to \R$ is computable in exponential time iff its belongs
to the class $f\in\gexpc$.
\end{theorem}



\begin{definition}[Definition of the complexity class $\gexpc$ for continuous systems]
We say that $f:\subseteq\R^n\rightarrow\R^m$ is in $\gexpc$  if and only if there exists
a  vector $p$ of polynomial functions with $d$ variables, a
vector $q$ of polynomial with $n$ variables, both with coefficients
in $\K$, an exponential function $\Omega: \Rp^2 \rightarrow\Rp$
such that for any $x\in\dom{f}$, there exists (a unique) $y:\Rp\rightarrow\R^d$ satisfying
for all $t\in\Rp$:
\begin{itemize}
\item $y(0)=q(x)$ and $y'(t)=p(y(t))$ for all $t\geqslant 0$\hfill$\blacktriangleright$ $y$ satisfies a PIVP
\item for any $\mu\in\Rp$, if $\glen{y}(0,t)\geqslant \Omega(\infnorm{x},\mu)$ then
$\infnorm{y_{1..m}(t)-f(x)}\leqslant e^{-\mu}$\hfill$\blacktriangleright$ $y_{1..m}$ converges
\item $\infnorm{y'(t)}\geqslant1$\hfill$\blacktriangleright$
technical
  condition: The
  length grows at least linearly with time\footnote{This is a
    technical condition required for the proof. This can be weakened,
    for example to $\infnorm{p(y(t))}\geqslant\frac{1}{\poly(t)}$. The
    technical issue is that  the speed of the system
becomes extremely small, it might take an exponential time to reach a
polynomial length, and we want to avoid such ``unatural'' cases.}
\end{itemize}
\end{definition}

\myparagraph{Applications to computational complexity}
We believe these characterizations to really
open  a new perspective on classical complexity, as we indeed provide a
natural definition (through previous definitions) of $\PTIME$ for
decision problems and of polynomial time for functions over the reals
using analysis only i.e.~ordinary differential equations and
polynomials, no need to talk about any (discrete) machinery like
Turing machines. 
This may open ways to characterize other
complexity classes like $\NP$ or $\PSPACE$. In the current settings of
course $\NP$ \label{dis:np} can be viewed as an existential quantification over our
definition\APPENDIX{(see Appendix~\ref{app:np})}, but we are obviously talking about ``natural''
characterizations, not involving unnatural quantifiers (for e.g.  a
concept of analysis like ordinary differential inclusions).

As a side effect, we also establish that solving
ordinary differential equations with polynomial right-hand side leads
to $\PTIME$- (or $\EXPTIME$-)complete problems, when the length of the solution curve is taken into
account. In an less formal way, this is stating that ordinary
differential equations can be solved by following
the solution curve (as most numerical analysis method do), but that for general (and even right-hand side polynomial)
ODEs, no better method can work, unless some famous complexity
questions do not hold. Note that our results only deal with ODEs with a polynomial right-hand side and that we do not know what happens for ODEs with analytic right-hand sides over unbounded domains. There are some results (see e.g.~\cite{MM93}) which show that ODEs with analytic right-hand sides can be computed locally in polynomial time. However these results do not apply to our setting since we need to compute the solution of ODEs over arbitrary large domains, and not only locally.


\myparagraph{Applications to continuous-time analog models}
\PIVP s are known to correspond to
functions that can be generated by the GPAC of Claude Shannon \cite{Sha41}. 

Defining a robust (time) complexity notion for continuous time systems is a
well known open problem \cite{CIEChapter2007} with no
generic solution provided to this day. In short, the difficulty is that the naive idea
of using the time variable of the ODE as measure of ``time complexity'' is problematic,
since time can be arbitrarily contracted in a continuous system due to the ``Zeno
phenomena'' (e.g.~by using functions like $\arctan$ which contract the
whole real line into a bounded set). It follows that all computable languages
can then be computed by a continuous system in time $O(1)$ (see e.g.~ \cite{Ruo93},
\cite{Ruo94}, \cite{Moo95b}, \cite{Bournez97}, \cite{Bou99},
\cite{AD91}, \cite{CP01}, \cite{Davies01}, \cite{Copeland98},
\cite{Cop02}). 

With that respect, we solve this open problem by stating that the ``time complexity''
should be measured by the length of the solution curve of the
ODE. Doing so, we get a robust notion of time complexity for PIVP
systems. 
 Indeed, the length is a geometric property of the
 curve and is thus ``invariant'' by rescaling. Notice that this is not sufficient
to get robustness: the fact that we restrict to \PIVP{} systems is crucial because more
general ODEs are usually hard to simulate (e.g. see \cite{Kaw10}).
This explains why all previous attempts of a general complexity for
general sytems failed in some sense
\cite{CIEChapter2007}. Super-Turing ``Zeno phenomena'' can still
happen with general ODEs, but not with \PIVP{s}.

\myparagraph{Applications to algorithms} We also believe that
transferring the notion of time complexity to a simple consideration
about length of curves allows for very elegant and nice proofs of
polynomiality of many methods for solving continuous but also discrete
problems. For example, the zero of a function $f$ can easily be computed
by considering the solution of $y'=- f(y)$ under reasonable hypotheses
on $f$. More interestingly, this may also covers many interior-point
methods or barrier methods where the problem can be transformed into
the optimization of some continuous function (see e.g.
\cite{karmarkar1984new,Ref6-BFFS03,BFFS03,kojima1991unified}). 

\myparagraphd{Related work}\APPENDIX{ is mainly discussed section by section,
with sometimes more details provided in Appendix
\ref{app:relatedwork}}. We believe no purely continuous-time definition of $\PTIME$ 
has ever been stated before. 
%
One direction of our characterization is based on a polynomial time algorithm (in the length of the curve)
to solve PIVPs over unbounded time domains, such a result strengthens all existings results on the complexity
of solving ODEs over unbounded time domains. In the
converse direction, our proof requires a way to simulate a Turing
machine using \PIVP{} systems with a polynomial length, a task whose
difficulty is discussed below, and still something that has never been
done up to date.

Attempts to derive a complexity theory for continous-time systems
include \cite{GM02}. However, the theory developped there is not
intended to cover generic dynamical systems but only specific systems
that are related to Lyapunov theory for dynamical systems. The global
minimizers of particular energy functions are supposed to give
solutions of the problem. The structure of such energy functions leads
to the introduction of problem classes $U$ and $NU$, with the
existence of complete problems for theses classes. 

Another attempt is \cite{BSF02}, also focussed on a very specific type
of systems: dissipative flow models. The proposed theory is nice but
non-generic. This theory has been used in several papers from the same
authors to study a particular class of flow dynamics \cite{BFFS03} for
solving linear programming problems.

Both approaches are not at all intended to cover generic ODEs, and
none of them is able to relate the obtained classes to classical
classes from computational complexity.

Up to our knowledge, the most up to date survey about continuous time
computation are \cite{CIEChapter2007,maclennan2009analog}. 

Relating computational complexity problems (like the $\PTIME$ vs $\NP$
question) to problems of analysis has already been the motivation of
series of works.  In particular, Félix Costa and Jerzy
Mycka have a series of work (see e.g. \cite{MC06}) relating the $\PTIME$ vs
$\NPTIME$ question to questions in the context of real and complex
analysis. 
Their approach is very different: they do so at the price of a whole
hierarchy of functions and operators over functions. In particular,
they can use multiple times an operator which solves ordinary
differential equations before defining an element of $DAnalog$ e
$NAnalog$ (the counterparts of $\PTIME$ and $\NPTIME$ introduced in their
paper), while in our case we do not need the multiple application of
this kind of operator: we only need to use \emph{one} application of
such operator (i.e.~we only need to solve one ordinary differential
equations with polynomial right-hand side).
%

We also mention that Friedman and Ko (see
\cite{Ko91}) proved that  polynomial time computable functions
are closed under maximization and integration if and only if some open
problems of computational complexity (like $\PTIME=\NP$ for the
maximization case) hold\APPENDIX{ ( see also the Appendix
\ref{app:relatedwork} for related work)}. The complexity of solving
Lipschitz continuous ordinary differential equation has been proved to
be polynomial-space complete by Kawamura \cite{Kaw10}.

All the results of this paper are fully developped in the PhD
thesis of Amaury Pouly \cite{TheseAmaury}.\APPENDIX{ For self-completeness, most proofs
are in appendix. Please
refer to \cite{TheseAmaury} for missing details. Results mentioned in
this paper have not yet been published, and are currently not
submitted\footnote{Preliminary results were submitted in the past but not with the
  strength of the current statements.},
with the exception of results on ODE solving (results of section~\ref{sec:computability}) but in a slightly
different and extended framework), which were very recently accepted
\cite{PoulyG16}.}

%

\section{Overview of the proof}\label{sec:overwiew_proof}

To show our main results (Theorem~\ref{main:th} and Theorem~\ref{th:p_gpac}),
we need to show two implications: (i) if a function $f:[a,b]\to\R$ (resp. a language $\mathcal{L}$) is polynomial time computable,
then it belongs to $\gpc$ (resp. it is analog-recognizable) and (ii) if a function $f:[a,b]\to\R$
belongs to $\gpc$ (resp. a language $\mathcal{L}$ is analog-recognizable) then it is polynomial time computable (resp. belongs to $\PTIME$).

The second implication (ii) is proved by computing the solution of a PIVP system
using some numerical algorithm. If a function $f:[a,b]\to\R$ in $\gpc$
can be computed (up to some given accuracy) by following the solution curve of its
associated ODE up to a reasonable (polynomial) amount of the length of the curve,
the numerical simulation of its associated ODE will use a reasonable (polynomial)
amount of resources to simulate this bounded portion of the solution curve.
Hence the function $f$ will be computed (up to some given accuracy, as usual in
Computable Analysis) by a Turing machine in polynomial time.
A similar idea can be used for showing the implication (ii) for $\PTIME$ and analog-recognizable languages.

The idea sketched above gives the intuition of the proof but the usual ODE solving algorithms cannot be used
here since (1) they are only guaranteed to compute the solution of an
ODE with a given accuracy over a bounded time domain, but here we need to
compute this solution over an unbounded time domain\footnote{Note that while $f$
has domain of definition $[a,b]$, from Definition~\ref{def:glc} $f$ is
approximated by a PIVP whose solution is defined over the unbounded
time domain $\R$} which introduce further complications 
and (2) we need polynomial complexity in the length of the curve, which is
not a classical measure of complexity. 

The first implication (i) is proved by simulating Turing machines with PIVPs and
by showing that these simulations can be performed by using a reasonable (polynomial)
amount of resources (length of the solution curve) if the Turing machine runs in polynomial time. 

Some simulation of Turing machines with PIVPs was already performed
e.g.~in \cite{BCGH07}, \cite{GCB08}. Basically one has to simulate the
behavior of a Turing machine with a continuous system. This is
problematic since Turing machines behave discretely (e.g.~``if $x$
happens then do $A$, otherwise do $B$'') and one only has access to
continuous (analytic) functions. This can be solved by
\emph{approximating} discontinuous functions with continuous functions
to obtain an approximation of the transition function of the Turing
machine. Then, by using special techniques, one can iterate the new
(now continuous) transition function to simulate the step-by-step
evolution of the Turing machine. Here we have one new difficult
problem to tackle (not covered in previous papers like \cite{BCGH07} and
\cite{GCB08}) because we must ensure that everything can be done
using only a reasonable (polynomial) amount of the length of the
solution curve of the PIVP. In particular, this constraint rules out particularly
simple techniques like integer encodings of the tape and error correction,
as used in the previously mentioned papers.

At a high level, our proof relies on considerations about (polynomial
length) \emph{ODE programming}: we prove that it is possible
to ``program'' with polynomial length ODE systems that keep some
variable fixed, do assignement, iterate some functions, compute limits, etc. 
We use those basic operations and basic functions with
PIVPs (e.g. $\min, \max$, continuous approximation of rounding, etc.)
to create more complex functions and operations that simulate the transition function of a given Turing
machine and its iterations. To be sure that the more complex functions still
satisfy all the properties we want (e.g.~that they belong to $\gpc$),
we prove several closure properties: in particular, we prove very
strong and elegant equivalent definitions of class $\gpc$.

For reasons of lack of space,
we do not detail all these operators and functions, but we sketch the proof of
a few properties and some key ideas of our techniques.
%
%
We use the following notation: when $p$ is a
polynomial, $\sigmap{p}$ is the sum of the absolute values of its
coefficients and $\degp{p}$ its degree. If $p$ is a vector of polynomials,
we extend those notions by taking the maximum for each component.

\subsection{Polytime analog computability implies polytime computability}
\label{sec:computability} 

We start by sketching the proof of the ``only if'' direction of Theorem~\ref{th:p_gpac}, and
then of Theorem~\ref{main:th}. Recall that a real function is polynomial time computable
if given arbitrary approximations of the input, we can produce arbitrary approximations
of the output in polynomial time. As it is customary, we proceed in two steps.
We first show that the function has a polynomial modulus of continuity. This allows us
to restrict the problem to rational inputs of controlled size.

\begin{theorem}[Modulus of continuity\APPENDIX{, Appendix~\ref{appendix:th:comp_implies_cont}}]\label{th:comp_implies_cont}
If $f\in\gpc$, then $f$ admits a polynomial modulus of continuity: there exists
a polynomial $\mho:\Rp^2\rightarrow\Rp$ such that for all $x,y\in\dom{f}$ and $\mu\in\Rp$:
\[\infnorm{x-y}\leqslant e^{-\mho(\infnorm{x},\mu)}\quad\Rightarrow\quad\infnorm{f(x)-f(y)}\leqslant e^{-\mu}.\]
\end{theorem}

We then show that the solution of a such a PIVP can be approximated in polynomial time.
For this, will need the following theorem to get the complexity
of numerically solving this PIVP. The idea of the proof is detailled below.

\begin{theorem}[Complexity of Solving
  \PIVP \cite{PoulyG16}]\label{th:pivp_comp_analysis}
If~\footnote{The existence of a solution $y$ up to a given time is undecidable \cite{GBC07} so we have to assume existence.}
$y:\R\rightarrow\R^d$ satisfies for all $t\geqslant 0$.
\begin{equation}\label{eq:ode}
y(0)=y_0\qquad y'(t)=p(y(t)).
\end{equation}
Then $y(t)$ can be computed with precision $2^{-\mu}$ in time bounded by
\begin{equation}\label{eq:pivp_comp_analysis_bound}
\poly(\degp{p},\glen{y}(0,t),\log\infnorm{y_0},\log\sigmap{p},\mu)^d.
\end{equation}
More precisely, there exists a Turing machine $\mathcal{M}$ such that for any oracle
$\mathcal{O}$ representing\footnote{See \cite{Ko91} for more details. In short, the machine can ask arbitrary approximation
of $y_0, p$ and $t$ to the oracle. The polynomial is represented by the finite list of coefficients.} $(y_0,p,t)$ and any $\mu\in\N$,
$\infnorm{\mathcal{M}^\mathcal{O}(\mu)-\PIVPmap(y_0,p,t)}\leqslant2^{-\mu}$
if $y(t)$ exists, and the number of steps of the machine is bounded by \eqref{eq:pivp_comp_analysis_bound}
for all such oracles.
\end{theorem}

\myparagraph{General Idea}
Assume that $\mathcal{L}$ is analog-recognizable in the sense of Definition
\ref{def:discrete_rec}, using corresponding notations
$d,q,p,\Omega$. 
  Let
$w\in\Gamma^*$ and consider the following system:
$y(0)=q(\psi(w))$, $y'(t)=p(y(t))$. 
We show that we can decide in time polynomial in $|w|$  whether
$w\in\mathcal{L}$ or not. Theorem~\ref{th:pivp_comp_analysis} can be used to conclude that we can compute $y(t)\pm e^{-\mu}$ in time
polynomial in $\log\infnorm{q(\psi(w))},\mu$ and $\glen{y}(0,t)$. Recall that $\infnorm{\psi(w)}=|w|$
and that the system is guaranteed to give an answer as soon as $\glen{y}(0,t)\geqslant\Omega(|w|)$.
This means that it is enough to compute $y(t^*)$, where $t^*$ satisfies $\glen{y}(0,t^*)\geqslant\Omega(|w|)$,
with precision $1/2$ to distinguish between $y_1(t)\geqslant1$ and $y_1(t)\leqslant-1$.
Since $\glen{y}(0,t)\geqslant t$, thanks to the technical condition of the definition, we know that we can find a $t^*\leqslant\Omega(|w|)$.
Note that $\glen{y}(0,\Omega(|w|))$ might not be polynomial in $|w|$ so we cannot
simply compute $y(\Omega(|w|))$.

Fortunately, the proof of Theorem~\ref{th:pivp_comp_analysis} provides us with an
algorithm that solves the PIVP by making small time steps, and at each step the length cannot increase by more
than a constant. This means that we can run algorithm to compute $y(\Omega(|w|))$ and
stop it as soon as the length is greater than $\Omega(|w|)$. Let $t^*$ be the time at which
the algorithm stops. Then the running time of the algorithm will be polynomial in
$t^*,\mu$ and $\glen{y}(0,t^*)\leqslant\Omega(|w|)+\bigO{1}$. Finally, thanks to the
technical condition, $t^*\leqslant\glen{y}(0,t^*)$, this algorithm has
running time polynomial in $|w|$. 

The proof of Theorem~\ref{main:th}\APPENDIX{ (Appendix
\ref{appendix:main:th})} is established using
the same principle based on Theorem
\ref{th:pivp_comp_analysis}, observing in addition 
that functions in $\gpc$ can easily be approximated by considering only
their value on  rationals, since they have a polynomial modulus of
continuity, as shown by the following theorem.

It thus appears that the true remaining difficulty lies in proving Theorem~\ref{th:pivp_comp_analysis}.
An important point is that none of the classical methods for solving
ordinary differential equations are polynomial time over unbounded time domains. Indeed,
no method of fixed order $r$ is polynomial in variable $t$ over the whole domain
$\R$.\footnote{This is
  why most studies restricts to a compact domain.} For more information, we refer
the reader to \cite{PoulyG16}.

\begin{remark}
Observe that the solution of the following \PIVP{ }
 $y'_1=y_1,  y'_2=y_1y_2, y'_3=y_2y_3, 
 \dots,  y'_n=y_{n-1} y_n$
is a tower of $n$ exponentials. Its solution can be computed in polynomial time over any fixed compact
$[a,b]$ \cite{MM93}. However, the solution cannot be computed
in polynomial time over $\R$, as just writing this value in binary cannot
ever been done in polynomial time. Hence, the solution of a \PIVP{}  cannot
  be computed in polynomial time, over $\R$, in the
  general case. A key feature of our
  method is that we are searching methods polynomial 
  in the
  length of the curve, which is not a classical framework.
\end{remark}

\subsection{Polytime computability implies polytime analog computability}
\label{sec:turing}

The idea of the proof of the ``if'' directions is to simulate a Turing
machine using a \PIVP. But this is far from trivial since we need
to do it with a polynomial length. 

\smallskip
\myparagraph{About generable functions} 
The following concept can be attributed to \cite{Sha41}: a function $f: \R \to \R$ is said to be a \PIVP{} function if there
 exists a system of the form \eqref{eq:gpac}  with $f(t)=y_1(t)$ for all $t$, where $y_1$
denotes first component of the vector $y$ defined in $\R^d$. 
We need in our proof to extend the concept to talk about (i)
multivariable functions and (ii) the growth of these functions. The following class and
closure properties can be seen as extensions of results from \cite{GBC09}. 

\begin{definition}[Polynomially bounded generable function]\label{def:gpac_generable_ext}
Let $d,e\in\N$, $I$ be an open and connected subset of $\R^d$
and $f:I\rightarrow\R^e$. We say that $f\in\gpval$ if and only if
there exists a polynomial  $\mtt{sp}:\R\rightarrow\Rp$, 
$n\geqslant e$, a $n\times d$ matrix $p$ consisting of polynomials with coefficients in $\K$
, $x_0\in\K^d$, $y_0\in\K^n$
and $y:I\rightarrow\R^n$ satisfying for all $x\in I$:
\begin{itemize}
\item $y(x_0)=y_0$ and $\jacobian{y}(x)=p(y(x))$ 
  \hfill$\blacktriangleright$ $y$ satisfies a differential
  equation\footnote{$\jacobian{y}$ denotes the Jacobian matrix of $y$.}
\item $f(x)=y_{1..e}(x)$\hfill$\blacktriangleright$ $f$ is a component of $y$
\item $\infnorm{y(x)}\leqslant
  \mtt{sp}(\infnorm{x})$\hfill$\blacktriangleright$ $y$ is
  polynomially bounded
\end{itemize}
\end{definition}

\begin{lemma}[Closure properties of $\gpval$\APPENDIX{, Appendix~\ref{app:closure}}] \label{lemma:closure}  Let 
$f:\subseteq\R^d\rightarrow\R^n\in \gpval$ and
$g:\subseteq\R^e\rightarrow\R^m\in\gpval$. Then $f+g$, $f-g$, $fg$ and
$f\circ g$ are in $\gpval$.
\end{lemma}
\begin{lemma}[Generable functions are closed under ODE\APPENDIX{, Appendix~\ref{app:cor:gpac_ext_ivp_stable}}]\label{cor:gpac_ext_ivp_stable}
Let $d\in\N$, $J\subseteq\R$ an interval, 
$f:\subseteq\R^d\rightarrow\R^d$ in \gpval, $t_0\in\K\cap J$ and $y_0\in\K^d\cap\dom{f}$.
Assume there exists $y:J\rightarrow\dom{f}$, and a polynomial
$\ovl{\mtt{sp}}:\Rp\rightarrow\Rp$ satisfying for all $t\in J$:
\[y(t_0)=y_0\qquad y'(t)=f(y(t))
\qquad\infnorm{y(t)}\leqslant\ovl{\mtt{sp}}(t)\]
Then $y\in\gpval$ and it is unique.
\end{lemma}

It follows that many 
polynomially bounded usual analytic\footnote{Functions from $\gpval$ are necessarily
analytic, as solutions of an analytic ODE are analytic.} functions are in the class
$\gpval$. The inclusion $\gpval \subset \gpc$ holds for functions whose domain is simple enough\APPENDIX{\footnote{For
example star domains with a rational vantage point.} (Appendix
\ref{app:stardomain})}.  However,  the inclusion $\gpval
\subset \gpc$ is strict\footnote{Even with functions
  with star domains with a vantage point.}, since
functions like the inverse of the Gamma function
$\Gamma(x)=\int_{0}%
^{\infty}t^{x-1}e^{-t}dt$ or Riemann's Zeta function
$\zeta(x)=\sum_{k=0}^\infty \frac1{k^x}$ are not differentially
algebraic \cite{Sha41} but belong to $\gpc$.

\smallskip
 \myparagraph{Robustness of AP} A 
very strong key argument of our proof is that the notion of
computability given by Definition~\ref{def:glc} is actually very
robust and can be stated in many equivalent ways. A key point is that the definition
can be weakened and strengthened. The following theorem shows that we weaken the
definition without changing the class. Since it might not be obvious to the reader,
we emphasize that this notion is a priori weaker (thus \gpc{} is a priori larger than \gpwc{}). Indeed, (i) the system accepts
errors in the input (ii) the system does not even converge, but merely approximates
the output, doing the best it can given the input error.

\begin{theorem}[Weak Computability]  \label{th:weak} $\gpc=\gpwc$ where $\gpwc$ corresponds to the class of functions 
$f:\subseteq\R^n\rightarrow\R^m$ such that there are some polynomials $\Omega:\Rp^2\rightarrow\Rp$
and $\Upsilon:\Rp^3\rightarrow\Rp$,  $d\in\N$,
$p,q \in \gpval$, 
such that for any $x\in\dom{f}$ and $\mu\in\Rp$, there exists (a unique) $y:\Rp\rightarrow\R^d$ satisfying
for all $t\in\Rp$:
\begin{itemize} 
\item $y(0)=q(x,\mu)$ and $y'(t)=p(y(t))$\hfill$\blacktriangleright$ $y$ satisfies a PIVP
\item if $t\geqslant\Omega(\infnorm{x},\mu)$ then $\infnorm{y_{1..m}(t)-f(x)}\leqslant e^{-\mu}$
\hfill$\blacktriangleright$ $y_{1..m}$ approximates $f(x)$ within $e^{-\mu}$
\item $\infnorm{y(t)}\leqslant\Upsilon(\infnorm{x},\mu,t)$
\hfill$\blacktriangleright$ $y(t)$ is polynomially bounded
\end{itemize}
\end{theorem}

The proof of Theorem~\ref{th:weak}, however, is quite involved: first
$p$ and $q$ can be equivalently assumed to be polynomials instead of functions in
$\gpval$ above, from Lemma~\ref{cor:gpac_ext_ivp_stable}. Then 
$\gpc \subset \gpwc$, follows from the fact that
this is possible to rescale the system using the length of the curve
as a new variable to make sure it does not grow faster than a
polynomial 
 time, we get what is needed\APPENDIX{ 
(Appendix~\ref{app:gpcsubsetgpwc})}. 
The other direction ($\gpwc \subset \gpc$) is really harder: 
the first
step is to transform a computation into a computation that tolerates
small perturbations of the dynamics ($\gpwc \subset \gprc$\APPENDIX{, Appendix
\ref{app:weak_to_robust}}). The second problem is to avoid that the system
explodes for inputs not in the domain of the function, or for too big perturbation of the dynamics
perturbations on inputs ($\gprc \subset \gpsc$\APPENDIX{, Appendix
\ref{app:robust_to_strong}}). As a third step, we allow the system to have its inputs (input and precision) changed
during the computation and the system has a maximum delay to react to these changes
($\gpsc \subset \gpuc$\APPENDIX{, Appendix~\ref{app:strong_to_unaware}}). 
Finally, as a fourth step, we add a mechanism that feeds the
system with the input and some precision. By continuously increasing
the precision with time, we ensure that the system will converge when
the input is stable. The result of these 4 steps is the following
lemma, yielding a
nice notion of online-computation ($\gpuc \subset \gpoc$\APPENDIX{, 
Appendix~\ref{app:unaware_to_online}}). Equality
$\gpc=\gpwc=\gpoc$ follows because time and length are related for polynomially bounded systems.
The notion of online computability is an example of a priori strengthening of our notion of computation;
yet it still corresponds to the same class of function. Intuitively, a function is online
computable if, on any (long enough) time interval where the input is almost constant,
the system converges (after some delay) the output of the function. Of course, the output
will have some error that is related to the input error (due to the input not being
exactly constant).

\begin{lemma}[Online computability] \label{lemma:online} $\gpwc \subset \gpoc$, where
  $\gpoc$ corresponds to the class of functions
  $f:\subseteq\R^n\rightarrow\R^m$
such that for polynomials
$\Upsilon,\Omega,\Lambda:\Rp^2\rightarrow\Rp$, there exists $\delta\geqslant0$, $d\in\N$ and
$p\in\K^d[\R^d\times\R^{n}]$ and $y_0\in\K^d$ such that for any $x\in C^0(\Rp,\R^n)$,
there exists (a unique) $y:\Rp\rightarrow\R^d$ satisfying for all $t\in\Rp$:
\begin{itemize} 
\item $y(0)=y_0$ and $y'(t)=p(y(t),x(t))$
\item $\infnorm{y(t)}\leqslant\Upsilon\big(\sup_{u\in[t-\delta,t]}\infnorm{x(u)},t\big)$
\item For any $I=[a,b]$, if there exists $\bar{x}\in\dom{f}$ and $\bar{\mu}\geqslant0$ such that
for all $t\in I$, $\infnorm{x(t)-\bar{x}}\leqslant e^{-\Lambda(\infnorm{\bar{x}},\bar{\mu})}$ then
$\infnorm{y_{1..m}(u)-f(\bar{x})}\leqslant e^{-\bar{\mu}}$ whenever
$a+\Omega(\infnorm{\bar{x}},\bar{\mu})\leqslant u\leqslant b$.
\end{itemize}
\end{lemma}

\myparagraph{ODE Programming} With the closure properties of $\gpc$,
programming with (polynomial length) ODE becomes a rather pleasant
exercise, once the logic is understood. For example, simulating the
assignement $y:=g_\infty$ corresponds to dynamics $y(0)=y_0$,
$y'(t)=\reach(\phi(t),y(t),g(t))+E(t)$, for a fixed function
$\reach\in\gpval$, tolerating bounded error $E(t)$ on dynamics, and
$g$ fluctuating around $g_\infty$\APPENDIX{ (Lemma
\ref{lem:reach},}\APPENDIX{ Appendix~\ref{app:unaware_to_online})}.  Other
example: from a $\gpc$ system computing $f$, just adding the
corresponding $\gpoc$-equations for $g$, yields a \PIVP~computing $g \circ f$\APPENDIX{(Lemma
\ref{th:gpac_comp_composition}, Appendix
\ref{app:th:real_setp_robust})}, by feeding output of the system computing $f$ to the (online)
input of $g$.

\bigskip
\myparagraph{Turing machines} Consider a 
Turing machine $\mathcal{M}=(Q,\Sigma,b,\delta,q_0,q_\infty)$. A
(instantaneous) {configuration} of $M$ can be seen as a tuple
$c=(x,\sigma,y,q)$ where $x\in\Sigma^*$ is the part of the tape at
left of the head, $y\in\Sigma^*$ is the part at the right,
$\sigma\in\Sigma$ is the symbol under the head and $q\in Q$ the
current state.  
Let $\machcfg{\mathcal{M}}$ be the set of configurations
of $\mathcal{M}$, and $\mathcal{M}$ denotes the function mapping a
configuration to its  next configuration. 
In order to simulate a machine, we encode
configurations with real numbers as follows.
Recall that $\Gamma=\{0,1,\dots,k-2\}$ and let 
$\realenc{c}=(0.x,\sigma,0.y,q)\in\Q\times\Sigma\times\Q\times Q$
where $0.x=x_1k^{-1}+x_2k^{-2}+\cdots+x_{|x|}k^{-|x|}\in\Q$ with 
$x=x_1x_2\dots x_{|x|}$.

\begin{theorem}[Robust Real Step\APPENDIX{, Appendix~\ref{app:th:real_setp_robust}}]\label{th:real_setp_robust}
For any machine $\mathcal{M}$, there is some function
$\realstep{\mathcal{M}}\in\gpc$ such that for all 
 $c\in\machcfg{\mathcal{M}}$, $\mu\in\Rp$
and $\bar{c}\in\R^4$, if $\infnorm{\realenc{c}-\bar{c}}\leqslant\frac{1}{2\cramped{k^2}}-e^{-\mu}$ then
$\infnorm{\realstep{\mathcal{M}}(\bar{c},\mu)-\realenc{\machstep{\mathcal{M}}(c)}}
\leqslant k\infnorm{\realenc{c}-\bar{c}}$. 
\end{theorem}

\vspace{-0.2cm}

The difficulty of the proof is that one step of Turing machine
with our encoding naturally involves computing the integer and fractional parts
of a number. These operations are discontinuous and thus cannot be done in $\gpc$
in full generality. This is solved by proving that a continuous and good enough
``fractional part'' like-function is in AP (and avoids constructions from \cite{GBC09}). 

\smallskip \myparagraph{Iterating Functions} A key point for proving the main result is to show
that it is possible to iterate a function using a \PIVP{} under some
specific hypotheses.
The proof consists in building by ODE programming an ordinary differential equation using
three variables $y$, $z$ and $w$ updating in a cycle to be repeated $n$
times. At all time, $y$ is an online
component of the system computing $f(w)$. During the first stage of the cycle, $w$
stays still and $y$ converges to $f(w)$.  During the second stage of
the cycle, $z$ copies $y$ while $w$ stays still. During the last
stage, $w$ copies $z$ thus effectively computing one iterate. This computes all the iterates $f(x),f^{[2]}(x),\ldots$.
The crucial point of this process is the error estimation, to guarantee that the system does not diverge,
while keeping polynomial length. One of the key assumption to ensure this is for $f$
to admit a specific kind of modulus of continuity. The other key assumption is an effective
``openness'' of the iteration domain.

\begin{theorem}[Closure by iteration\APPENDIX{, Appendix~\ref{app:th:gpac_comp_iter}}]\label{th:gpac_comp_iter}
Let $I\subseteq\R^m$, $(f:I\rightarrow \R^m)\in\gpc$, $\eta\in\left[0,1/2\right[$ and
assume that there exists a family of subsets $I_n\subseteq I$, for all $n\in\N$ and polynomials
$\mho:\Rp\rightarrow\Rp$ and $\Pi:\Rp^2\rightarrow\Rp$ such that:
\begin{itemize}
\item for all $n\in\N$,  $I_{n+1}\subseteq I_n$ and $f(I_{n+1})\subseteq I_n$
\item for all $x\in\ I_n$,
$\infnorm{f^{[n]}(x)}\leqslant\Pi(\infnorm{x},n)$
\item for all $x\in I_n$, $y\in\R^m, \mu\in\Rp$, if
    $\infnorm{x-y}\leqslant e^{-\mho(\infnorm{x})-\mu}$
    then $y\in I$ and $\infnorm{f(x)-f(y)}\leqslant e^{-\mu}$.
\end{itemize}
Define $f_\eta^*(x,u)=f^{[n]}(x)$ for $x\in I_n$, $u\in[n-\eta,n+\eta]$ and $n\in\N$. Then
$f_\eta^*\in\gpc$.
\end{theorem}

The iteration of the (transition) functions given by
Theorem~\ref{th:real_setp_robust} leads to a way to emulate
any function computable in polynomial time. 

At a high level, the ``if'' direction of Theorem~\ref{th:p_gpac} then
follows. Indeed\APPENDIX{ (Appendix~\ref{app:th:p_gpac})}, decidability can be seen as the computability of some
particular function with boolean output.

For the ``if''
direction of Theorem~\ref{main:th}\APPENDIX{ (Appendix~\ref{app:th:gpac_comp_iter})}, there
are further nontrivial obstacles to overcome. Given $x\in[a,b]$ and $\mu\in\N$,
we want to compute an approximation of $f(x)\pm 2^{-\mu}$ and take the
limit when $\mu\rightarrow\infty$. 
To compute $f$, we will use a polynomial time computable function $g$
that computes $f$ over rationals, and $m$ a modulus of continuity. All
we have to do is simulate $g$ with input $\tilde{x}$ and $\mu$, where
$\tilde{x}=x\pm2^{-m(\mu)}$ because we can only feed the machine with
a finite input of course.  The remaining nontrivial part of the proof is how to
obtain the encoding of $\tilde{x}$ from $x$ and $\mu$. Indeed, the
encoding is a discrete quantity whereas $x$ is real number, so by a
simple continuity argument, one can see that no such function can
exist.  The trick is the following: from $x$ and $\mu$, we can compute
two encodings $\psi_1$ and $\psi_2$ such that at least one of them is
valid, and we know which one it is. So we are going to simulate $g$ on
both inputs and then select the result. Again, the select operation
cannot be done continuously unless we agree to ``mix'' both results,
i.e. we will compute $\alpha g(\psi_1)+(1-\alpha) g(\psi_2)$. The
trick is to ensure that $\alpha=1$ or $0$ when only one encoding is
valid, $\alpha\in]0,1[$ when both are valid (by ``when'' we mean with
respect to $x$). This way, a mixing of both will ensure continuity but
in fact when both encodings are valid, the outputs are nearly the same
so we are still computing $f$. Obtaining such encodings $\psi_1$ and $\psi_2$ is
also nontrivial and requires more uses of the closure by iteration property.


\bibliography{extracted}

\NOAPPENDIX{\end{document}}
\newpage

\newpage
\appendix

\section{Table of Contents}

\tableofcontents
\newpage

\section{Complements on Related Works}
\label{app:relatedwork}

Attempts to derive a complexity theory for continous-time systems
include \cite{GM02}: However, the theory developped there is not
intended to cover generic dynamical systems but only specific systems
that are related to Lyapunov theory for dynamical systems: The global
minimizers of particular energy functions are supposed to give
solutions of the problem. The structure of such energy functions leads
to the introduction of problem classes $U$ and $NU$, with the
existence of complete problems for theses classes. 

Another attempt is \cite{BSF02}, also focussed on a very specific type
of systems: dissipative flow models. The proposed theory is nice but
non-generic. This theory has been used in several papers from same
authors to study a particular class of flow dynamics \cite{BFFS03} for
solving linear programming problems.

Both approaches are not at all intended to cover generic ODEs, and
none of them is able to relate the obtained classes to classical
classes from computational complexity.

Up to our knowledge, the most up to date survey about continuous time
computation is \cite{CIEChapter2007}. 

Relating computational complexity problems (like the $\PTIME$ vs $\NP$
question) to problems of analysis has already been the motivation of
series of works:  In particular, Felix Costa and Jerzy
Mycka have a series of work (see e.g. \cite{MC06}) relating the $\PTIME$ vs
$\NPTIME$ question to questions in the context of real and complex
analysis. 

We give some arguments here, in case this is needed, to state that
their approach is very different: they do so at the price of a whole
hierarchy of functions and operators over functions. In particular,
they can use multiple times an operator which solves ordinary
differential equations before defining an element of $DAnalog$ e
$NAnalog$ (the counterparts of $\PTIME$ and $\NPTIME$ introduced in their
paper), while in our case we do not need the multiple application of
this kind of operator: we only need to use \emph{one} application of
such operator (i.e.~we only need to solve one ordinary differential
equations with polynomial right-hand side).

It its true that one can sometimes convert the
multiple use of operators solving ordinary differential equations into
a single application \cite{GC03}, but this happens only in very
specific cases, which do not seem to include the classes $DAnalog$ e
$NAnalog$. In particular, the application of nested continuous
recursion (i.e. nested use of solving ordinary differential equations)
may be needed using their constructions, whereas we define $\PTIME$ using only a simple notion of acceptance and only \emph{one}
system of ordinary differential equations.

\section{Some Formal Statements About Facts Mentioned in the Discussion}

\subsection{A Characterization of $\EXPTIME$}
\label{app:exptime}

\begin{theorem} 
Let $a$ and $b$ in $\Rpoly$.
A function $f: [a,b] \to \R$ is computable in exponential time iff its belongs
to the class $f\in\gexpc$.
\end{theorem}

\begin{theorem}[An implicit characterization of $\PTIME$]
Let $\mathcal{L}$ be any decision problem (language).

$\mathcal{L}\in\PTIME$ if and only if $\mathcal{L}$ is
exponential-length analog-recognizable.
\end{theorem}

\begin{definition}[Definition of the complexity class $\gexpc$ for continuous systems]
We say that $f:\subseteq\R^n\rightarrow\R^m$ is in $\gexpc$  if and only if there exists
a  vector $p$ of polynomial functions with $d$ variables, a
vector $q$ of polynomial with $n$ variables, both with coefficients
in $\K$, an exponential function $\Omega: \Rp^2 \rightarrow\Rp$
such that for any $x\in\dom{f}$, there exists (a unique) $y:\Rp\rightarrow\R^d$ satisfying
for all $t\in\Rp$:
\begin{itemize}
\item $y(0)=q(x)$ and $y'(t)=p(y(t))$ for all $t\geqslant 0$\hfill$\blacktriangleright$ $y$ satisfies a PIVP
\item for any $\mu\in\Rp$, if $\glen{y}(0,t)\geqslant \Omega(\infnorm{x},\mu)$ then
$\infnorm{y_{1..m}(t)-f(x)}\leqslant e^{-\mu}$\hfill$\blacktriangleright$ $y_{1..m}$ converges
\item $\infnorm{y'(t)}\geqslant1$\hfill$\blacktriangleright$
technical
  condition: The
  length grows at least linearly with time\footnote{This is a
    technical condition required for the proof. This can be weakened,
    for example to $\infnorm{p(y(t))}\geqslant\frac{1}{\poly(t)}$. The
    technical issue is that  the speed of the system
becomes extremely small, it might take an exponential time to reach a
polynomial length, and we want to avoid such ``unatural'' cases.}
\end{itemize}
\end{definition}

\begin{definition}[Discrete recognizability] 
A language $\mathcal{L}\subseteq\Gamma^*$  is called \emph{exponential-length analog-recognizable} if there
exists a vector $q$ of polynomials 
with two variables,
a vector $p$ of polynomials with $d$ variables, both with coefficients
in $\Rpoly$, and an exponential
function $\Omega: \Rp \to \Rp$,
  such that for all
$w\in\Gamma^*$
there is a (unique) $y:\Rp\rightarrow\R^d$ such that for all $t\in\Rp$:
\begin{itemize}
\item $y(0)=q(\psi(w))$ and $y'(t)=p(y(t))$
\hfill$\blacktriangleright$ $y$ satisfies a differential equation
\item if $|y_1(t)|\geqslant1$ then 
$|y_1(u)|\geqslant1$ for all $u\geqslant t$
\hfill$\blacktriangleright$ the decision is stable
\item if $w\in\mathcal{L}$ (resp. $\notin\mathcal{L}$) and
  $\glen{y}(0,t)\geqslant \Omega(|w|)$
then $y_1(t)\geqslant1$ (resp. $\leqslant-1$)
\hfill$\blacktriangleright$ decision
\item $\glen{y}(0,t)\geqslant t$
\hfill$\blacktriangleright$ technical condition\footnote{Same remarks
  as above.}
\end{itemize}
\end{definition}

\subsection{A (Too simple) Characterization of $\NP$}
\label{app:np}

Following the discussion page \pageref{dis:np}, here is a trivial way
to get a characterization of $\NP$.



\begin{definition}[Discrete $\NP$-recognizability] 
A language $\mathcal{L}\subseteq\Gamma^*$  is called \emph{ $\NP$-analog-recognizable} if there
exists a vector $q$ of polynomials 
in
with two variables,
a vector $p$ of polynomials with $d$ variables, both with coefficients
in $\Rpoly$, and a polynomial $\Omega: \Rp \to \Rp$,
  such that for all
$w\in\Gamma^*$, for some $s \in \Gamma^*$ of size polynomial in
$|w|$, 
there is a (unique) $y:\Rp\rightarrow\R^d$ such that for all
$t\in\Rp$, 
\begin{itemize}
\item $y(0)=q(\psi(<w,s>))$ and $y'(t)=p(y(t))$
\hfill$\blacktriangleright$ $y$ satisfies a differential equation
\item if $|y_1(t)|\geqslant1$ then 
$|y_1(u)|\geqslant1$ for all $u\geqslant t$
\hfill$\blacktriangleright$ the decision is stable
\item if $w\in\mathcal{L}$ (resp. $\notin\mathcal{L}$) and
  $\glen{y}(0,t)\geqslant \Omega(|w|)$
then $y_1(t)\geqslant1$ (resp. $\leqslant-1$)
\hfill$\blacktriangleright$ decision
\item $\glen{y}(0,t)\geqslant t$
\hfill$\blacktriangleright$ technical condition\footnote{Same remarks
  as above.}
\end{itemize}
\end{definition}

\begin{theorem}[An implicit characterization of $\NP$]
Let $\mathcal{L}$ be any decision problem (language).

$\mathcal{L}\in\NP$ if and only if $\mathcal{L}$ is $\NP$-analog-recognizable.
\end{theorem}

Following the discussion page \pageref{dis:np}, the purpose would be to get
something more ``natural'', not involving logic like quantifiers (for e.g.  a
concept of analysis like ordinary differential inclusions instead of
ordinary differential equations).

\section{Notations}

\renewcommand{\arraystretch}{1.2}

\begin{longtable}{@{}p{4cm}lp{9cm}@{}}
\multicolumn{3}{@{}l}{\textbf{\Large Notations for sets}}\\
Concept & Notation & Comment \\
\midrule
\endhead
Real interval & $[a,b]$ & $\{x\in\R | \thinspace a \leqslant x \leqslant b \}$ \\
    & $[a,b[$ & $\{x\in\R | \thinspace a \leqslant x < b \}$ \\
    & $]a,b]$ & $\{x\in\R | \thinspace a < x \leqslant b \}$ \\
    & $]a,b[$ & $\{x\in\R | \thinspace a < x < b \}$ \\
Line segment & $[x,y]$ & $\{(1-\alpha)x+\alpha y,\alpha\in[0,1]\}$\\
    & $[x,y[$ & $\{(1-\alpha)x+\alpha y,\alpha\in[0,1[\}$\\
    & $]x,y]$ & $\{(1-\alpha)x+\alpha y,\alpha\in]0,1]\}$\\
    & $]x,y[$ & $\{(1-\alpha)x+\alpha y,\alpha\in]0,1[\}$\\
Integer interval & $\intinterv{a}{b}$ & $\{a,a+1,\ldots,b\}$ \\
Natural numbers & $\N$ & $\{0, 1, 2, \ldots\}$\\
 & $\Ns$ & $\N\setminus\{0\}$\\
Integers & $\Z$ & $\{\ldots,-2,-1,0,1,2,\ldots\}$\\
Rational numbers & $\Q$ & \\
Real numbers & $\R$ & \\
Non-negative numbers & $\Rp$ & $\Rp=[0,+\infty[$\\
Non-zero numbers & $\Rs$ & $\Rs=\R\setminus\{0\}$ \\
Positive numbers & $\Rps$ & $\Rps=]0,+\infty[$ \\
Set shifting & $x+Y$ & $\{x+y,y\in Y\}$\\
Set addition & $X+Y$ & $\{x+y,x\in X, y\in Y\}$\\
Matrices & $\MAT{n}{m}{\K}$ & Set of $n\times m$ matrices over field $\K$\\
& $\MAT{n}{}{\K}$ & Shorthand for $\MAT{n}{n}{\K}$\\
& $\MAT{n}{m}{}$ & Set of $n\times m$ matrices over a field is deduced from the context\\
Polynomials & $\K[X_1,\ldots,X_n]$ & Ring of polynomials with variables $X_1,\ldots,X_n$
    and coefficients in $\K$\\
& $\K[\A^n]$ & Polynomial functions with $n$ variables, coefficients in $\K$
    and domain of definition $\A^n$ \\
Fractions & $\K(X)$ & Field of rational fractions with coefficients in $\K$\\
Power set & $\powerset{X}$ & The set of all subsets of $X$\\
Domain of definition & $\dom{f}$ & If $f:I\rightarrow J$ then $\dom{f}=I$ \\
Cardinal & $\card{X}$ & Number of elements\\
Polynomial vector & $\K^n[\A^d]$ & Polynomial in $d$ variables with coefficients in $\K^n$ \\
    & $\K[\A^d]^n$ & Isomorphic $\K^n[\A^d]$\\
Polynomial matrix & $\MAT{n}{m}{\K}[\A^n]$ & Polynomial in $n$ variables with matrix coefficients \\
    & $\MAT{n}{m}{\K[\A^n]}$ & Isomorphic $\MAT{n}{m}{\K}[\A^n]$\\
Smooth functions & $C^k$ & Partial derivatives of order $k$ exist and are continuous\\
    & $C^\infty$ & Partial derivatives exist at all orders\\
\bottomrule
\end{longtable}

\begin{longtable}{@{}p{5cm}lp{8.5cm}@{}}
\multicolumn{3}{@{}l}{\textbf{\Large Complexity classes}}\\
Concept & Notation & Comment \\
\midrule
\endhead
Polynomial Time & $\PTIME$ & Class of decidable languages\\
    & $\FP$ & Class of computable functions\\
Polynomial time computable numbers & $\Rpoly$ & 
\\
Polynomial time computable real functions & $P(\R)$ &\\
\bottomrule
\end{longtable}

\begin{longtable}{@{}p{5cm}lp{9cm}@{}}
\multicolumn{3}{@{}l}{\textbf{\Large Metric spaces and topology}}\\
Concept & Notation & Comment \\
\midrule
\endhead
$p$-norm & $\inorm{x}{p}$ & $\displaystyle\left(\sum_{i=1}^{n}|x_i|^p\right)^{\frac{1}{p}}$ \\
Infinity norm & $\infnorm{x}$ & $\max(|x_1|,\ldots,|x_n|)$ \\
\bottomrule
\end{longtable}

\begin{longtable}{@{}p{4.5cm}lp{9cm}@{}}
\multicolumn{3}{@{}l}{\textbf{\Large Notations for polynomials}}\\
Concept & Notation & Comment \\
\midrule
\endhead
Univariate polynomial & $\displaystyle\sum_{i=0}^{d}a_iX^i$ &\\
Multi-index & $\alpha$& $(\alpha_1,\ldots,\alpha_k)\in\N^k$ \\
& $|\alpha|$ & $\alpha_1+\cdots+\alpha_k$ \\
& $\alpha!$ & $\alpha_1!\alpha_2!\cdots\alpha_k!$ \\
Multivariate polynomial & $\displaystyle\sum_{|\alpha|\leqslant d}a_\alpha X^\alpha$ & where $X^\alpha=X_1^{\alpha_1}\cdots X_k^{\alpha_k}$\\
Degree & $\degp{P}$ & Maximum degree of a monomial, $X^\alpha$ is of degree $|\alpha|$, conventionally $\degp{0}=-\infty$\\
& $\degp{P}$ & $\max(\degp{P_i})$ if $P=(P_1,\ldots,P_n)$ \\
& $\degp{P}$ & $\max(\degp{P_{ij}})$ if $P=(P_{ij})_{i\in\intinterv{1}{n},j\in\intinterv{1}{m}}$ \\
Sum of coefficients & $\sigmap{P}$ & $\sigmap{P}=\sum_{\alpha}|a_\alpha|$ \\
& $\sigmap{P}$ & $\max(\sigmap{P_1},\ldots,\sigmap{P_n})$ if $P=(P_1,\ldots,P_n)$ \\
& $\sigmap{P}$ & $\max(\sigmap{P_{ij}})$ if $P=(P_{ij})_{i\in\intinterv{1}{n},j\in\intinterv{1}{m}}$ \\
A polynomial & $\poly$ & An unspecified polynomial \\
\bottomrule
\end{longtable}

\begin{longtable}{@{}p{5cm}lp{9cm}@{}}
\multicolumn{3}{@{}l}{\textbf{\Large Miscellaneous functions}}\\
Concept & Notation & Comment \\
\midrule
\endhead
Sign function & $\sgn{x}$ & Conventionally $\sgn{0}=0$ \\
Ceiling function & $\ceil{x}$ & $\min\{n\in\Z\thinspace|\thinspace x\leqslant n\}$ \\
Rounding function & $\round{x}$ & $\argmin_{n\in\Z}|n-x|$, undefined for $x=n+\frac{1}{2}$ \\
Integer part function & $\intp(x)$ & $\max(0,\lfloor x\rfloor)$\\
    & $\intp_n(x)$ & $\min(n,\intp(x))$ \\
Fractional part function & $\fracp(x)$ & $x-\intp{x}$ \\
    & $\fracp_n(x)$ & $x-\intp_n(x)$\\
Composition operator & $f\circ g$ & $(f\circ g)(x)=f(g(x))$ \\
Identity function & $\idfun$ & $\idfun(x)=x$ \\
Indicator function & $\indicator{X}$ & $\indicator{X}(x)=1$ if $x\in X$ and $\indicator{X}(x)=0$ otherwise \\
$n^{th}$ iterate & $\fiter{f}{n}$ & $\fiter{f}{0}=\idfun$ and $\fiter{f}{n+1}=\fiter{f}{n}\circ f$\\
\bottomrule
\end{longtable}

\begin{longtable}{@{}p{4cm}lp{8.5cm}@{}}
\multicolumn{3}{@{}l}{\textbf{\Large Calculus}}\\
Concept & Notation & Comment \\
\midrule
\endhead
Derivative & $f'$ &\\
$n^{th}$ derivative & $f^{(n)}$ & $f^{(0)}=f$ and $f^{(n+1)}={f^{(n)}}'$\\
Partial derivative & $\partial_if,\frac{\partial f}{\partial x_i}$ & with respect to the $i^{th}$ variable\\
Scalar product & $\scalarprod{x}{y}$ & $\sum_{i=1}^nx_iy_i$ in $\R^n$\\
Gradient & $\grad{f}(x)$ & $(\partial_1 f(x), \ldots, \partial_nf(x))$ \\
Jacobian matrix & $\jacobian{f}(x)$ & $(\partial_jf_i(x))_{i\in\intinterv{1}{n},j\in\intinterv{1}{m}}$\\
Taylor approximation & $\taylor{a}{n}{f}(t)$ & $\displaystyle\sum_{k=0}^{n-1}\frac{f^{(k)}(a)}{k!}(t-a)^k$ \\
Big O notation & $f(x)=\bigO{g(x)}$ & $\exists M,x_0\in\R$, $|f(x)|\leqslant M|g(x)|$ for all $x\geqslant x_0$ \\
Soft O notation & $f(x)=\softO{g(x)}$ & Means $f(x)=\bigO{g(x)\log^kg(x)}$ for some $k$\\
Subvector & $x_{i..j}$ & $(x_i,x_{i+1},\ldots,x_j)$\\
Matrix transpose & $\transpose{M}$ &\\
Past supremum & $\pastsup{\delta}{f}(t)$ & $\sup_{u\in[t,t-\delta]\cap\Rp}f(t)$\\
Partial function & $f:\subseteq X\rightarrow Y$ & $\dom{f}\subseteq X$\\
Restriction & $\frestrict{f}{I}$ & $\frestrict{f}{I}(x)=f(x)$ for all $x\in\dom{f}\cap I$\\
\bottomrule
\end{longtable}

\begin{longtable}{@{}p{4cm}lp{10cm}@{}}
\multicolumn{3}{@{}l}{\textbf{\Large Words}}\\
Concept & Notation & Comment \\
\midrule
\endhead
Alphabet & $\Sigma,\Gamma$ & A finite set\\
Words & $\Sigma^*$ & $\bigcup_{n\geqslant0}\Sigma^n$\\
Empty word & $\emptyword$ &\\
Letter & $w_i$ & $i^{th}$ letter, starting from one\\
Subword & $w_{i..j}$ & $w_iw_{i+1}\cdots w_j$\\
Length & $|w|$ &\\
Repetition & $w^k$ & $\underbrace{ww\cdots w}_{k\text{ times}}$\\
\bottomrule
\end{longtable}

\section{Polytime analog computability implies polytime computability}

\subsection{Proof of Theorem~\ref{main:th} (AP implies P)}
\label{appendix:main:th}

Let us introduce the following definition. 

\begin{definition}[Analog computability]\label{def:gc}
Let $n,m\in\N$, $f:\subseteq\R^n\rightarrow\R^m$ and $\Upsilon,\Omega:\Rp^2\rightarrow\Rp$.
We say that $f$ is $(\Upsilon,\Omega)$-computable if and only if there exists $d\in\N$,
and $p\in\K^d[\R^d],q\in\K^d[\R^n]$
such that for any $x\in\dom{f}$, there exists (a unique) $y:\Rp\rightarrow\R^d$ satisfying
\begin{itemize}
\item $y(0)=q(x)$ and $y'(t)=p(y(t))$ for all $t\geqslant 0$\hfill$\blacktriangleright$ $y$ satisfies a PIVP
\item for all $\mu\in\Rp$, if $t\geqslant\Omega(\infnorm{x},\mu)$ then $\infnorm{y_{1..m}(t)-f(x)}\leqslant e^{-\mu}$
\hfill$\blacktriangleright$ $y_{1..m}$ converges to $f(x)$
\item $\infnorm{y(t)}\leqslant\Upsilon(\infnorm{x},t)$, for all $t\geqslant0$\hfill$\blacktriangleright$ $y(t)$ is bounded
\end{itemize}
We denote by $\gc{\Upsilon}{\Omega}$ the set of $(\Upsilon,\Omega)$-computable functions.
\end{definition}



A function $f \in \gpc$ must belong to $\gc{\Upsilon}{\Omega}$ where $\Upsilon,\Omega$
are polynomials which we can assume to be increasing function. This
follows from the fact that any system can be rescaled using
the length of the curve to make sure it does not grow faster than a
polynomial. A formal proof of this fact can be found in Appendix~\ref{app:gpcsubsetgpwc}
(the proof that AP implies AWP implies exactly that). 

 Apply Definition~\ref{def:gc} to get $d,p$ and $q$.
Apply Theorem~\ref{th:comp_implies_cont} to $f$ to get $\mho$ and define:
\[m(n)=\tfrac{1}{\ln2}\mho(\max(|a|,|b]),n\ln2)\]
It follows from the definition that $m$ is a modulus of continuity of $f$ since for any $n\in\N$ and $x,y\in[a,b]$
such that $|x-y|\leqslant2^{-m(n)}$ we have:
\[|x-y|\leqslant2^{-\frac{1}{\ln2}\mho(\max(|a|,|b|),n\ln2)}\\
    =e^{-\mho(\max(|a|,|b|),n\ln2)}\\
    \leqslant e^{-\mho(|x|,n\ln2)}\]
Thus $|f(x)-f(y)|\leqslant e^{-n\ln2}=2^{-n}$. We will now see how to approximate $f$ in
polynomial time. Let $r\in\Q$ and $n\in\N$, we would like to compute $f(r)\pm2^{-n}$.
By definition of $f$, there exists a unique $y:\Rp\rightarrow\R^d$ such that for all $t\in\Rp$:
\[y(0)=q(r)\qquad y'(t)=p(y(t)\]
Furthermore, $|y_1(\Omega(|r|,\mu))-f(r)|\leqslant e^{-\mu}$ for any $\mu\in\Rp$ and
$\infnorm{y(t)}\leqslant\Upsilon(|r|,t)$ for all $t\in\Rp$. Note that the coefficients
of $p$ and $q$ belongs to $\Rpoly$. 
One can compute a rational $r'$ such that $|y(t)-r'|\leqslant2^{-n}$ in time:
\[\poly(\degp{p},\glen{y}(0,t),\log\infnorm{y(0)},\log\sigmap{p},-\log2^{-n})^d\]
Recall that in this case, all the parameters $d,\sigmap{p},\degp{p}$ only depend on $f$ and thus fixed
and that $|r|$ is bounded by a constant. Thus these are all considered constants.
So in particular, we can compute $r'$ such that $|y(\Omega(|r|,(n+1)\ln2)-r'|\leqslant2^{-n-1}$ in time:
\[\poly(\glen{y}(0,\Omega(|r|,(n+1)\ln2)),\log\infnorm{q(r)},(n+1)\ln2)\]
Note that $|r|\leqslant\max(|a|,|b|)$ and since $a$ and $b$ are constants and $q$
is a polynomial, $\infnorm{q(r)}$ is bounded by a constant. Furthermore,
\begin{align*}
\glen{y}(0,\Omega(|r|,(n+1)\ln2))&=\int0^{\Omega(|r|,(n+1)\ln2)}\max(1,\infnorm{y(t)})^{\degp{p}}dt\\
    &\leqslant \int0^{\Omega(|r|,(n+1)\ln2)}\poly(\Upsilon(\infnorm{r},t))dt\\
    &\leqslant \Omega(|r|,(n+1)\ln2)\poly(\Upsilon(|r|,\Omega(|r|,(n+1)\ln2)))dt\\
    &\leqslant\poly(|r|,n)\leqslant\poly(n)
\end{align*}
Thus $r'$ can be computed in time:
\[\poly(n)\]
Finally:
\begin{align*}
|f(r)-r'|&\leqslant|f(r)-y(\Omega(|r|,(n+1)\ln2))|+|y(\Omega(|r|,(n+1)\ln2))-r'|\\
    &\leqslant e{-(n+1)\ln2}+2^{-n-1}\\
    &\leqslant 2^{-n}
\end{align*}
This show that $f$ is polytime computable.

\subsection{Proof of Theorem~\ref{th:comp_implies_cont}}
\label{appendix:th:comp_implies_cont}

A way to get a short proof is to use Lemma~\ref{lemma:online}: 
%
%
Let  $\Omega,\delta,d,p$ and $y_0$ be corresponding to the statement
of Lemma~\ref{lemma:online}. 
Without loss of generality,
we assume $\Omega$  to be an increasing function.

Let $u,v\in\dom{f}$
and $\mu\in\Rp$. Assume that $\infnorm{u-v}\leqslant e^{-\Lambda(\infnorm{u}+1,\mu+\ln2)}$
and consider the following system:
\[y(0)=y_0\qquad y'(t)=p(y(t),u)\]
By definition, $\infnorm{y_{1..m}(t)-f(u)}\leqslant e^{-\mu-\ln2}$ for all $t\geqslant\Omega(\infnorm{u},\mu+\ln2)$.
For the same reason, $\infnorm{y_{1..m}(t)-f(v)}\leqslant e^{-\mu-\ln2}$ for all $t\geqslant\Omega(\infnorm{v},\mu+\ln2)$
because $\infnorm{u-v}\leqslant e^{-\Lambda(\infnorm{u}+1,\mu+ln2)}\leqslant e^{-\Lambda(\infnorm{v},\mu+\ln2)}$.
Apply both result to $t=\Omega(\infnorm{u}+1,\mu+\ln2)$ to get that $\infnorm{f(u)-f(v)}\leqslant2e^{-\mu-\ln2}$.

\section{Polytime computability implies polytime analog computability}

\subsection{Proof of Lemma~\ref{lemma:closure}}
\label{app:closure}

\begin{remark}[Uniqueness]
The uniqueness of $y$ in Definition~\ref{def:gpac_generable_ext} can be seen as follows:
consider $x\in I$ and $\gamma$ a smooth curve\footnote{see Remark~\ref{rem:gpac_ext_dom}}
from $x_0$ to $x$ with values in $I$ and consider $z(t)=y(\gamma(t))$ for $t\in[0,1]$.
It can be seen that $z'(t)=\jacobian{y}(\gamma(t))\gamma'(t)=p(y(\gamma(t))\gamma'(t)=p(z(t))\gamma'(t)$,
$z(0)=y(x_0)=y_0$ and $z(1)=y(x)$.
The initial value problem $z(0)=y_0$ and $z'(t)=p(z(t))\gamma'(t)$ satisfies the hypothesis
of the Cauchy-Lipschitz theorem and as such admits a unique solution. Since this
IVP is independent of $y$, it shows that $y(x)$ must be unique. Note that
the existence of $y$ (and thus the domain of definition) is an hypothesis of the definition.
\end{remark}

\begin{remark}[Regularity]\label{rem:gpac_ext_regularity}
In the euclidean space $\R^n$, $C^k$ smoothness is equivalent to the smoothness
of the order $k$ partial derivatives. Consequently, the equation $\jacobian{y}=p(y)$
on the open set $I$ immediately proves that $y$ is $C^\infty$. As
solutions of analytic ODE are analytic,  $y$ is in fact real analytic.
\end{remark}

\begin{remark}[Domain of definition]\label{rem:gpac_ext_dom}
Definition~\ref{def:gpac_generable_ext} requires the domain of definition of $f$ to be connected,
otherwise it would not make sense. Indeed, we can only define the value of $f$
at point $u$ if there exists a path from $x_0$ to $u$ in the domain of $f$.
It could seem, at first sight, that the domain being ``only''
connected may be too weak to work with. This is not the case, because
in the euclidean space $\R^d$, \emph{open} connected subsets are
always smoothly arc connected, that is any two points can be connected using a
smooth $C^1$ (and even $C^\infty$) arc. 
\end{remark}

\begin{remark}[Multidimensional output]\label{rem:multidim_out_ext}
The following is true: $f:\subseteq\R^d\rightarrow\R^n$
is generable if and only if each of its component is generable
(\emph{i.e.} $f_i$ is generable for all $i$).
\end{remark}

\begin{remark}[Definition consistency]
Definition~\ref{def:gpac_generable_ext} for
$d=e=1$ corresponds to \PIVP~ functions. 
\end{remark}

Lemma~\ref{lemma:closure} follows clearly from the following more precise
statement: 

\begin{lemma}[Arithmetic on generable functions]\label{lem:gpac_ext_class_stable}
Let $d,e,n,m\in\N$, $\mtt{sp},\ovl{\mtt{sp}}:\R\rightarrow\Rp$,
$f:\subseteq\R^d\rightarrow\R^n\in\gval{\mtt{sp}}$ and
$g:\subseteq\R^e\rightarrow\R^m\in\gval{\ovl{\mtt{sp}}}$. Then:
\begin{itemize}
\item $f+g, f-g\in\gval{\mtt{sp}+\ovl{\mtt{sp}}}$ over $\dom{f}\cap\dom{g}$ if $d=e$ and $n=m$
\item $fg\in\gval{\max(\mtt{sp},\ovl{\mtt{sp}},\mtt{sp}\thinspace\ovl{\mtt{sp}})}$ if $d=e$ and $n=m$
\item $f\circ g\in\gval{\max(\ovl{\mtt{sp}},\mtt{sp}\circ\ovl{\mtt{sp}})}$ if $m=d$ and $g(\dom{g})\subseteq \dom{f}$
\end{itemize}
\end{lemma}

\begin{proof}
We focus on the case of the composition, the other cases are very similar.

Apply Definition~\ref{def:gpac_generable_ext} to $f$ and $g$ to respectively get
$l,\bar{l}\in\N$, $p\in\MAT{l}{d}{\K}[\R^l]$, $\bar{p}\in\MAT{\bar{l}}{e}{\K}[\R^{\bar{l}}]$,
$x_0\in\dom{f}\cap\K^d$, $\bar{x}_0\in\dom{g}\cap\K^e$, $y_0\in\K^l$, $\bar{y}_0\in\K^{\bar{l}}$,
$y:\dom{f}\rightarrow\R^l$ and $\bar{y}:\dom{g}\rightarrow\R^{\bar{l}}$.
Define $h=y\circ g$, then $\jacobian{h}=\jacobian{y}(g)\jacobian{g}=p(h)\bar{p}_{1..m}(\bar{y})$
and $h(\bar{x_0})=y(\bar{y}_0)\in\K^l$. 
In other words $(\bar{y},h)$ satisfy:
\[\left\{\begin{array}{@{}r@{}l}\bar{y}(\bar{x}_0)&=y_0\in\K^{\bar{l}}\\h(\bar{x}_0)&=y(\bar{y}_0)\in\K^l\end{array}\right.
\qquad\left\{\begin{array}{@{}r@{}l}\bar{y}'&=\bar{p}(\bar{y})\\h'&=p(h)\bar{p}_{1..m}(\bar{y})\end{array}\right.\]
This shows that $f\circ g=z_{1..m}\in\gval{}$. Furthermore,
$\infnorm{(\bar{y}(x),h(x))}\leqslant\max(\infnorm{\bar{y}(x)},\infnorm{y(g(x))})
\leqslant\max(\mtt{\ovl{sp}}(\infnorm{x}),\mtt{sp}(\infnorm{g(x)}))
\leqslant\max(\mtt{\ovl{sp}}(\infnorm{x}),\mtt{sp}(\mtt{\ovl{sp}}(\infnorm{x})))$.
\end{proof}

\subsection{Proof of Lemma~\ref{cor:gpac_ext_ivp_stable}}
\label{app:cor:gpac_ext_ivp_stable}

Lemma~\ref{cor:gpac_ext_ivp_stable} follows from the following more
general statement: 

\begin{theorem}[Generable ODE rewriting]\label{prop:gpac_ext_ivp_stable_pre}
Let $d,n\in\N$, $I\subseteq\R^n$, $X\subseteq\R^d$, $\mtt{sp}:\Rp\rightarrow\Rp$ and
$(f:I\times X\rightarrow\R^n)\in\gval[\K]{\mtt{sp}}$. Define $\ovl{\mtt{sp}}=\max(\idfun,\mtt{sp})$.
Then there exists $m\in\N$, $(g:I\times X\rightarrow\R^m)\in\gval[\K]{\ovl{\mtt{sp}}}$
and $p\in\K^m[\R^m\times\R^d]$ such that for any interval $J$,
$t_0\in\K\cap J$, $y_0\in\K^n\cap J$, $y\in C^1(J,I)$ and $x\in C^1(J,X)$,
if $y$ satisfies:
\[\left\{\begin{array}{@{}r@{}l@{}}y(t_0)&=y_0\\y'(t)&=f(y(t),x(t))\end{array}\right.
\qquad\forall t\in J\]
then there exists $z\in C^1(J,\R^m)$ such that:
\[\left\{\begin{array}{@{}r@{}l@{}}z(t_0)&=g(y_0,x(t_0))
\\z'(t)&=p(z(t),x'(t))\end{array}\right.
\qquad \left\{\begin{array}{@{}r@{}l@{}}y(t)&=z_{1..d}(t)\\\infnorm{z(t)}&\leqslant\ovl{\mtt{sp}}(\infnorm{y(t),x(t)})\end{array}\right.
\qquad \forall t\in J\]
\end{theorem}

\begin{proof}
Apply Definition~\ref{def:gpac_generable_ext} to $f$ get $m\in\N$, $p\in\MAT{m}{n+d}{\K}[\R^m]$,
$f_0\in\dom{f}\cap\K^d$, $w_0\in\K^m$ and $w:\dom{f}\rightarrow\R^m$ such that $w(f_0)=w_0$,
$\jacobian{w(v)}=p(w(v))$, $\infnorm{w(v)}\leqslant\mtt{sp}(\infnorm{v})$ and $w_{1..n}(v)=f(v)$ for all $v\in\dom{f}$.
Define $u(t)=w(y(t),x(t))$, then:
\begin{align*}
u'(t)&=\jacobian{w}(y(t),x(t))(y'(t),x'(t))\\
&=p(w(y(t),x(t)))(f(y(t),x(t)),x'(t))\\
&=p(u(t))(u_{1..n}(t),x'(t))\\
&=q(u(t),x'(t))
\end{align*}
where $q\in\K^m[\R^{m+d}]$ and $u(t_0)=w(y(t_0))=w(y_0,x(t_0))$. Note that $w$ itself
is a generable function and more precisely $w\in\gpval[\K]{\mtt{sp}}$ by definition.
Finally, note that $y'(t)=u_{1..d}(t)$ so that we get for all $t\in J$:
\[\left\{\begin{array}{@{}r@{}l@{}}y(t_0)&=y_0\\
y'(t)&=u_{1..d}(t)\end{array}\right.
\qquad \left\{\begin{array}{@{}r@{}l@{}}u(t_0)&=w(y_0,x(t_0))\\
u'(t)&=q(u(t),x'(t))\end{array}\right.\]
Define $z(t)=(y(t),u(t))$, then $z(t_0)=(y_0,w(y_0,x(t_0)))=g(y_0,x(t_0))$
where  $y_0\in\K^n$ and $w\in\gval[\K]{\mtt{sp}}$ so $g\in\gval[\K]{\ovl{\mtt{sp}}}$.
And clearly $z'(t)=r(z(t),x'(t))$ where $r\in\K^{n+m}[\R^{n+m}]$.
Finally, $\infnorm{z(t)}=\infnorm{y(t),w(y(t),x(t))}
\leqslant\max(\infnorm{y(t)},\mtt{sp}(\infnorm{y(t),x(t)}))
\leqslant\ovl{\mtt{sp}}(\infnorm{y(t),x(t)})$.
\end{proof}

\subsection{Proof of $\gpval \subset \gpc$ under conditions on the
  domain}
\label{app:stardomain}

Generable functions are continuous
and continuously differentiable, so locally Lipschitz continuous. We can give a
precise expression for the modulus of continuity in the case where the domain of
definition is simple enough.


\begin{lemma}[Modulus of continuity]\label{prop:generable_mod_cont}
Let $\mtt{sp}:\Rp\rightarrow\Rp$, $f\in\gval{\mtt{sp}}$. There exists $q\in\K[\R]$ such that
for any $x_1,x_2\in\dom{f}$, if $[x_1,x_2]\subseteq\dom{f}$ then
    $\infnorm{f(x_1)-f(x_2)}\leqslant\infnorm{x_1-x_2}q(\mtt{sp}(\max(\infnorm{x_1},\infnorm{x_2})))$.
In particular, if $f\in\gpval[\K]$ then there exists $q\in\K[\R]$ such that
if $[x_1,x_2]\subseteq\dom{f}$ then $\infnorm{f(x_1)-f(x_2)}\leqslant\infnorm{x_1-x_2}q(\max(\infnorm{x_1},\infnorm{x_2}))$.
\end{lemma}

The following can be established: See \cite{TheseAmaury}. 

\begin{lemma}[Generable field stability, \cite{TheseAmaury}]\label{cor:generable_field_ext_stab}
Let $(f:\subseteq\R^d\rightarrow\R^e)\in\gval{}$, then $f(\K^d\cap\dom{f})\subseteq\K^e$.
\end{lemma}

We can go to the proof of $\gpval \subset \gpc$ under conditions on the
  domain: 

\begin{definition}[Star domain]A set $X\subseteq\R^n$ is called a \emph{star domain}
if there exists $x_0\in X$ such that for all $x\in U$ the line segment from $x_0$ to $x$
is in $X$, i.e $[x_0,x]\subseteq X$. Such an $x_0$ is called a \emph{vantage point}.
\end{definition}

\begin{theorem}[$\gpval\subseteq\gpc$ over star domains]\label{th:gpac_gen_implies_comp}If
$f\in\gpval$ has a star domain with a vantage point with coordinates
in $\Rpoly$ then $f\in\gpc$.
\end{theorem}

\begin{proof}
Let $(f:\subseteq\R^n\rightarrow\R^m)\in\gpval$ and $z_0\in\dom{f}\cap\K^n$ a generable vantage point.
Apply Definition~\ref{def:gpac_generable_ext} to get $d,p,x_0,y_0$ and $y$. Since $y$ is generable
and $z_0\in\K^d$, apply Lemma~\ref{cor:generable_field_ext_stab}
to get that $y(z_0)\in\K^d$.
Let $x\in\dom{f}$ and consider the following system:
\[
\left\{\begin{array}{@{}r@{}l@{}}
x(0)&=x\\\gamma(0)&=x_0\\z(0)&=y(z_0)
\end{array}\right.
\qquad
\left\{\begin{array}{@{}r@{}l@{}}
x'(t)&=0\\\gamma'(t)&=x(t)-\gamma(t)\\z'(t)&=p(z(t))(x(t)-\gamma(t))
\end{array}\right.
\]
First note that $x(t)$ is constant and check that $\gamma(t)=x+(x_0-x)e^{-t}$ and note that $\gamma(\Rp)\subseteq[x_0,x]\subseteq\dom{f}$
because it is a star domain.
Thus $z(t)=y(\gamma(t))$ since $\gamma'(t)=x(t)-\gamma(t)$ and $\jacobian{y}=p$.
It follows that $\infnorm{f(x)-z_{1..m}(t)}=\infnorm{f(x)-f(\gamma(t))}$ since $z_{1..m}=f$.
Apply Lemma~\ref{prop:generable_mod_cont} to $f$ to get $q$, and since $\infnorm{\gamma(t)}\leqslant\infnorm{x_0,x}$
we have:
\[\infnorm{f(x)-z_{1..m}(t)}\leqslant\infnorm{x-x_0}e^{-t}q(\infnorm{x_0,x})\leqslant e^{-t}\poly(\infnorm{x})\]
Finally, $\infnorm{z(t)}\leqslant\poly(\infnorm{x})$ because
$\infnorm{z(t)}$ is polynomially bounded. This implies that the length
of the curve is also polynomially bounded. 

As a final remark, one can observe that the issue of the domain is in fact reduced to
the problem of building $\gamma$. In the case of a star domain, this is trivial.
In the general case, one would need to show that there is a ``generic'' such $\gamma$
that given a point $x$ goes from $x_0$ to $x$ and stays in the domain of $f$.
\end{proof}

\subsection{Proof that AP implies AWP} 
\label{app:gpcsubsetgpwc}

We start by a remark:

\begin{lemma}[Norm function, \cite{TheseAmaury}]\label{lem:norm}
There is a family of function $\norm_{\infty,\delta}\in\gpval$ 
such that:
For any $x\in\R^n$ and $\delta\in]0,1]$ we have:
\[\infnorm{x}\leqslant\norm_{\infty,\delta}(x)\leqslant\infnorm{x}+\delta.\]
\end{lemma}

To prove $\gpc \subset \gpwc$, the kea idea is to rescale the system
using the length of the curve to make sure it does not grow faster
than a polynomial. This is then ensured by the technical condition.

More precisely:

Let $f\in\gplc$. 
Apply Definition~\ref{def:glc} to get $\Omega, d,p,q$. Also assume that
polynomial  $\Omega$ is an increasing function. Let $k=\degp{p}$.
Apply Lemma~\ref{lem:norm} to get that $g(x)=\norm_{\infty,1}(p(x))$ belongs to $\gpval$.
Apply Definition~\ref{def:gpac_generable_ext} to get corresponding $m,r,x_0$ and $z_0$.
Let $x\in\dom{f}$. For the analysis, it will useful to consider the following systems:
\[\left\{\begin{array}{@{}r@{}l}
y(0)&=q(x)\\z(x_0)&=z_0
\end{array}\right.\qquad
\left\{\begin{array}{@{}r@{}l}
y'(t)&=p(y(t))\\
\jacobian{z}(x)&=r(z(x))\end{array}\right.
\]
Note that by definition $z_1(x)=g(x)$. Define $\psi(t)=g(y(t))$
and $\hat{\psi}(u)=\int_0^u\psi(t)dt$. Now define the following system:
\[\left\{\begin{array}{@{}r@{}l}
\hat{y}(0)&=q(x)\\\hat{z}(0)&=z(q(x))\\\hat{w}(0)&=\frac{1}{g(q(x))}
\end{array}\right.\qquad
\left\{\begin{array}{@{}r@{}l}
\hat{y}'(u)&=\hat{w}(u)p(\hat{y}(u))\\
\hat{z}'(u)&=\hat{w}(u)r(\hat{z}(u))p(\hat{y}(u))\\
\hat{w}'(u)&=-\hat{w}(u)^3r_1(\hat{z}(u))p(\hat{y}(u))
\end{array}\right.
\]
where by $r_1$ we mean the first line of $r$. We will check that $\hat{y}(u)=y(\hat{\psi}^{-1}(u))$,
$\hat{z}(u)=z(\hat{y}(u))$ and $\hat{w}(u)=(\hat{\psi}^{-1})'(u)$. We will use the fact that
for any $h\in C^1$, $(g^{-1})'=\frac{1}{g'\circ g^{-1}}$. Also note that $\hat{\psi}'=\psi$.
\begin{itemize}
\item $\hat{y}(0)=y(\hat{\psi}^{-1}(0))=y(0)=q(x)$
\item $\hat{y}'(u)=(\hat{\psi}^{-1})'(u)y'(\hat{\psi}^{-1}(u))=\hat{w}(u)p(y(\hat{\psi}^{-1}(u)))=\hat{w}(u)p(\hat{y}(u))$
\item $\hat{z}(0)=z(\hat{y}(0))=z(q(x))$
\item $\hat{z}'(u)=\jacobian{z}(\hat{y}(u))\hat{y}'(u)=\hat{w}(u)r(z(\hat{y}(u)))p(\hat{y}(u))=\hat{w}(u)r(\hat{z}(u))p(\hat{y}(u))$
\item $\hat{w}(0)=\frac{1}{\hat{\psi}'(\hat{\psi}^{-1}(0))}=\frac{1}{\psi(0)}=\frac{1}{g(q(x))}$
\item $\hat{w}'(u)=\frac{-(\hat{\psi}^{-1})'(u)\hat{\psi}''(\hat{\psi}^{-1}(u))}{\cramped{(\hat{\psi}'(\hat{\psi}^{-1}(u)))^2}}
    =-\hat{w}(u)^3\psi'(\hat{\psi}^{-1}(u))=\scalarprod{\grad{g}(y(\hat{\psi}^{-1}(u)))}{y'(\hat{\psi}^{-1})}$
    and since $\grad{g}(x)=\transpose{r_1(z(x))}$ (transpose of the first line of the jaocibian
matrix of $z$ because $g=z_1$) then $\hat{w}'(u)
    =-\hat{w}(u)^3\scalarprod{\transpose{r_1(z(y(\hat{\psi}^{-1}(u))))}}{p(y(\hat{\psi}^{-1}(u)))}
    =-\hat{w}(u)^3r_1(\hat{z}(u))p(\hat{y}(u))$
\end{itemize}

We now claim that this system computes $f$ quickly and has polynomial bound.
First note that by Lemma~\ref{lem:norm}, $\infnorm{y'(t)}\leqslant g(y(t))\leqslant\infnorm{y'(t)}+1$
thus $\glen{y}(0,t)\leqslant\hat{\psi}(t)\leqslant\glen{y}(0,t)+t$. Thus
\begin{multline} \glen{\hat{y}}(0,u)=\int_0^u\infnorm{\hat{y}'(\xi)}d\xi=
\int_0^{\hat{\psi}^{-1}(u)}\infnorm{\hat{w}(\hat{\psi}(t))p(\hat{y}(\hat{\psi}(t)))}\hat{\psi}'(t)dt
\\=\int_0^{\hat{\psi}^{-1}(u)}\infnorm{(\hat{\psi}^{-1})'(\hat{\psi}(t))\hat{\psi}'(t)p(y(t))}dt
=\int_0^{\hat{\psi}^{-1}(u)}\infnorm{p(y(t))}dt=\glen{y}(0,\hat{\psi}^{-1}(u))\leqslant\hat{\psi}(\hat{\psi}^{-1}(u))\leqslant
u.
\end{multline}
It follows that $\infnorm{\hat{y}(u)}\leqslant\infnorm{\hat{y}(0)}+u\leqslant\infnorm{q(x)}+u\leqslant\poly(\infnorm{x},u)$.
Similarly, $\infnorm{\hat{z}(u)}=\infnorm{z(\hat{y}(u))}\leqslant\poly(\infnorm{x},u)$
because $z\in\gpval$ and thus is polynomially bounded. Finally,
$\infnorm{\hat{w}}=\frac{1}{\psi(\hat{\psi}^{-1}(u)}=\frac{1}{g(\hat{y}(u))}\leqslant\frac{1}{\infnorm{y'(\hat{\psi}^{-1}(u))}}\leqslant1$
because by hypothesis, $\infnorm{y'(t)}\geqslant1$ for all $t\in\Rp$. This shows that
indeed $\infnorm{(\hat{y},\hat{z},\hat{w})(u)}$ is polynomially bounded in $\infnorm{x}$ and $u$.
Now let $\mu\in\Rp$ and $t\geqslant1+\Omega(\infnorm{x},\mu)$ then
$\glen{\hat{y}}(0,t)=\glen{y}(0,\hat{\psi}^{-1}(t))\geqslant\hat{\psi}(\hat{\psi}^{-1}(t))-\hat{\psi}^{-1}(t)
\geqslant t-\hat{\psi}^{-1}(t)\geqslant 1+\Omega(\infnorm{x},\mu)-\frac{1}{\psi(\hat{\psi}^{-1}(t))}
\geqslant \Omega(\infnorm{x},\mu)$ because, as we already saw, $\infnorm{\psi(\hat{\psi}^{-1}(t))}\geqslant1$.
Thus by definition, $\infnorm{\hat{y}_{1..m}(t)-f(x)}\leqslant e^{-\mu}$ because $\hat{y}(t)=y(\hat{\psi}^{-1}(t))$.
This shows that $f\in\gpwc$.

\subsection{Proof that AWP implies ARP}
\label{app:weak_to_robust} 

The purpose is to state that one can tolerate small errors on the
dynamic.

Formally:

\begin{theorem}[Weak $\subseteq$ robust]\label{th:weak_to_robust}
$\gpwc\subseteq\gprc$.
\end{theorem}

where 

\begin{definition}[Analog robust computability]\label{def:grc}
Let $n,m\in\N$, $f:\subseteq\R^n\rightarrow\R^m$, $\Theta,\Omega:\Rp^2\rightarrow\Rp$
and $\Upsilon:\Rp^3\rightarrow\Rp$.
We say that $f$ is $(\Upsilon,\Omega,\Theta)$-robustly-computable if and only if there exists $d\in\N$,
and $(h:\R^d\rightarrow\R^d),(g:\R^n\times\Rp\rightarrow\R^d)\in\gpval$
such that for any $x\in\dom{f}$, $\mu\in\Rp$, $e_0\in\R^d$ and $e\in C^0(\Rp,\R^d)$
satisfying $\infnorm{e_0}+\int_0^\infty\infnorm{e(t)}dt\leqslant e^{-\Theta(\infnorm{x},\mu)}$,
there exists (a unique) $y:\Rp\rightarrow\R^d$ satisfying for all $t\in\Rp$:
\begin{itemize}
\item $y(0)=g(x,\mu)+e_0$ and $y'(t)=h(y(t))+e(t)$\hfill$\blacktriangleright$ $y$ satisfies a generable IVP
\item if $t\geqslant\Omega(\infnorm{x},\mu)$ then $\infnorm{y_{1..m}(t)-f(x)}\leqslant e^{-\mu}$
\hfill$\blacktriangleright$ $y_{1..m}$ converges to $f(x)$
\item $\infnorm{y(t)}\leqslant\Upsilon(\infnorm{x},\mu,t)$
\hfill$\blacktriangleright$ $y(t)$ is bounded
\end{itemize}
We denote by $\grc{\Upsilon}{\Omega}{\Theta}$ the set of
$(\Upsilon,\Omega,\Theta)$-robustly-computable functions, and by $\gprc$ the set of
$(\poly,\poly,\poly)$-robustly-computable functions.
\end{definition}

The following Lemma can be proved: See \cite{TheseAmaury} for its
motivation, and for an explanation of the proof of the next theorem on
simple cases. 

\begin{lemma}[PIVP Slow-Stop,\cite{TheseAmaury}]\label{lem:pivp_slowstop}
Let $d\in\N$, $y_0\in\R^d$, $T,\theta\in\Rp$, $(e_{0,y},e_{0,A})\in\R^{d+1}$,
$(e_y,e_A)\in C^0(\Rp,\R^{d+1})$ and $p\in\K^d[\R^d]$.
Assume that $\infnorm{e_0}+\int_0^\infty\infnorm{e(t)}dt\leqslant e^{-\theta}$
and consider the following system:
\[
\left\{\begin{array}{@{}r@{}l}y(0)&=y_0+e_{0,y}\\
A(0)&=T+2+e_{0,A}\end{array}\right.
\quad
\left\{\begin{array}{@{}r@{}l}y'(t)&=\frac{1+\tanh(A(t))}{2}p(y(t))+e_{y}(t)\\
A'(t)&=-1+e_{A}(t)\end{array}\right.
\]
Then there exists an increasing function $\psi\in C^0(\Rp,\Rp)$
and $z:\psi(\Rp)\rightarrow\R^d$ such that:
\[\psi(0)=0\qquad z(0)=y_0+e_{0,y}\qquad z'(t)=p(z(t))+(\psi^{-1})'(t)e_{y}(\psi^{-1}(t))\]
and $y(t)=z(\psi(t))$. Furthermore $\psi(T+1)\geqslant T$
and $\psi(t)\leqslant T+4$ for all $t\in\Rp$. Furthermore, $|A(t)|\leqslant T+3$ for
all $t\in\Rp$.
\end{lemma}

We will also need the following small theorem about PIVP.

\begin{theorem}[Parameter dependency of PIVP,\cite{PoulyG16}]\label{th:dependency_right_side}
Let $I=[a,b]$, $p\in\R^n[\R^{n+d}]$, $k=\degp{p}$,
$e\in C^0(I,\R^d)$, $x,\delta\in C^0(I,\R^n)$ and $y_0,z_0\in\R^d$.
Assume that $y,z:I\rightarrow\R^d$ satisfy:
\[\left\{\begin{array}{@{}r@{}l}y(a)&=y_0\\y'(t)&=p(y(t),x(t))\end{array}\right.\qquad
\left\{\begin{array}{@{}r@{}l}z(a)&=z_0\\z'(t)&=e(t)+p(z(t),x(t)+\delta(t))\end{array}\right.\qquad t\in I\]
Assume that there exists $\varepsilon>0$ such that for all $t\in I$,
\[
\mu(t):=\left(\infnorm{z_0-y_0}+\int_{a}^t\infnorm{e(u)}+k\sigmap{p}M^{k-1}(u)\infnorm{\delta(u)}du\right)
\exp\left(k\sigmap{p}\int_{a}^tM^{k-1}(u)du\right)<\varepsilon
\]
where $M(t)=\varepsilon+\infnorm{y(t)}+\infnorm{x(t)}+\infnorm{\delta(t)}$. Then for all $t\in I$,
\[\infnorm{z(t)-y(t)}\leqslant\mu(t)\]
\end{theorem}

Recall:

\begin{definition}[Analog weak computability]\label{def:gwc}
Let $n,m\in\N$, $f:\subseteq\R^n\rightarrow\R^m$, $\Omega:\Rp^2\rightarrow\Rp$
and $\Upsilon:\Rp^3\rightarrow\Rp$.
We say that $f$ is $(\Upsilon,\Omega)$-weakly-computable if and only if there exists $d\in\N$,
$p\in\K^d[\R^d],q\in\K^d[\R^{n+1}]$
such that for any $x\in\dom{f}$ and $\mu\in\Rp$, there exists (a unique) $y:\Rp\rightarrow\R^d$ satisfying
for all $t\in\Rp$:
\begin{itemize}
\item $y(0)=q(x,\mu)$ and $y'(t)=p(y(t))$\hfill$\blacktriangleright$ $y$ satisfies a PIVP
\item if $t\geqslant\Omega(\infnorm{x},\mu)$ then $\infnorm{y_{1..m}(t)-f(x)}\leqslant e^{-\mu}$
\hfill$\blacktriangleright$ $y_{1..m}$ converges to $f(x)$
\item $\infnorm{y(t)}\leqslant\Upsilon(\infnorm{x},\mu,t)$
\hfill$\blacktriangleright$ $y(t)$ is bounded
\end{itemize}
We denote by $\gwc{\Upsilon}{\Omega}$ the set of $(\Upsilon,\Omega)$-weakly-computable functions.
\end{definition}

The proof of Theorem~\ref{th:weak_to_robust} is then the following.

\begin{proof}
Let $\Upsilon^*,\Omega^*$ be polynomials such that $f\in\gwc{\Upsilon^*}{\Omega^*}$.
Without loss of generality, we assume they are increasing functions of both arguments.
Apply Definition~\ref{def:gwc} to get $d\in\N$, $p\in\K^d[\R^d]$, $q\in\K^d[\R^{n+1}]$ and let $k=\degp{p}$.
Define:
\begin{align*}
T(\alpha,\mu)&=\Omega^*(\alpha,\mu+\ln2)\\
\Theta(\alpha,\mu)&=k\sigmap{p}(T(\alpha+1,\mu)+4)(\Upsilon^*(\alpha,\mu,T(\alpha+1,\mu)+4)+1)^{k-1}+\mu+\ln2\\
\Omega(\alpha,\mu)&=T(\alpha+1,\mu)+1
\end{align*}
Let $x\in\dom{f}$, $(e_{0,y},e_{0,A})\in\R^{d+1}$, $(e_y,e_A)\in C^0(\Rp,\R^{d+1})$ and $\mu\in\Rp$ such that
$\infnorm{e_0}+\int_0^\infty\infnorm{e(t)}dt\leqslant e^{-\Theta(\infnorm{x},\mu)}$.
Apply Lemma~\ref{lem:pivp_slowstop} and consider the following systems (where $\psi$ is given
by the lemma):
\[
\left\{\begin{array}{@{}r@{}l}y(0)&=q(x,\mu)+e_{0,y}\\
A(0)&=T(\norm_{\infty,1}(x),\mu)+2+e_{0,A}\end{array}\right.
\quad
\left\{\begin{array}{@{}r@{}l}y'(t)&=\frac{1+\tanh(A(t))}{2}p(y(t))+e_{y}(t)\\
A'(t)&=-1+e_{A}(t)\end{array}\right.
\]
\[
\left\{\begin{array}{@{}r@{}l}z(0)&=q(x,\mu)+e_{0,y}\\
z'(t)&=p(z(t))+(\psi^{-1})'(t)e_{y}(\psi^{-1}(t))\end{array}\right.
\quad
\left\{\begin{array}{@{}r@{}l}w(0)&=q(x,\mu)\\
w'(t)&=p(w(t))\end{array}\right.
\]
By definition of $p$ and $q$, if $t\geqslant\Omega^*(\infnorm{x},\mu)$ then
$\infnorm{w_{1..m}(t)-f(x)}\leqslant e^{-\mu}$. Furthermore,
$\infnorm{w(t)}\leqslant\Upsilon^*(\infnorm{x},\mu,t)$ for all $t\in\Rp$.
Define $T^*=T(\norm_{\infty,1}(x),\mu)$.
Apply Lemma~\ref{lem:norm} to get
that $\infnorm{x}\leqslant\norm_{\infty,1}(x)\leqslant\infnorm{x}+1$
and thus $T(\infnorm{x},\mu)\leqslant T^*\leqslant T(\infnorm{x}+1,\mu)$.
By construction, $\psi(t)\leqslant T^*+4$ for all $t\in\Rp$. Let $t\in\Rp$,
apply Theorem~\ref{th:dependency_right_side} by checking that:
\begin{align*}
&\left(\infnorm{e_{0,y}}+\int_0^{\psi(t)}\infnorm{(\psi^{-1})'(u)e_{y}(\psi^{-1}(u))du}\right)
e^{k\sigmap{p}\int_0^{\psi(t)}(\infnorm{w(u)}+1)^{k-1}du}\\
&\leqslant\left(\infnorm{e_{0,y}}+\int_0^t\infnorm{e_{y}(u)}du\right)
e^{k\sigmap{p}\int_0^{\psi(t)}(\Upsilon^*(\infnorm{x},\mu,u)+1)^{k-1}du}\tag*{by a change of variable}\\
&\leqslant e^{k\sigmap{p}\psi(t)(\Upsilon^*(\infnorm{x},\mu,\psi(t))+1)^{k-1}-\Theta(\infnorm{x},\mu)}\tag*{by hypothesis on the error}\\
&\leqslant e^{k\sigmap{p}(T(\infnorm{x}+1,\mu)+4)(\Upsilon^*(\infnorm{x},\mu,T(\infnorm{x}+1,\mu)+4)+1)^{k-1}-\Theta(\infnorm{x},\mu)}\tag*{because $\psi$ is bounded}\\
&\leqslant e^{-\mu-\ln2}\leqslant1\tag*{by definition of $\Theta$}
\end{align*}
Thus $\infnorm{z(\psi(t))-w(\psi(t))}\leqslant e^{-\mu-\ln2}$ for all $t\in\Rp$.
Furthermore, if $t\geqslant\Omega(\infnorm{x},\mu)$ then $\psi(t)\geqslant
\psi(T(\infnorm{x}+1,\mu)+1)\geqslant\psi(T^*+1)\geqslant T^*$. By construction $\psi(T^*)\geqslant T^*$
so $\psi(t)\geqslant T^*\geqslant T(\infnorm{x},\mu)=\Omega^*(\infnorm{x},\mu+\ln2)$
thus $\infnorm{z(\psi(t))-f(x)}\leqslant e^{-\mu-\ln2}$. Consequently,
$\infnorm{y(t)-f(x)}\leqslant\infnorm{z(\psi(t))-w(\psi(t))}+\infnorm{w(\psi(t))-f(x)}
\leqslant 2e^{-\mu-\ln2}\leqslant e^{-\mu}$.

Let $t\in\Rp$, then $\infnorm{y(t)}=\infnorm{z(\psi(t))}\leqslant\infnorm{w(\psi(t))}+e^{-\mu}
\leqslant\Upsilon^*(\infnorm{x},\mu,\psi(t))+1\leqslant\Upsilon^*(\infnorm{x},\mu,T(\infnorm{x}+1,\mu)+4)+1
\leqslant\Upsilon^*(\infnorm{x},\mu,\Omega^*(\infnorm{x}+1,\mu+\ln2)+4)+1$
which is polynomially bounded in $\infnorm{x}$ and $\mu$.
Furthermore $|A(t)|\leqslant T^*+4\leqslant\Omega^*(\infnorm{x}+1,\mu+\ln2)+4$
which are both polynomially bounded in $\infnorm{x}$, $\mu$.

Finally, $(y,A)(0)=g(x,\mu)+e_0$ and $(y,A)'(t)=h(y(t),A(t))+e(t)$
where $g$ and $h$ belong to $\gpval[\K]{}$ because $\tanh,\norm_{\infty,1}\in\gpval$.
\end{proof}

\begin{remark}[Polynomial versus generable]\label{rem:weak_to_robust_gen}
The proof of Theorem~\ref{th:weak_to_robust} also works if $q$ is generable (i.e. $q\in\gpval$)
instead of polynomial in Definition~\ref{def:gc} or Definiinition~\ref{def:gwc}.
\end{remark}

\subsection{Proof that ARP implies ASP}
\label{app:robust_to_strong}

The purpose is to state that one can tolerate small errors on the
dynamic + on inputs. 

Formally: 

\begin{theorem}[Robust $\subseteq$ strong]\label{th:robust_to_strong}
$\gprc\subseteq\gpsc$.
\end{theorem}

where

\begin{definition}[Analog strong computability]\label{def:gsc}
Let $n,m\in\N$, $f:\subseteq\R^n\rightarrow\R^m$, $\Theta,\Omega:\Rp^2\rightarrow\Rp$
and $\Upsilon:\Rp^4\rightarrow\Rp$.
We say that $f$ is $(\Upsilon,\Omega,\Theta)$-strongly-computable if and only if there exists $d\in\N$,
and $(h:\R^d\rightarrow\R^d),(g:\R^n\times\Rp\rightarrow\R^d)\in\gpval$
such that for any $x\in\R^n$, $\mu\in\Rp$, $e_0\in\R^d$ and $e\in C^0(\Rp,\R^d)$,
there is exists (a unique) $y:\Rp\rightarrow\R^d$ satisfying for all $t\in\Rp$
and $\hat{e}(t)=\infnorm{e_0}+\int_0^t\infnorm{e(u)}du$:
\begin{itemize}
\item $y(0)=g(x,\mu)+e_0$ and $y'(t)=h(y(t))+e(t)$\hfill$\blacktriangleright$ $y$ satisfies a generable IVP
\item if $x\in\dom{f}$,  $t\geqslant\Omega(\infnorm{x},\mu)$ and $\hat{e}(t)\leqslant e^{-\Theta(\infnorm{x},\mu)}$
then $\infnorm{y_{1..m}(t)-f(x)}\leqslant e^{-\mu}$
\item $\infnorm{y(t)}\leqslant\Upsilon(\infnorm{x},\mu,\hat{e}(t),t)$\hfill$\blacktriangleright$ $y(t)$ is bounded
\end{itemize}
We denote by $\gsc{\Upsilon}{\Omega}{\Theta}$ the set of
$(\Upsilon,\Omega,\Theta)$-strongly-computable functions, and by $\gpsc$ the set of
$(\poly,\poly,\poly)$-strongly-computable functions.
\end{definition}

The following Lemma can be proved by providing explicitely such a
function: 

\begin{lemma}[Max function, \cite{TheseAmaury}]\label{lem:max}
There is a family of functions $\mx_\delta\in\gpval$ such that: 
For any $x,y\in\R$ and $\delta\in]0,1]$ we have:
\[\max(x,y)\leqslant\mx_\delta(x,y)\leqslant\max(x,y)+\delta\]
For any $x\in\R^n$ and $\delta\in]0,1]$ we have:
\[\max(x_1,\ldots,x_n)\leqslant\mx_\delta(x)\leqslant\max(x_1,\ldots,x_n)+\delta\]
\end{lemma}

The following lemmas can also be established:

\begin{lemma}[Bounds on $\tanh$, \cite{TheseAmaury}]\label{lem:bounds_tanh}
$1-\sgn{t}\tanh(t)\leqslant e^{-|t|}$ for all $t\in\R$.
\end{lemma}

\begin{lemma}[Perturbed time-scaling]\label{lem:perturbed_time_scale_gpac}
Let $d\in\N$, $x_0\in\R^d$, $p\in\R^d[\R^d]$, $e\in C^0(\Rp,\R^d)$ and $\phi\in C^0(\Rp,\Rp)$.
Let $\psi(t)=\int_0^t\phi(u)du$.
Assume that $\psi$ is an increasing function and that $y,z:\Rp\rightarrow\R^d$ satisfy for all $t\in\Rp$:
\[\left\{\begin{array}{@{}r@{}l}y(0)&=x_0\\y'(t)&=p(y(t))+(\psi^{-1})'(t)e(\psi^{-1}(t))\end{array}\right.\qquad
\left\{\begin{array}{@{}r@{}l}z(0)&=x_0\\z'(t)&=\phi(t)p(z(t))+e(t)\end{array}\right.\]
Then $z(t)=y\left(\psi(t)\right)$ for all $t\in\Rp$. In particular,
$\int_0^{\psi(t)}\infnorm{(\psi^{-1})'(u)e(\psi^{-1}(u))}du=\int_0^t\infnorm{e(u)}du$
and $\sup_{u\in[0,\psi(t)]}\infnorm{(\psi^{-1})'(u)e(\psi^{-1}(u))}=\sup_{u\in[0,t]}\frac{\infnorm{e(u)}}{\phi(u)}$.
\end{lemma}

\begin{proof}
Use that $\phi=\psi'$, $\psi'\cdot(\psi^{-1})'\circ\psi=1$ and that $\psi'\geqslant0$.
\end{proof}

On a more technical side, we will need to ``apply'' Definition~\ref{def:grc} over finite intervals
and we need the following lemma to do so.

\begin{lemma}[Finite time robustness]\label{lem:finite_time_robust}
Let $f\in\grc{\Upsilon}{\Omega}{\Theta}$, $I=[0,T]$, $x\in\dom{f}$, $\mu\in\Rp$, $e_0\in\R^d$
and $e\in C^0(I,\R^d)$ such that $\infnorm{e_0}+\int_I\infnorm{e(t)}dt<e^{-\Theta(\infnorm{x},\mu)}$.
Assume that $y:I\rightarrow\R^d$ satisfies for all $t\in I$:
\[y(0)=g(x,\mu)+e_0\qquad y'(t)=h(y(t))+e(t)\]
where $g,h$ come from Definition~\ref{def:grc} applied to $f$. Then for all $t\in I$:
\begin{itemize}
\item $\infnorm{y(t)}\leqslant\Upsilon(\infnorm{x},\mu,t)$
\item if $t\geqslant\Omega(\infnorm{x},\mu)$ then $\infnorm{y_{1..m}-f(x)}\leqslant e^{-\mu}$
\end{itemize}
\end{lemma}

\begin{proof}
The trick is simply to extend $e$ so that it is defined over $\Rp$ and such that:
\[\infnorm{e_0}+\int_0^\infty\infnorm{e(u)}du\leqslant e^{-\Theta(\infnorm{x},\mu)}\]
This is always possible because the truncated integral is stricer smaller than the bound.
Formally, define for $t\in\Rp$:
\[\bar{e}(t)=\begin{cases}e(t)&\text{if }t\leqslant T\\
    e(T)e^{\frac{e(T)}{\varepsilon}(T-t)}&\text{otherwise}\end{cases}
\qquad\text{where }\varepsilon=e^{-\Theta(\infnorm{x},\mu)}-\infnorm{e_0}-\int_I\infnorm{e(t)}>0\]
One easily checks that $\bar{e}\in C^0(\Rp,\R^d)$ and that:
\begin{align*}
\infnorm{e_0}+\int_0^\infty\infnorm{\bar{e}(t)}dt&=\infnorm{e_0}+\int_0^T\infnorm{e(t)}dt+\int_T^\infty e(T)e^{\frac{e(T)}{\varepsilon}(T-t)}dt\\
    &=e^{-\Theta(\infnorm{x},\mu)}-\varepsilon+\left[-\varepsilon e(T)e^{\frac{e(T)}{\varepsilon}(T-t)}\right]_T^\infty\\
    &=e^{-\Theta(\infnorm{x},\mu)}
\end{align*}
Assume that $z:\Rp\rightarrow\R^d$ satisfies for $t\in\Rp$:
\[z(0)=g(x,\mu)\qquad z'(t)=g(z(t))+\bar{e}(t)\]
Then $z$ satisfies Definition~\ref{def:grc} so $\infnorm{z}(t)\leqslant\Upsilon(\infnorm{x},\mu)$
and if $t\geqslant\Omega(\infnorm{x},\mu)$ then $\infnorm{z_{1..m}(t)-f(x)}\leqslant e^{-\mu}$.
Conclude by noting that $z(t)=y(t)$ for all $t\in[0,T]$ since $e(t)=\bar{e}(t)$.
\end{proof}

The proof of Theorem~\ref{th:robust_to_strong} is then the following.

\begin{proof}
Let $\Omega,\Theta,\Upsilon$ be polynomials and $(f:\subseteq\R^n\rightarrow\R^m)\in\grc{\Upsilon}{\Omega}{\Theta}$.
Without loss of generality, we assume that $\Omega$, $\Theta$, $\Upsilon$ are
increasing functions of their arguments.
Apply Definition~\ref{def:grc} to get $d$, $h$ and $g$.
Let $x\in\R^n$, $\mu\in\Rp$, $(e_{0,y},e_{0,\ell})\in\R^{d+1}$ and $(e_y,e_\ell)\in C^0(\Rp,\R^{d+1})$.
Define $\hat{e}(t)=\infnorm{e_0}+\int_0^t\infnorm{e(u)}du$,
and consider the following system for $t\in\Rp$:
\[
\left\{\begin{array}{@{}r@{}l}
y(0)&=g(x,\mu)+e_{0,y}\\
y'(t)&=\psi(t)h(y(t))+e_{y}(t)\\
\ell(0)&=\mx_1(\norm_{\infty,1}(x),\mu)+1+e_{0,\ell}\\
\ell'(t)&=1+e_{\ell}(t)
\end{array}\right.
\]
\[\psi(t)=\frac{1+\tanh(\Delta(t))}{2}\qquad
\Delta(t)=\Upsilon(\ell(t),\ell(t),\ell(t))+1-\norm_{\infty,1}(y(t))\]
We will first show that the system remains polynomially bounded. Apply
Lemma~\ref{lem:max} and Lemma~\ref{lem:norm} to get that:
\begin{align*}
\infnorm{\ell(0)}&\leqslant\max(\infnorm{x}+1,\mu)+1+\infnorm{e_{0,\ell}}\\
    &\leqslant\poly(\infnorm{x},\mu)+\infnorm{e_{0,\ell}}
\end{align*}
Consequently:
\begin{align}
\infnorm{\ell(t)}&\leqslant\infnorm{\ell(0)}+\int_0^t1+\infnorm{e_{\ell}(u)}du\nonumber\\
    &\leqslant \poly(\infnorm{x},\mu)+t+\infnorm{e_{0,\ell}}+\int_0^t\infnorm{e_{\ell}(u)}du\nonumber\\
    &\leqslant\poly(\infnorm{x},\mu)+t+\hat{e}(t)\nonumber\\
    &\leqslant\poly(\infnorm{x},\mu,t,\hat{e}(t))\label{eq:robust_to_strong:ell}
\end{align}
Since $g,h\in\gpval$,
there exists $\mtt{sp}$ and $\ovl{\mtt{sp}}$ polynomials such that $\infnorm{g(x)}\leqslant\mtt{sp}(\infnorm{x})$
and $\infnorm{h(x)}\leqslant\ovl{\mtt{sp}}(\infnorm{x})$
for all $x\in\R^d$ and without loss of generability, we assume that $\mtt{sp}$ and $\ovl{\mtt{sp}}$ are
increasing functions. Let $t\in\Rp$, there are two possibilities:
\begin{itemize}
\item If $\Delta(t)\geqslant0$ then $\norm_{\infty,1}(y(t))\leqslant1+\Upsilon(\ell(t),\ell(t),\ell(t))$
so apply  Lemma~\ref{lem:norm} and use \eqref{eq:robust_to_strong:ell} to conclude that $\infnorm{y(t)}\leqslant\poly(\infnorm{x},\mu,t,\hat{e}(t))$
and thus:
\begin{align}\infnorm{\psi(t)h(y(t))}&\leqslant\ovl{\mtt{sp}}(\infnorm{y(t)})\tag*{use that $\tanh<1$}\\
    &\leqslant\poly(\infnorm{x},\mu,t,\hat{e}(t))\label{eq:robust_to_strong:rhs_one}
\end{align}
\item If $\Delta(t)<0$ then apply Lemma~\ref{lem:bounds_tanh} to get that $\psi(t)\leqslant\frac{1}{2}e^{\Delta(t)}\leqslant e^{\Delta(t)}$.
Apply Lemma~\ref{lem:norm} to get that $\Delta(t)\leqslant\Upsilon(\ell(t),\ell(t),\ell(t))+1-\infnorm{y(t)}$
and thus $\infnorm{y(t)}\leqslant \Upsilon(\ell(t),\ell(t),\ell(t))+1-\Delta(t)$ and thus:
\begin{align}
\infnorm{\psi(t)h(y(t))}&\leqslant e^{\Delta(t)}\ovl{\mtt{sp}}(\infnorm{y(t)})\tag*{use the bound on $\psi$}\\
    &\leqslant e^{\Delta(t)}\ovl{\mtt{sp}}(\Upsilon(\ell(t),\ell(t),\ell(t))+1-\Delta(t))\tag*{use the bound on $\infnorm{y(t)}$}\\
    &\leqslant\poly(\ell(t))e^{\Delta(t)}\poly(-\Delta(t))\tag*{use that $\Upsilon$ is polynomial}\\
    &\leqslant\poly(\ell(t))\tag*{use that $e^{-x}\poly(x)=\bigO{1}$ for $x\geqslant0$ and fixed $\poly$}\\
    &\leqslant\poly(\infnorm{x},\mu,t,\hat{e}(t))\label{eq:robust_to_strong:rhs_two}
\end{align}
\end{itemize}
Putting \eqref{eq:robust_to_strong:rhs_one} and \eqref{eq:robust_to_strong:rhs_two} together, we get that:
\begin{align*}
\infnorm{y(t)}&\leqslant\infnorm{g(x,\mu)}+\infnorm{e_{0,y}}+\int_0^t\infnorm{\psi(u)h(y(u))}+\infnorm{e_y(u)}du\\
    &\leqslant\mtt{sp}(\infnorm{x,\mu})+\int_0^t\poly(\infnorm{x},\mu,u,\hat{e}(u))du+\hat{e}(t)\\
    &\leqslant\poly(\infnorm{x},\mu,t,\hat{e}(t))
\end{align*}
We will now analyze the behavior of the system when the error is bounded.
Define $\Theta^*(\alpha,\mu)=\Theta(\alpha,\mu)+1$. Define $\hat{\psi}(t)=\int_0^t\psi(u)du$ and note that it is a diffeomorphism
since $\psi>0$. Apply Lemma~\ref{lem:perturbed_time_scale_gpac} to get that $y(t)=z(\hat{\psi}(t))$
for all $t\in\Rp$, where $z$ satisfies for $\xi\in\hat{\psi}(\Rp)$:
\[z(0)=g(x,\mu)+e_{0,y}\qquad z'(\xi)=h(z(\xi))+\tilde{e}(\xi)\qquad\text{where}
\int_0^{\hat{\psi}(t)}\infnorm{\tilde{e}(\xi)}d\xi=\int_0^t\infnorm{e_y(u)}du\]
Assume that $x\in\dom{f}$ and let $T\in\Rp$ such that $\hat{e}(T)\leqslant e^{-\Theta^*(\infnorm{x},\mu)}$.
Then $\hat{e}(T)<e^{-\Theta(\infnorm{x},\mu)}$ and for all $t\in[0,T]$:
\begin{align*}
\infnorm{e_{0,y}}+\int_0^{\hat{\psi}(t)}\infnorm{\tilde{e}}(u)du&=\infnorm{e_{0,y}}+\int_0^t\infnorm{e_y(u)}du\\
    &\leqslant\hat{e}(t)\leqslant e^{-\Theta(\infnorm{x},\mu)}
\end{align*}
Apply Lemma~\ref{lem:finite_time_robust} to get for all $u\in[0,\hat{\psi}(T)]$:
\begin{equation}\infnorm{z(u)}\leqslant\Upsilon(\infnorm{x},\mu,u)\end{equation}
\begin{equation}\text{if }u\geqslant\Omega(\infnorm{x},\mu)\text{ then }\infnorm{z_{1..m}(u)-f(x)}\leqslant e^{-\mu}\end{equation}
Apply Lemmas~\ref{lem:max} and~\ref{lem:norm} to get for all $t\in[0,T]$:
\begin{align*}
\ell(t)&\geqslant \mx_1(\norm_{\infty,1}(\infnorm{x},\mu))+1-\infnorm{e_{0,\ell}}+t-\int_0^t\infnorm{e_\ell(u)}du\\
    &\geqslant \max(\infnorm{x},\mu)+1+t-\hat{e}(t)\\
    &\geqslant\max(\infnorm{x},\mu,t)\tag*{using that $\hat{e}(t)\leqslant1$}
\end{align*}
Consequently, using Lemma~\ref{lem:norm}, for all $t\in[0,T]$:
\begin{align*}\label{eq:robust_to_strong:ineq_Delta}
\Delta(t)&\geqslant\Upsilon(\ell(t),\ell(t),\ell(t))-\infnorm{y(t)}\\
    &\geqslant\Upsilon(\infnorm{x},\mu,t)-\infnorm{y(t)}\tag*{using that $\ell(t)\geqslant\max(\infnorm{x},\mu,t)$}\\
    &=\Upsilon(\infnorm{x},\mu,t)-\infnorm{z(\hat{\psi}(t))}\tag*{using that $y(t)=z(\hat{\psi}(t))$}\\
    &\geqslant0\tag*{because $\hat{\psi}(t)\in[0,\hat{\psi}(T)]$}
\end{align*}
Consequently for all $t\in[0,T]$:
\[\hat{\psi}(t)=\int_0^t\psi(u)du=\int_0^t\frac{1+\tanh(\Delta(u))}{2}du\geqslant\frac{t}{2}\]
Define $\Omega^*(\alpha,\mu)=2\Omega(\alpha,\mu)$.
Assume that $T\geqslant \Omega^*(\infnorm{x},\mu)$ then $\hat{\psi}(T)\geqslant\Omega(\infnorm{x},\mu)$
and thus $\infnorm{y_{1..m}(T)-f(x)}=\infnorm{z(\hat{\psi}(T))-f(x)}\leqslant e^{-\mu}$.

Finally, $(y,\ell)(0)=g^*(x,\mu)+e_0$ where $g^*\in\gpval$. Similarly $(y,\ell)'(t)=h^*((y,\ell)(t))+e(t)$
where $h^*\in\gpval$. Note again that both $h^*$ and $g^*$ are defined over the entire space.
This concludes the proof that $f\in\gsc{\Omega^*}{\poly}{\Theta^*}$.
\end{proof}

\subsection{Proof that ASP implies AXP}
\label{app:strong_to_unaware}

The purpose is to deal with the fact a system could explode
(i.e. behave uncorrectly) for inputs not in the domain of the
function, or for too big perturbation of the dynamics, by adding a
mechanism to forbid explosions in these cases

Formally:

\begin{theorem}[Strong $\subseteq$ \unaware{}]\label{th:strong_to_unaware}$\gpsc\subseteq\gpuc$.
If $f\in\gpsc$ then there exists polynomials $\Upsilon,\Lambda,\Theta$
and a constant polynomial $\Omega$ such that $f\in\guc{\Upsilon}{\Omega}{\Lambda}{\Theta}$.
\end{theorem}

where

\begin{definition}[\Unaware{} computability]\label{def:guc}
Let $n,m\in\N$, $f:\subseteq\R^n\rightarrow\R^m$, $\Upsilon:\Rp^3\rightarrow\Rp$
and $\Omega,\Lambda,\Theta:\Rp^2\rightarrow\Rp$.
We say that $f$ is $(\Upsilon,\Omega,\Lambda,\Theta)$-\unawarely{}-computable
if and only if there exists $\delta\geqslant0$, $d\in\N$ and
$(g:\R^{d}\times\R^{n+1}\rightarrow\R^d)\in\gpval[\K]{}$  such that for any
$x\in C^0(\Rp,\R^n)$, $\mu\in C^0(\Rp,\Rp)$, $y_0\in\R^d$, $e\in C^0(\Rp,\R^d)$
there exists (a unique) $y:\Rp\rightarrow\R^d$ satisfying for all $t\in\Rp$:
\begin{itemize}
\item $y(0)=y_0$ and $y'(t)=g(t,y(t),x(t),\mu(t))+e(t)$
\item $\infnorm{y(t)}\leqslant
\Upsilon\left(\pastsup{\delta}{\infnorm{x}}(t),\pastsup{\delta}{\mu}(t),
    \infnorm{y_0}\indicator{[1,\delta]}(t)+\int_{\max(0,t-\delta)}^t\infnorm{e(u)}du\right)$
\item For any $I=[a,b]$, if there exists $\bar{x}\in\dom{f}$ and $\check{\mu},\hat{\mu}\geqslant0$ such that
for all $t\in I$, $\mu(t)\in[\check{\mu},\hat{\mu}]$, $\infnorm{x(t)-\bar{x}}\leqslant e^{-\Lambda(\infnorm{\bar{x}},\hat{\mu})}$ and
$\int_{a}^b\infnorm{e(u)}du\leqslant e^{-\Theta(\infnorm{\bar{x}},\hat{\mu})}$ then
$\infnorm{y_{1..m}(u)-f(\bar{x})}\leqslant e^{-\check{\mu}}$ whenever
$a+\Omega(\infnorm{\bar{x}},\hat{\mu})\leqslant u\leqslant b$.
\end{itemize}
We denote by $\guc{\Upsilon}{\Omega}{\Lambda}{\Theta}$ the set of
$(\Upsilon,\Omega,\Lambda,\Theta)$-\unawarely{}-computable functions
and by $\gpuc$ the set of
$(\poly,\poly,\poly,\poly)$-\unawarely{}-computable functions.
\end{definition}

A very common pattern in signal processing is known as ``sample and hold'',
where we have a variable signal and we would like to apply some process to it. Unfortunately,
the processor often assumes (almost) constant input and does not work in real time (analog-to-digital
converters are typical example).
In this case, we cannot feed the signal directly to the processor so we need some black
box that samples the signal to capture its value, and hold this value long enough
for the processor to compute its output. This process is usually used in a $\tau$-periodic
fashion: the box samples for time $\delta$ and holds for time
$\tau-\delta$.

The following is proved in \cite{TheseAmaury}


\begin{lemma}[Sample and hold, \cite{TheseAmaury}]\label{lem:sample}
There is a family of functions  $\sample_{I,\tau}(t,\mu,x,g) \in \gpval$, where
$t\in\R,\mu,\tau\in\Rp,x,g\in\R,I=[a,b]\subsetneq[0,\tau]$,  such that: 
Let $\tau\in\Rp$, $I=[a,b]\subsetneq[0,\tau]$, $y:\Rp\rightarrow\R$,
$y_0\in\R$, $x,e\in C^0(\Rp,\R)$ and
$\mu:\Rp\rightarrow\Rp$ an increasing function. Suppose that for all $t\in\Rp$:
\[y(0)=y_0\qquad y'(t)=\sample_{I,\tau}(t,\mu(t),y(t),x(t))+e(t)\]
Then:
\[|y(t)|\leqslant2+\smashoperator{\int_{\max(0,t-\tau-|I|)}^t}|e(u)|du+\max\left(|y(0)|\indicator{[0,b]}(t),\pastsup{\tau+|I|}|x|(t)\right)\]
Furthermore:
\begin{itemize}
\item if $t\notin I\pmod{\tau}$ then $|y'(t)|\leqslant e^{-\mu(t)}+|e(t)|$
\item for $n\in\N$, if there exists $\bar{x}\in\R$ and $\nu,\nu'\in\Rp$ such that $|\bar{x}-x(t)|\leqslant e^{-\nu}$ and $\mu(t)\geqslant\nu'$ for
all $t\in n\tau+I$ then $|y(n\tau+b)-\bar{x}|\leqslant\int_{n\tau+I}|e(u)|du+e^{-\nu}+e^{-\nu'}$
\item for $n\in\N$, if there exists $\check{x},\hat{x}\in\R$ and $\nu\in\Rp$ such that $x(t)\in[\check{x},\hat{x}]$ and $\mu(t)\geqslant\nu$ for
all $t\in n\tau+I$ then $y(n\tau+b)\in[\check{x}-\varepsilon,\hat{x}+\varepsilon]$ where $\varepsilon=2e^{-\nu}+\int_{n\tau+I}|e(u)|du$
\item for any $J=[c,d]\subseteq\Rp$, if there exists $\nu,\nu'\in\Rp$ and $\bar{x}\in\R$ such that
$\mu(t)\geqslant\nu'$ for all $t\in J$ and $|x(t)-\bar{x}|\leqslant e^{-\nu}$ for all $t\in J\cap(n\tau+I)$ for some $n\in\N$,
then $|y(t)-\bar{x}|\leqslant e^{-\nu}+e^{-\nu'}+\int_{t-\tau-|I|}^t|e(u)|du$ for all $t\in[c+\tau+|I|,d]$
\item if there exists $\Omega:\Rp\rightarrow\Rp$ such that for any $J=[a,b]$ and $\bar{x}\in\R$ such that for all
$\nu\in\Rp$, $n\in\N$ and $t\in(n\tau+I)\cap[a+\Omega(\nu),b]$, $|\bar{x}-x(t)|\leqslant e^{-\nu}$; then
$|y(t)-\bar{x}|\leqslant e^{-\nu}$ for all $t\in[a+\Omega^*(\nu),b]$ where
$\Omega^*(\nu)=\max(\Omega(\nu+\ln3),\mu^{-1}(\nu+\ln3))+\tau+|I|$
\end{itemize}
\end{lemma}

\begin{lemma}[``periodic low-integral-low'']\label{lem:plil}
There is a family of functions $\plil_{I,\tau}\in\gpval$  where
$\mu,\tau\in\Rp$, $I=[a,b]\subsetneq[0,\tau]$ and $x\in\R$ such that: 
there exists a constant $K$ and $\phi$ such that $\plil_{I,\tau}(t,\mu,x)=\phi(t,\mu,x)x$ and:
\begin{itemize}
\item $\plil_{I,\tau}(\cdot,\mu,x)$ is $\tau$-periodic
\item $\forall t\notin I$, $|\plil_{I,\tau}(t,\mu,x)|<e^{-\mu}$
\item for any $\alpha:I\rightarrow\Rp,\beta:I\rightarrow\R$:
\[1\leqslant\int_a^b \phi(t,\alpha(t),\beta(t))dt\leqslant K\]
\end{itemize}
\end{lemma}

We then get to the proof of Theorem~\ref{th:strong_to_unaware}

\begin{proof}
Let $(f:\subseteq\R^n\rightarrow\R^m)\in\gsc{\Upsilon}{\Omega}{\Theta}$ where
$\Upsilon$, $\Omega$ $\Theta$ are polynomials which we assume, without loss of generability,
to be increasing functions of theirs inputs. Apply Definition~\ref{def:gsc} to get
$d$, $h$ and $g$.

Let $e=1+d+m$, $x\in C^0(\Rp,\R^n)$, $\mu\in C^0(\Rp,\Rp)$, $(\nu_0,y_0,z_0)\in\R^e$,
$(e_\nu,e_y,e_z)\in C^0(\Rp,\R^e)$
and consider the following system:
\[
\left\{\begin{array}{@{}r@{}l}
\nu(0)&=\nu_0\\y(0)&=y_0\\z(0)&=z_0
\end{array}\right.\quad
\left\{\begin{array}{@{}r@{}l}
\nu'(t)&=\sample_{[0,1],4}(t,\mu^*(t),\nu(t),\mu(t)+\ln\Delta+7)+e_\nu(t)\\
y'(t)&=\sample_{[1,2],4}(t,\mu^*(t),y(t),g(x(t),\nu(t)))\\
        &\hspace{1em}+\plil_{[2,3],4}(t,\mu^*(t),A(t)h(y(t)))+e_y(t)\\
z'(t)&=\sample_{[3,4],4}(t,\mu^*(t),z(t),y_{1..m}(t))+e_z(t)
\end{array}\right.
\]
where
\[\Delta=5\qquad\Delta'=\ln\Delta+10\]
\[\mu^*(t)=\mho^*(1+\norm_{\infty,1}(x(t)),\nu(t)+4)\]
\[A(t)=1+\Omega(1+\norm_{\infty,1}(x(t)),\nu(t))\]
\[\Lambda^*(\alpha,\mu)=\Theta^*(\alpha,\mu)=\mho^*(\alpha,\mu+\Delta')\]
\[\mho^*(\alpha,\mu)=\mu+\ln\Delta+\Theta(\alpha,\mu)+\ln q(\alpha+\mu)\]

Let $I=[a,b]$ and assume there exists $\bar{x}\in\dom{f}$ and $\check{\mu},\hat{\mu}\in\Rp$ such that
for all $t\in I$, $\mu(t)\in[\check{\mu},\hat{\mu}]$, $\infnorm{x(t)-\bar{x}}\leqslant e^{-\Lambda^*(\infnorm{\bar{x}},\hat{\mu})}$
and $\int_a^b\infnorm{e(u)}du\leqslant e^{-\Theta^*(\infnorm{\bar{x}},\hat{\mu})}$.
Apply Theorem~\ref{prop:generable_mod_cont} to $g$ to get $q\in\K[\R]$, without loss of generality
we can assume that $q$ is an increasing function and $q\geqslant1$.
We will use Lemma~\ref{lem:norm} to get that $\norm_{\infty,1}(x(t))+1\geqslant\infnorm{\bar{x}}$
because $\infnorm{x(t)-\bar{x}}\leqslant1$. Also note that $\mu^*,\Theta^*,\Lambda^*$
are increasing functions of their arguments.
Let $n\in\N$ such that $[4n,4n+4]\subseteq I$ and $t\in[4n,4n+4]$. We will first analyse the variable $\nu$,
note that the analysis is extremely rough to simplify the proof.
\begin{itemize}
\item \textbf{if $t\in[4n,4n+1]$ then} $\mu^*(t)\geqslant0$ so apply
  Lemma~\ref{lem:sample} to get that
$\nu(4n+1)\in[\check{\mu}+\ln\Delta+7-\varepsilon,\hat{\mu}+\ln\Delta+7+\varepsilon]$ where
$\varepsilon\leqslant2e^{-0}+\int_{4n}^{4n+1}|e_\nu(u)|du\leqslant3$
because $\int_a^b\infnorm{e(t)}\leqslant1$. Define $\bar{\nu}=\nu(4n+1)$, then $\bar{\nu}\in[\check{\mu}+\ln\Delta+4,\hat{\mu}+\underbrace{\ln\Delta+10}_{=\Delta'}]$
\item \textbf{if $t\in[4n+1,4n+4]$ then} $\mu^*(t)\geqslant0$ so apply
  Lemma~\ref{lem:sample} to get that
$|\nu'(t)|\leqslant e^{-0}+\int_{4n+1}^t|e_\nu(u)|du$ and thus $|\nu(t)-\bar{\nu}|\leqslant(t-4n-1)+\int_{4n+1}^t\infnorm{e(u)}du\leqslant4$
because $\int_a^b\infnorm{e(t)}\leqslant1$. In other words $\nu(t)\in[\bar{\nu}-4,\bar{\nu}+4]$.
\end{itemize}
Furthermore for $t\in[4n+1,4n+4]$ we have:
\[
\mu^*(t)\geqslant\Theta^*(1+\norm_{\infty,1}(x(t)),\nu(t)+4)
    \geqslant\mho^*(\infnorm{\bar{x}},\bar{\nu})
\]
It will also be useful to note that:
\begin{align*}
\Lambda^*(\infnorm{\bar{x}},\hat{\mu})
    =\Theta^*(\infnorm{\bar{x}},\hat{\mu})
    &\geqslant\mho^*(\infnorm{\bar{x}},\hat{\mu}+\Delta')\\
    &\geqslant\mho^*(\infnorm{\bar{x}},\bar{\nu})
\end{align*}
We can now analyze $y$ using this property:
\begin{itemize}
\item \textbf{if $t\in[4n+1,4n+2]$ then} $|\nu'(t)|\leqslant e^{-\mu^*(t)}+|e_\nu(t)|$
thus $|\nu(t)-\bar{\nu}|\leqslant e^{-\mho^*(\infnorm{\bar{x}},\bar{\nu})}+\int_{4n+1}^{4n+2}|e_\nu(u)|du$.
Furthermore $\sup_{[4n+1,4n+2]}\infnorm{x}\leqslant\infnorm{\bar{x}}+1$, thus:
\begin{align*}
\infnorm{g(\bar{x},\bar{\nu})-g(x(t),\nu(t))}
&\leqslant\max(|\nu(t)-\bar{\nu}|,\infnorm{x(t)-\bar{x}})q(\max(\infnorm{\bar{x}},|\bar{\nu}|))\\
&\leqslant\max\left(e^{-\Theta^*(\infnorm{\bar{x}},\hat{\mu})}+
    e^{-\mho^*(\infnorm{\bar{x}},\bar{\nu})},e^{-\Lambda^*(\infnorm{\bar{x}},\hat{\mu})}\right)q(\infnorm{\bar{x}}+\bar{\nu})\\
&\leqslant 2e^{-\Theta(\infnorm{\bar{x}},\bar{\nu})-\ln\Delta}
\end{align*}
Also note that $\infnorm{y'(t)-\sample_{[1,2],4}(t,\mu^*(t),y(t),g(x(t),\nu(t)))}\leqslant e^{-\mu^*(t)}$
by Lemma~\ref{lem:plil}. So we can apply Lemma~\ref{lem:sample} to get that
$\infnorm{y(4n+2)-g(\bar{x},\bar{\nu})}\leqslant
2e^{-\Theta(\infnorm{\bar{x}},\bar{\nu})-\ln\Delta}+e^{-\mho^*(\infnorm{\bar{x}},\bar{\nu})}+\int_{4n+1}^{4n+2}\infnorm{e(u)}du
\leqslant 4e^{-\Theta(\infnorm{\bar{x}},\bar{\nu})-\ln\Delta}$.
\item \textbf{if $t\in[4n+2,4n+3]$ then} apply Lemmas~\ref{lem:sample} and~\ref{lem:plil}
to get $\phi$ such that $\int_{4n+2}^{4n+3}\phi(u)du\geqslant1$ and
$\infnorm{y'(t)-\phi(t)A(t)h(y(t))}\leqslant e^{-\mu^*(t)}+\infnorm{e_y(t)}$.
Define $\psi(t)=\int_{4n+2}^t\phi(u)A(u)du$ then
$\psi(4n+3)\geqslant\Omega(\infnorm{\bar{x}},\bar{\nu})$ since
$A(u)\geqslant\Omega(\infnorm{\bar{x}},\bar{\nu})$ for $u\in[4n+2,4n+3]$.
Apply Lemma~\ref{lem:perturbed_time_scale_gpac} over $[4n+2,4n+3]$ to get that
$y(t)=w(\psi(t))$ where $w$ satisfies $w(0)=y(4n+2)$
and $w'(\xi)=h(w(\xi))+\tilde{e}(\xi)$ where $\tilde{e}\in C^0(\Rp,\R^d)$ satisfies
$\int_{0}^{\psi(t)}\infnorm{\tilde{e}(\xi)}d\xi=\int_{4n+2}^t\infnorm{e_y(u)}du
\leqslant e^{-\Theta^*(\infnorm{\bar{x}},\hat{\mu})}\leqslant e^{-\Theta(\infnorm{\bar{x}},\bar{\nu})-\ln\Delta}$.
Furthermore, $\infnorm{w(0)-g(\bar{x},\bar{\nu})}\leqslant4e^{-\Theta(\infnorm{\bar{x}},\bar{\nu})-\ln\Delta}$
from the result above. In other words, $w(0)=g(\bar{x},\bar{\nu})+\tilde{e}_0$
and $w'(t)=g(w(t))+\tilde{e}(t)$ where
$\infnorm{\tilde{e}_0}+\int_0^{\psi(t)}\infnorm{e(u)}du\leqslant 5e^{-\Theta(\infnorm{\bar{x}},\bar{\nu})-\ln\Delta}
\leqslant e^{-\Theta(\infnorm{\bar{x}},\bar{\nu})}$ because $\Delta\geqslant5$.
Apply Definition~\ref{def:gsc} to get that $\infnorm{w_{1..m}(\psi(4n+3))-f(\bar{x})}\leqslant e^{-\bar{\nu}}$
since $\psi(4n+3)\geqslant\Omega(\infnorm{\bar{x}},\bar{\nu})$.
\item \textbf{if $t\in[4n+3,4n+4]$ then} $\infnorm{y'(t)}\leqslant e^{-\mu^*(t)}+\infnorm{e_y(t)}$
thus $\infnorm{y(t)-y(4n+3)}\leqslant e^{-\mho^*(\infnorm{\bar{x}},\bar{\nu})}+\int_{4n+3}^t\infnorm{e_y(u)}du
\leqslant 2e^{-\bar{\nu}}$
so $\infnorm{y_{1..m}(t)-f(\bar{x})}\leqslant3e^{-\bar{\nu}}$.
\end{itemize}
Note that the above reasoning is also true for the last segment $[4n,b]\subseteq I$
in which case the result only applies up to time $b$ of course. In other words,
the results apply as long as $t\in[4n,4+4]\cap I$ and $4n\geqslant a$.
From this we conclude that if $t\in[a+4,b]\cap[4n+3,4n+3]$ for some $n\in\N$ then
$\infnorm{y_{1..m}(t)-f(\bar{x})}\leqslant3e^{-\bar{\nu}}$.
Apply Lemma~\ref{lem:sample} to get, using that $\bar{\nu}\geqslant\check{\mu}+\ln\Delta$ and $\Delta\geqslant5$,
that for all $t\in[a+5,b]$:
\begin{align*}
\infnorm{z(t)-f(\bar{x})}
    &\leqslant 3e^{-\bar{\nu}}+e^{-\mho^*(\infnorm{\bar{x}},\bar{\nu})}+\int_{t-5}^t\infnorm{e(u)}du
    \leqslant 5e^{-\bar{\nu}}\\
    &\leqslant e^{-\check{\mu}}
\end{align*}

To complete the proof, we must also analyze the norm of the system. As a shorthand, we introduce
the following notation:
\[\operatorname{int}_\delta^+\alpha(t)=\int_{\max(0,t-\delta)}^t\alpha(u)du\]
Apply Lemma~\ref{lem:sample}
to get that:
\begin{align*}
|\nu(t)|&\leqslant 2+\smashoperator{\int_{\max(0,t-5)}^t}|e_\nu(u)|du
        +\max\left(|\nu_0|\indicator{[0,4]}(t),\pastsup{5}{|\mu+\ln\Delta+7|}(t)\right)\\
        &\leqslant\poly\left(|\nu_0|\indicator{[0,5]}(t)+\operatorname{int}_{5}^+|e_\nu|(t),\pastsup{5}\mu(t)\right)
\end{align*}
The analysis of $y$ is a bit more painful, as it uses both results about the
sampling function and the strongly-robust system we are simulating. Let $n\in\N$,
and $t\in[4n,4n+4]$:
\begin{itemize}
\item \textbf{if $t\in[4n,4n+1]$ then} apply Lemmas~\ref{lem:sample} and~\ref{lem:plil}
to get, using that $\mu(t)\geqslant0$, that $\infnorm{y'(t)}\leqslant2+\infnorm{e(t)}$ and thus
$\infnorm{y(t)-y(4n)}\leqslant2+\int_{4n}^t\infnorm{e(u)}du$.
\item \textbf{if $t\in[4n+1,4n+2]$ then} using the result on $\nu$,
$\infnorm{g(x(t),\nu(t))}\leqslant\sup_{[4n+1,t]}\poly(\infnorm{x},\nu)
\leqslant \poly\left(|\nu_0|\indicator{[0,5]}(t)+\operatorname{int}_{6}^+\infnorm{e}(t),\pastsup{6}\mu(t),\pastsup{1}\infnorm{x}(t)\right)$.
Apply Lemmas~\ref{lem:sample} and~\ref{lem:plil}
to get, using that $\mu(t)\geqslant0$ and the result on $\nu$, that:
\begin{align*}
\infnorm{y(4n+2)}&\leqslant\sup_{[4n+1,4n+2]}\infnorm{g(x,\nu)}+2+\int_{4n+1}^{4n+2}\infnorm{e(u)}du\\
    &\leqslant\poly\left(|\nu_0|\indicator{[0,5]}(4n+2)+\operatorname{int}_{6}^+\infnorm{e}(4n+2),\pastsup{6}\mu(4n+2),
                         \pastsup{1}\infnorm{x}(4n+2)\right)
\end{align*}
and also that:
\begin{align*}
\infnorm{y(t)}&\leqslant\max\left(\sup_{[4n+1,t]}\infnorm{g(x,\nu)}+2,\infnorm{y(4n+1)}\right)+\int_{4n+1}^t\infnorm{e(u)}du\\
    &\leqslant\poly\left(|\nu_0|\indicator{[0,5]}(t)+\operatorname{int}_{6}^+\infnorm{e}(t),\pastsup{6}\mu(t),
                         \pastsup{1}\infnorm{x}(t),\infnorm{y(4n)}\right)
\end{align*}
\item \textbf{if $t\in[4n+2,4n+3]$ then} apply Lemma~\ref{lem:sample},
  Lemmas~\ref{lem:plil},
\ref{lem:perturbed_time_scale_gpac} and~\ref{def:gsc} to get that
$\infnorm{y(t)}\leqslant \Upsilon(0,0,\hat{e}(\hat{A}(t)),\hat{A}(t))$
where $\hat{A}(t)=\int_{4n+2}^tA(u)du$ and $\hat{e}(\hat{A}(t))=\infnorm{y(4n+2)-g(0,0)}+\int_{4n+2}^t1+\infnorm{e(u)}du$.
Since $\Omega$ is a polynomial, and using the result on $\nu$, we get that:
\begin{align*}
\hat{A}(t)&\leqslant\sup_{[4n+2,t]}\poly(\infnorm{x},|\nu|)\\
    &\leqslant\poly\left(|\nu_0|\indicator{[0,5]}(t)+\operatorname{int}_{6}^+\infnorm{e},\pastsup{6}\mu(t),\pastsup{1}{\infnorm{x}}(t)\right)
\end{align*}
and using that $4n+2\leqslant t\leqslant4n+3$:
\begin{align*}
\infnorm{y(4n+2)-g(0,0)}&\leqslant\infnorm{y(4n+2)}+\infnorm{g(0,0)}\\
    &\leqslant\poly\left(|\nu_0|\indicator{[0,5]}(t)+\operatorname{int}_{6}^+\infnorm{e},\pastsup{7}\mu(t),
                         \pastsup{2}\infnorm{x}(t)\right)
\end{align*}
And since $\Upsilon$ is a polynomial, we conclude that:
\[
\infnorm{y(t)}\leqslant\poly\left(|\nu_0|\indicator{[0,5]}(t)+\operatorname{int}_{6}^+\infnorm{e}(t),\pastsup{7}\mu(t),
                         \pastsup{2}\infnorm{x}(t)\right)
\]
\item \textbf{if $t\in[4n+3,4n+4]$ then} apply Lemmas~\ref{lem:sample} and~\ref{lem:plil}
to get, using that $\mu(t)\geqslant0$, that $\infnorm{y'(t)}\leqslant2+\infnorm{e(t)}$ and thus
$\infnorm{y(t)-y(4n+3)}\leqslant2+\int_{4n+3}^t\infnorm{e(u)}du$.
\end{itemize}
From this analysis we can conclude that for all $t\in[0,2]$:
\begin{align*}
\infnorm{y(t)}&\leqslant\poly\left(|\nu_0|\indicator{[0,5]}(t)+\operatorname{int}_{6}^+\infnorm{e}(t),\pastsup{6}\mu(t),
                         \pastsup{1}\infnorm{x}(t),\infnorm{y(0)}\right)\\
                &\leqslant\poly\left(|\nu_0|+\operatorname{int}_{6}^+\infnorm{e}(t),\pastsup{6}\mu(t),
                         \pastsup{1}\infnorm{x}(t),\infnorm{y_0}\right)
\end{align*}
and for all $n\in\N$ and $t\in[4n+2,4n+6]$:
\[
\infnorm{y(t)}\leqslant\poly\left(|\nu_0|\indicator{[0,5]}(t)+\operatorname{int}_{9}^+\infnorm{e}(t),\pastsup{9}\mu(t),
                         \pastsup{4}\infnorm{x}(t)\right)
\]
Putting everything together, we get for all $t\in\Rp$:
\[
\infnorm{y(t)}\leqslant\poly\left(\infnorm{y_0,\nu_0}\indicator{[0,5]}(t)+\operatorname{int}_{9}^+\infnorm{e}(t),\pastsup{9}\mu(t),
                         \pastsup{4}\infnorm{x}(t)\right)
\]
Finally apply Lemma~\ref{lem:sample} to get the a similar bound on $z$ and thus
on the entire system.
\end{proof}

\subsection{Proof that AXP implies AOP}
\label{app:unaware_to_online}

The purpose is now to go to a notion of online computation, i.e. to
Lemma~\ref{lemma:online}




We start by the following lemmas:

\begin{lemma}[$\gpuc$ time rescaling]\label{lem:gpuc_time_rescaling}
If $f\in\gpuc$ then there exists polynomials $\Upsilon,\Lambda,\Theta$
and a constant polynomial $\Omega$ such that $f\in\guc{\Upsilon}{\Omega}{\Lambda}{\Theta}$.
\end{lemma}

\begin{proof}
We go for the shortest proof: we will show that $\gpuc\subseteq\gpwc$
and use Theorem~\ref{th:weak_to_robust} then Theorem
\ref{th:robust_to_strong} followed by Theorem~\ref{th:strong_to_unaware}
which proves exactly our statement.

The proof that $\gpuc\subseteq\gpwc$ is next to trivial because the \unaware{} system and some given input and precision,
we can simply store the input and precision into some variables and feed them into the system.
We make the system autonomous by using a variable to store the time.

Let
$(f:\subseteq\R^n\rightarrow\R^m)\in\guc{\Upsilon}{\Omega}{\Lambda}{\Theta}$,
apply Definition~\ref{def:guc} to get
$\delta, d$ and $g$.
Let $x\in\dom{f}$ and $\mu\in\Rp$, and consider the following system:
\[
\left\{\begin{array}{@{}r@{}l@{}}
x(0)&=x\\\mu(0)&=\mu\\\tau(0)&=0\\y(0)&=0
\end{array}\right.
\qquad
\left\{\begin{array}{@{}r@{}l@{}}
x'(t)&=0\\\mu'(t)&=0\\\tau'(t)&=1\\
y'(t)&=g(t,y(t),x(t),\mu(t))
\end{array}\right.
\]
Clearly he system of the form $z(0)=h(x,\mu)$ and $z'(t)=H(z(t))$
where $h$ and $H$ belong to $\gpval$ (and are defined over the entire space).
Apply the definition to get that:
\[\infnorm{y(t)}\leqslant\Upsilon(\infnorm{x},\mu,0)\]
And thus the entire system in bounded by a polynomial in $\infnorm{x},\mu$
and $t$.
Furthermore, if $t\geqslant\Omega(\infnorm{x},\mu)$ then $\infnorm{y_{1..m}(t)-f(x)}\leqslant e^{-\mu}$.
To conclude the proof, we need to rewrite the system as a PIVP using
Theorem~\ref{prop:gpac_ext_ivp_stable_pre}.
\end{proof}

The following is established in \cite{TheseAmaury}

\begin{lemma}[Reach]\label{lem:reach}
There exists $\reach\in\gpval$ such that: 
For any $I=[a,b]$, any $\phi\in C^0(I,\Rp)$, any $g,E\in C^0(I,\R)$, any $y_0,g_\infty\in\R$ and $\eta>0$
such that for all $t\in I$, $|g(t)-g_\infty|\leqslant\eta$.
Assume that $y:I\rightarrow\R$ satisfies
\[
\left\{\begin{array}{@{}r@{}l}y(0)&=y_0\\y'(t)&=\reach(\phi(t),y(t),g(t))+E(t)\end{array}\right.
\]
Then for any $t\in I$,
\[|y(t)-g_\infty|\leqslant\eta+\int_a^t|E(u)|du+\exp\left(-\int_a^t\phi(u)du\right)
\qquad\text{whenever }\int_a^t\phi(u)du\geqslant1\]
And for any $t\in I$,
\[|y(t)-g_\infty|\leqslant\max(\eta,|y(0)-g_\infty|)+\int_0^t|E(u)|du\]
\end{lemma}

We then get to the proof of Lemma~\ref{lemma:online}.

\begin{proof}
Apart from the issue of the input, the system is quite intuitive: we constantly feed
the \unaware{} system with the (smoothed) input and some precision. By increasing
the precision with time, we ensure that the system will converge when the input is stable.
However there is a small catch: over a time interval $I$, if we change the precision
within a range $[\check{\mu},\hat{\mu}]$ then we must provide the \unaware{} system
with precision based on $\hat{\mu}$ in order to get precision $\check{\mu}$.
Since the \unaware{} system takes time $\Omega(\infnorm{x},\hat{\mu})$
to compute, we need arrange so that the requested precision doesn't change too much over periods
of this duration to make things simpler. We will use to our advantage
that $\Omega$ can always be assumed to be a constant.

Let $(f:\subseteq\R^n\rightarrow\R^m)\in\guc{\Upsilon}{\Omega}{\Lambda}{\Theta}$
where $\Upsilon,\Omega,\Lambda$ and $\Theta$ are polynomials, which we can assume
to be increasing functions of their arguments. Apply Lemma~\ref{lem:gpuc_time_rescaling} to
get $\omega>0$ such that for all $\alpha\in\R^n,\mu\in\Rp$:
\[\Omega(\alpha,\mu)=\omega\]
Apply Definition~\ref{def:guc} to get
$\delta,d$ and $g$.
Define:
\[\tau=\omega+2\qquad\delta'=\max(\delta,\tau+1)\]
Let $x\in C^0(\Rp,\R^n)$ and consider the following systems:
\[
\left\{\begin{array}{@{}r@{}l@{}}
x^*(0)&=0\\
y(0)&=0\\
z(0)&=0\\
\end{array}\right.
\qquad
\left\{\begin{array}{@{}r@{}l@{}}
{x^*}'(t)&=\reach(\phi(t),x^*(t),x(t))\\
y'(t)&=g(t,y(t),x^*(t),\mu(t))\\
z'(t)&=\sample_{[\omega+1,\omega+2],\tau}(t,\mu(t),z(t),y_{1..m}(t))
\end{array}\right.
\]
where
\[\phi(t)=\ln2+\mu(t)+\Lambda^*(2+x_1(t)^2+\cdots+x_n(t)^2,\mu(t))\qquad\mu(t)=\frac{t}{\tau}\]
Let $t\geqslant1$,
since $\phi\geqslant1$ then Lemma~\ref{lem:reach} gives:
\[
\infnorm{x^*(t)}\leqslant\pastsup{1}{\infnorm{x}}(t)+e^{-\int_{t-1}^t\phi(u)du}
    \leqslant\pastsup{1}{\infnorm{x}}(t)+1
\]
Also for $t\in[0,1]$ we get that:
\[\infnorm{x^*(t)}\leqslant\sup_{[0,t]}\infnorm{x}\]
This proves that $\infnorm{x^*(t)}\leqslant\pastsup{1}{\infnorm{x}}(t)+1$ for all $t\in\Rp$.
From this we deduce that:
\begin{align*}
\infnorm{y(t)}&\leqslant\Upsilon(\pastsup{\delta}\infnorm{x^*}(t),\pastsup{\delta}{\mu}(t),0)\\
    &\leqslant\poly(\pastsup{\delta}\infnorm{x}(t),t)
\end{align*}
Apply Lemma~\ref{lem:sample} to get that:
\begin{align*}
\infnorm{z(t)}&\leqslant2+\pastsup{\tau+1}\infnorm{y}(t)\\
    &\leqslant\poly(\pastsup{\delta'}\infnorm{x}(t),t)
\end{align*}
Let $I=[a,b]$ and assume there exists $\bar{x}\in\dom{f}$ and $\bar{\mu}$ such that
for all $t\in I$, $\infnorm{x(t)-\bar{x}}\leqslant e^{-\Lambda(\infnorm{\bar{x}},\bar{\mu})}$.
Note that $2+\sum_{i=1}^nx_i(t)^2\geqslant1+\infnorm{x(t)}\geqslant\infnorm{\bar{x}}$
for all $t\in I$.
Let $n\in\N$ such that $n\geqslant\bar{\mu}+\ln2$ and $[n\tau,(n+1)\tau]\subseteq I$.
Note that $\mu(t)\in[n,n+1]$ for all $t\in I_n$.
Apply Lemma~\ref{lem:reach}, using that $\phi\geqslant1$,
to get that for all $t\in[n\tau+1,(n+1)\tau]$:
\begin{align*}
\infnorm{x^*(t)-\bar{x}}
    &\leqslant e^{-\Lambda^*(\infnorm{\bar{x}},n)}+e^{-\int_{n\tau}^t\phi(u)du}
    \leqslant 2e^{-\Lambda^*(\infnorm{\bar{x}},n)}\\
    &\leqslant e^{-\Lambda(\infnorm{\bar{x}},\bar{\mu}+\ln2)}
\end{align*}
Using the definition of \unaware{} computability, we get that for all
$t\in[n\tau+1+\omega,(n+1)\tau]=[n\tau+\omega+1,n\tau+\omega+2]$:
\[\infnorm{y_{1..m}-f(\bar{x})}\leqslant e^{-\bar{\mu}+\ln2}\]
Define $J=[a+(1+\bar{\mu}+\ln2)\tau,b]\subseteq I$. Assume that $t\in J\cap[n\tau+1,(n+1)\tau]$
for some $n\in\N$, then we must have $(n+1)\tau\geqslant (1+\bar{\mu}+\ln2)\tau$
and thus $n\geqslant \bar{\mu}+\ln2$ so we can apply the above reasoning to get that
$\infnorm{y_{1..m}(t)-f(x)}\leqslant e^{-\bar{\mu}+\ln2}$.
Furthermore, we also have $\mu(t)\geqslant\frac{(1+\bar{\mu}+\ln2)\tau}{\tau}\geqslant \bar{\mu}+\ln2$
for all $t\in J$.
Apply Lemma~\ref{lem:sample} to conclude that for any $t\in [a+\tau+\bar{\mu}+\ln2+\tau+1,b]$,
we have $\infnorm{z(t)-f(x)}\leqslant 2e^{-\bar{\mu}+\ln2}\leqslant e^{-\bar{\mu}}$.

To conclude the proof, we need to rewrite the system as a PIVP using
Lemma~\ref{prop:gpac_ext_ivp_stable_pre}.
Note that this works because we only rewrite the variable $y$, and doing so we
require that $x^*$ be a $C^1$ function (which is the case) and the new initial
variable will depend on $x^*(0)=0$ which is constant.
\end{proof}

\subsection{Proof of Theorem~\ref{th:real_setp_robust}}
\label{app:th:real_setp_robust}





We first precise some concepts.

\subsubsection{More on Turing Machines}

First the  \emph{step} function of a Turing machine $\mathcal{M}$
corresponds to the function
defined by:
\[\mathcal{M}(x,\sigma,y,q)=\begin{cases}
(\emptyword,b,\sigma'y,q') & \text{if }d=L\text{ and }x=\emptyword\\
(x_{2..|x|},x_1,\sigma'y,q') & \text{if }d=L\text{ and }x\neq\emptyword\\
(x,\sigma',y,q') & \text{if }d=S\\
(\sigma'x,b,\emptyword,q') & \text{if }d=R\text{ and }y=\emptyword\\
(\sigma'x,y_1,y_{2..|y|},q') & \text{if }d=R\text{ and }y\neq\emptyword\\
\end{cases}\quad\text{where }
\left\{\begin{array}{@{}l@{}}q'=\delta_1(q,\sigma)\\
\sigma'=\delta_2(q,\sigma)\\
d=\delta_3(q,\sigma)
\end{array}\right.
\]

\begin{definition}[Result of a computation]\label{def:tm_compute}
The \emph{result of a computation} of $\mathcal{M}$ on a word $w\in\Sigma^*$
is defined by:
\[\mathcal{M}(w)=\begin{cases}
x&\text{if }\exists n\in\N, \fiter{\mathcal{M}}{n}(c_0(w))=c_\infty(x)\\
\bot&\text{otherwise}
\end{cases}\]
\end{definition}

\begin{remark}
The result of a computation is well-defined because we imposed that when a machine
reaches a halting state, it does not move, change state or change the symbol
under the head.
\end{remark}

\subsubsection{Polynomial interpolation}

In order to implement the transition function of the Turing Machine,
we will use a polynomial interpolation scheme (Lagrange interpolation).
But since our simulation may have to deal with some amount of error in inputs, 
we have to investigate how this error propagates through the interpolating polynomial.

\begin{definition}[Lagrange polynomial]\label{def:lagrange_poly}
Let $d\in\N$ and $f:G\rightarrow\R$ where $G$ is a finite subset of $\R^d$, we define
\[\lagrange{f}(x)=\sum_{\bar{x}\in G}f(\bar{x})\prod_{\substack{y\in G\\y\neq\bar{x}}}\prod_{i=1}^d\frac{x_i-y_i}{\bar{x}_i-y_i}\]
\end{definition}

\begin{lemma}[Lagrange interpolation]\label{lem:lagrange_interp}
For any finite $G\subseteq\K^d$ and $f:G\rightarrow\K$, $\lagrange{f}\in\gpc$
and $\frestrict{\lagrange{f}}{G}=f$.
\end{lemma}

\begin{proof}
The fact that $\lagrange{f}$ matches $f$ on $G$ is a classical calculation. Also
$\lagrange{f}$ is a polynomial with coefficients in $\K$ so clearly it belongs to $\gpc$.
\end{proof}


%
%

We will often need to interpolate characteristic functions,
that is polynomials that value $1$ when $f(x)=a$ and $0$ otherwise.
For convenience we define a special notation for it.

\begin{definition}[Characteristic interpolation]
Let $d\in\N$, $f:G\rightarrow\R$ where $G$ is a finite subset of $\R^d$, $\alpha\in\R$, and define:
\[\lagreq{f}{\alpha}(x)=\lagrange{f_\alpha}(x)\qquad
\lagrneq{f}{\alpha}(x)=\lagrange{1-f_\alpha}(x)\qquad
f_\alpha(x)=\begin{cases}1&\text{if }f(x)=\alpha\\0&\text{otherwise}\end{cases}\]
\end{definition}

\begin{lemma}[Characteristic interpolation]\label{lem:char_interp}
For any finite $G\subseteq\K^d$, $f:G\rightarrow\K$ and $\alpha\in\K$, $\lagreq{f}{\alpha},\lagrneq{f}{\alpha}\in\gpc$.
\end{lemma}

\begin{proof}
Observe that $f_\alpha:G\rightarrow\{0,1\}$ and
$\{0,1\}\subseteq\K$. Apply Lemma~\ref{lem:lagrange_interp}.
\end{proof}

\subsubsection{Specific Functions and Operations}

We need some specific adhoc functions:


\begin{definition}[Round]\label{def:comp:round}
Let $\crnd\in C^0(\R,\R)$ be the unique function such that:
\begin{itemize}
\item $\crnd(x,\mu)=n$ for all $x\in\left[n-\frac{1}{2}+e^{-\mu},n+\frac{1}{2}-e^{-\mu}\right]$
for all $n\in\Z$
\item $\crnd(x,\mu)$ is affine over $\left[n+\frac{1}{2}-e^{-\mu},n+\frac{1}{2}+e^{-\mu}\right]$
for all $n\in\Z$
\end{itemize}
\end{definition}

\begin{theorem}[Round, \cite{TheseAmaury}]\label{th:comp:round}
$\crnd\in\gpc$.
\end{theorem}

The idea of the proof of above theorem is to build a function computing the ``fractional part'' function,
by this we mean a $1$-periodic function that maps $x$ to $x$ over $[-1+e^{-\mu},1-e^{-\mu}]$
and is affine at the border to be continuous. The rounding function immediately
follows by subtracting the fractional of $x$ to $x$. In the details, building this
function is not immediate. The intuition is that $\frac{1}{2\pi}\arccos(\cos(2\pi x))$
works well over $[0,1/2-e^{-\mu}]$ but needs to be fixed at the border (near $1/2$),
and also its parity needs to be fixed based on the sign of $\sin(2\pi x)$.

\begin{theorem}[Closure by arithmetic operations]\label{th:gpac_comp_arith}
If $f,g\in\gpc$ then $f\pm g,fg\in\gpc$, with the obvious restrictions on the domains
of definition.
\end{theorem}
\begin{proof}
We do the proof in the case of $f+g$ in details. Let $\Omega,\Upsilon,\Omega',\Upsilon'$
polynomials such that $f\in\gc{\Upsilon}{\Omega}$ and $g\in\gc{\Upsilon'}{\Omega'}$.
Apply Definition~\ref{def:gc} to $f$ and $g$
to get $d,p,q$ and $d',p',q'$ respectively. Let $x\in\dom{f}\cap\dom{g}$ and consider
the following system:
\[
\left\{\begin{array}{r@{}l}y(0)&=q(x)\\z(0)&=q'(x)\\w(0)&=q(x)+q'(x)\end{array}\right.
\qquad
\left\{\begin{array}{r@{}l}y'(t)&=p(y(t))\\z'(t)&=p'(z(t))\\w'(t)&=y'(t)+z'(t)\end{array}\right.
\]
Let $\Omega^*(\alpha,\mu)=\max(\Omega(\alpha,\mu+\ln2),\Omega'(\alpha,\mu+\ln2))$ and
$\Upsilon^*(\alpha,t)=\Upsilon(\alpha,t)+\Upsilon'(\alpha,t)$.
Since, by construction, $w(t)=y(t)+z(t)$, if $t\geqslant\Omega^*(\alpha,\mu)$
then $\infnorm{y_{1..m}(t)-f(x)}\leqslant e^{-\mu-\ln2}$ and $\infnorm{z_{1..m}(t)-g(x)}\leqslant e^{-\mu-\ln2}$
thus $\infnorm{w_{1..m}(t)-f(x)-g(x)}\leqslant e^{-\mu}$. Furthermore,
$\infnorm{y(t)}\leqslant\Upsilon(\infnorm{x},t)$ and $\infnorm{z(t)}\leqslant\Upsilon'(\infnorm{x},t)$
thus $\infnorm{w(t)}\leqslant\Upsilon^*(\infnorm{x},t)$.

The case of $f-g$ is exactly the same. The case of $fg$ is slightly more involved:
one need to take $w'(t)=y_1'(t)z_1(t)+y_1(t)z_1'(t)=p_1(y(t))z_1(t)+y_1(t)p_1'(z(t))$
so that $w(t)=y(t)z(t)$. The error analysis is a bit more complicated.
First note that $\infnorm{f(x)}\leqslant1+\Upsilon(\infnorm{x},\Omega(\infnorm{x},0))$
and $\infnorm{g(x)}\leqslant1+\Upsilon'(\infnorm{x},\Omega'(\infnorm{x},0))$,
and denote by $\ell(\infnorm{x})$ and $\ell^*(\infnorm{x})$ those two bounds respectively.
Let $t\geqslant\Omega(\infnorm{x},\mu+\ln2\ell^*(\infnorm{x}))$ then $\infnorm{y_1(t)-f(x)}\leqslant e^{-\mu-\ln2\infnorm{g(x)}}$
and similarly if $t\geqslant\Omega'(\infnorm{x},\mu+\ln2(1+\ell^*(\infnorm{x})))$ then
$\infnorm{z_1(t)-g(x)}\leqslant e^{-\mu-\ln2(1+\infnorm{f(x)})}$. Thus for $t$ greater
than the maximum of both bounds, $\infnorm{y_1(t)z_1(t)-f(x)g(x)}\leqslant\infnorm{(y_1(t)-f(x))g(x)}+\infnorm{y_1(t)(z_1(t)-g(x))}
\leqslant e^{-\mu}$ because $\infnorm{y_1(t)}\leqslant1+\infnorm{f(x)}\leqslant1+\ell(\infnorm{x})$.
\end{proof}

\begin{theorem}[Closure by composition]\label{th:gpac_comp_composition}
If $f,g\in\gpc$ and $f(\dom{f})\subseteq\dom{g}$ then $g\circ f\in\gpc$.
\end{theorem}
\begin{proof}
Let $f:I\subseteq\R^n\rightarrow J\subseteq\R^m$ and $g:J\rightarrow K\subseteq\R^l$.
We will show that $g\circ f$ is computable by using the fact that both $f$ and $g$ are
online-computable. We could show directly that $g\circ f$ is online-computable but this would
only complicated the proof for no apparent gain.

Apply Lemma~\ref{lemma:online} to get that $g$ is
$(\Upsilon,\Omega,\Lambda)$-online-computable,

where
\begin{definition}[Online computability]\label{def:goc}
Let $n,m\in\N$, $f:\subseteq\R^n\rightarrow\R^m$ and $\Upsilon,\Omega,\Lambda:\Rp^2\rightarrow\Rp$.
We say that $f$ is $(\Upsilon,\Omega,\Lambda)$-online-computable if and only if there exists $\delta\geqslant0$, $d\in\N$ and
$p\in\K^d[\R^d\times\R^{n}]$ and $y_0\in\K^d$ such that for any $x\in C^0(\Rp,\R^n)$,
there exists (a unique) $y:\Rp\rightarrow\R^d$ satisfying for all $t\in\Rp$:
\begin{itemize}
\item $y(0)=y_0$ and $y'(t)=p(y(t),x(t))$
\item $\infnorm{y(t)}\leqslant\Upsilon\big(\pastsup{\delta}{\infnorm{x}}(t),t\big)$
\item For any $I=[a,b]$, if there exists $\bar{x}\in\dom{f}$ and $\bar{\mu}\geqslant0$ such that
for all $t\in I$, $\infnorm{x(t)-\bar{x}}\leqslant e^{-\Lambda(\infnorm{\bar{x}},\bar{\mu})}$ then
$\infnorm{y_{1..m}(u)-f(\bar{x})}\leqslant e^{-\bar{\mu}}$ whenever
$a+\Omega(\infnorm{\bar{x}},\bar{\mu})\leqslant u\leqslant b$.
\end{itemize}
We denote by $\goc{\Upsilon}{\Omega}{\Lambda}$ the set of $(\Upsilon,\Omega,\Lambda)$-online-computable
functions.
\end{definition}

Apply Definition~\ref{def:goc} to get $e,\Delta,z_0$ for $g$.
Assume that $f$ is $(\Upsilon',\Omega')$-computable.
Apply Definition~\ref{def:gc} to get $d,p,q$ for $f$.
Let $x\in I$ and consider the following system:
\begin{equation*}
\left\{\begin{array}{@{}r@{}l}y(0)&=q(x)\\y'(t)&=p(y(t))\end{array}\right.
\qquad
\left\{\begin{array}{@{}r@{}l}z(0)&=z_0\\z'(t)&=q(z(t),y_{1..m}(t))\end{array}\right.
\end{equation*}

Define $v(t)=(x(t),y(t),z(t))$ then it immediately follows that $v$ satisfies a PIVP of the form
$v(0)=\poly(x)$ and $v'(t)=\poly(v(t))$. Furthermore, by definition:
\begin{align*}
\infnorm{v(t)}&\leqslant\max(\infnorm{x},\infnorm{y(t)},\infnorm{z(t)})\\
    &\leqslant\max\left(\infnorm{x},\infnorm{y(t)},\Upsilon\left(\sup_{u\in[t,t-\Delta]\cap\Rp}\infnorm{y_{1..m}(t)},t\right)\right)\\
    &\leqslant\poly\left(\infnorm{x},\sup_{u\in[t,t-\Delta]\cap\Rp}\infnorm{y(t)},t\right)\\
    &\leqslant\poly\left(\infnorm{x},\sup_{u\in[t,t-\Delta]\cap\Rp}\Upsilon'\left(\infnorm{x},u\right),t\right)\\
    &\leqslant\poly\left(\infnorm{x},t\right)\\
\end{align*}

Define $\bar{x}=f(x)$, $\Upsilon^*(\alpha)=1+\Upsilon'(\alpha,0)$ and
$\Omega''(\alpha,\mu)=\Omega'(\alpha,\Lambda(\Upsilon^*(\alpha),\mu))+\Omega(\Upsilon^*(\alpha),\mu)$.
By definition of $\Upsilon'$, $\infnorm{\bar{x}}\leqslant1+\Upsilon'(\infnorm{x},0)=\Upsilon^*(\infnorm{x})$.
Let $\mu\geqslant0$ then by definition of $\Omega'$,
if $t\geqslant\Omega'(\infnorm{x},\Lambda(\Upsilon^*(\infnorm{x}),\mu))$ then
$\infnorm{y_{1..m}(t)-\bar{x}}\leqslant e^{-\Lambda(\Upsilon^*(\infnorm{x}),\mu)}
\leqslant e^{-\Lambda(\infnorm{\bar{x}},\mu)}$. Apply Defiintion~\ref{def:goc}
for $a=\Omega'(\infnorm{x},\Lambda(\Upsilon^*(\infnorm{x}),\mu))$ to get that
$\infnorm{z_{1..l}(t)-g(f(x))}\leqslant e^{-\mu}$ for any $t\geqslant a+\Omega(\bar{x},\mu)$.
And since $t\geqslant a+\Omega(\bar{x},\mu)$ whenever $t\geqslant\Omega''(\infnorm{x},\mu)$,
we get that $g\circ f$ is computable.
\end{proof}

\subsubsection{On Encoding and Ideal Step Function}

Finaly, a remark on our selected encoding: 

Recall: 
\begin{definition}[Real encoding]\label{def:tm_config_real_enc}
Let $c=(x,\sigma,y,q)$ be a configuration of $\mathcal{M}$, the \emph{real encoding} of $c$ is
$\realenc{c}=(0.x,\sigma,0.y,q)\in\Q\times\Sigma\times\Q\times Q$ where $0.x=x_1k^{-1}+x_2k^{-2}+\cdots+x_{|w|}k^{-|w|}\in\Q$.
\end{definition}

We have:

\begin{lemma}[Encoding range]\label{lem:tm_config_range}
For any word $x\in\intinterv{0}{k-2}^*$, $0.x\in\left[0,\frac{k-1}{k}\right]$.
\end{lemma}

\begin{proof}
$0\leqslant0.x=\sum_{i=1}^{|x|}x_ik^{-i}
\leqslant\sum_{i=1}^{\infty}(k-2)k^{-i}\leqslant\frac{k-2}{k-1}\leqslant\frac{k-1}{k}$.
\end{proof}

The same way we considered the step function for Turing machines on configurations,
we have to define a step function that works directly the encoding of configuration.
This function is ideal in the sense that it is only defined over real numbers that
are encoding of configurations.

\begin{definition}[Ideal real step]\label{def:tm_config_ideal_real_step}
The \emph{ideal real step} function of a Turing machine $\mathcal{M}$ is the function
defined over $\realenc{\machcfg{\mathcal{M}}}$ by:
\[\idealrealstep{\mathcal{M}}(\tilde{x},\sigma,\tilde{y},q)=\begin{cases}
\left(\fracp(k\tilde{x}),\intp(k\tilde{x}),\frac{\sigma'+\tilde{y}}{k},q'\right) & \text{if }d=L\\
\left(\tilde{x},\sigma',\tilde{y},q'\right) & \text{if }d=S\\
\left(\frac{\sigma'+\tilde{x}}{k},\intp(k\tilde{y}),\fracp(k\tilde{y}),q'\right) & \text{if }d=R\\
\end{cases}\quad\text{where }
\left\{\begin{array}{@{}c@{}l@{}}q'&=\delta_1(q,\sigma)\\
\sigma'&=\delta_2(q,\sigma)\\
d&=\delta_3(q,\sigma)
\end{array}\right.
\]
\end{definition}

\begin{lemma}[$\idealrealstep{\mathcal{M}}$ is correct]\label{lem:real_step_correct}
For any machine $\mathcal{M}$ and configuration $c$,
$\idealrealstep{\mathcal{M}}(\realenc{c})=\realenc{\machstep{\mathcal{M}}(c)}$.
\end{lemma}

\begin{proof}
Let $c=(x,\sigma,y,q)$ and $\tilde{x}=0.x$.
The proof boils down to a case analysis (the analysis is the same for $x$ and $y$):
\begin{itemize}
\item If $x=\emptyword$ then $\tilde{x}=0$ so $\intp(k\tilde{x})=b$ and $\fracp(k\tilde{x})=0=0.\emptyword$ because $b=0$.
\item If $x\neq\emptyword$, $\intp(k\tilde{x})=x_1$ and $\fracp(k\tilde{x})=0.x_{2..|x|}$ because $k\tilde{x}=x_1+0.x_{2..|x|}$
and Lemma~\ref{lem:tm_config_range}.
\end{itemize}
\end{proof}

\subsubsection{Proof of Theorem~\ref{th:real_setp_robust}}

We consider the following function.

\begin{definition}[Real step]\label{def:tm_config_real_step}
For any $\bar{x},\bar{\sigma},\bar{y},\bar{q}\in\R$ and $\mu\in\Rp$, define the \emph{real step} function
of a Turing machine $\mathcal{M}$ by:
\[\realstep{\mathcal{M}}(\bar{x},\bar{\sigma},\bar{y},\bar{q},\mu)=
\realstep{\mathcal{M}}^*(\bar{x},\crnd(\bar{\sigma},\mu),\bar{y},\crnd(\bar{q},\mu),\mu)\]
where:
\[\realstep{\mathcal{M}}^*(\bar{x},\bar{\sigma},\bar{y},\bar{q},\mu)=
\realstep{\mathcal{M}}^{\star}\big(\bar{x},\bar{y},\lagrange{\delta_1}(\bar{q},\bar{\sigma}),
\lagrange{\delta_2}(\bar{q},\bar{\sigma}),\lagrange{\delta_2}(\bar{q},\bar{\sigma}),\mu\big)\]
where:
\[\realstep{\mathcal{M}}^{\star}\big(\bar{x},\bar{y},\bar{q},\bar{\sigma},\bar{d},\mu\big)=
\begin{pmatrix}
\myop{choose}\left[\fracp^*(k\bar{x}),\bar{x},\frac{\bar{\sigma}+\bar{x}}{k}\right]\\
\myop{choose}\left[\intp^*(k\bar{x}),\bar{\sigma},\intp^*(k\bar{y})\right]\\
\myop{choose}\left[\frac{\bar{\sigma}+\bar{y}}{k},\bar{y},\fracp^*(k\bar{y})\right]\\
\bar{q}
\end{pmatrix}\]
where:
\[\myop{choose}[l,s,r]=\lagreq{\idfun}{L}(\bar{d})l+\lagreq{\idfun}{S}(\bar{d})s+\lagreq{\idfun}{R}(\bar{d})r\]
\[\intp^*(x)=\crnd\left(x-\tfrac{1}{2}+\tfrac{1}{2k},\mu+\ln k\right)\qquad\fracp^*(x)=x-\intp^*(x)\]
\[\crnd\text{ is defined in Definition~\ref{def:comp:round}}.\]
\end{definition}

We can now prove  Theorem~\ref{th:real_setp_robust}:

\begin{proof}[Proof Of Theorem~\ref{th:real_setp_robust}]
We begin by a small result about $\intp^*$ and $\fracp^*$:
if $\infnorm{\bar{x}-0.x}\leqslant\frac{1}{2\cramped{k^2}}-e^{-\mu}$
then $\intp^*(k\bar{x})=\intp(k0.x)$ and $\infnorm{\fracp^*(k\bar{x})-\fracp(k0.x)}\leqslant k\infnorm{\bar{x}-0.x}$.
Indeed, by Lemma~\ref{lem:tm_config_range}, $k0.x=n+\alpha$ where $n\in\N$ and $\alpha\in\left[0,\frac{k-1}{k}\right]$.
Thus $\intp^*(k\bar{x})=\crnd\left(k\bar{x}-\frac{1}{2}+\frac{1}{2k},\mu\right)=n$
because $\alpha+k\infnorm{\bar{x}-0.x}-\frac{1}{2}+\frac{1}{2k}\in
\left[-\frac{1}{2}+ke^{-\mu},\frac{1}{2}-ke^{-\mu}\right]$. Also,
$\fracp^*(k\bar{x})=k\bar{x}-\intp^*(k\bar{x})=k\infnorm{\bar{x}-0.x}+kx-\intp(kx)=\fracp(kx)+k\infnorm{\bar{x}-0.x}$.

Write $\realenc{c}=(x,\sigma,y,q)$ and $\bar{c}=(\bar{x},\bar{\sigma},\bar{y},\bar{q})$.
Apply Definition~\ref{def:comp:round} to get that $\crnd(\bar{\sigma},\mu)=\sigma$ and $\crnd(\bar{q},\mu)=q$
because $\infnorm{(\bar{\sigma},\bar{q})-(\sigma,q)}\leqslant\frac{1}{2}-e^{-\mu}$.
Consequently, $\lagrange{\delta_i}(\bar{q},\bar{\sigma})=\delta_i(q,\sigma)$
and $\realstep{\mathcal{M}}(\bar{c},\mu)=\realstep{\mathcal{M}}^{\star}(\bar{x},\bar{y},q',\sigma',d')$
where $q'=\delta_1(q,\sigma)$, $\sigma'=\delta_2(q,\sigma)$ and $d'=\delta_3(q,\sigma)$.
In particular $d'\in\{L,S,R\}$ so there are three cases to analyze.
\begin{itemize}
\item \textbf{If $d'=L$ then} $\myop{choose}[l,s,r]=l$, $\intp^*(k\bar{x})=\intp(kx)$,
$\infnorm{\fracp^*(k\bar{x})-\fracp(kx)}\leqslant k\infnorm{\bar{x}-x}$ and
$\infnorm{\frac{\sigma'+\bar{y}}{k}-\frac{\sigma'+y}{k}}\leqslant\infnorm{\bar{x}-x}$.
Thus $\infnorm{\realstep{\mathcal{M}}(\bar{c},\mu)-\idealrealstep{\mathcal{M}}(\realenc{c})}\leqslant
k\infnorm{\bar{c}-\realenc{c}}$. Conclude using Lemma~\ref{lem:real_step_correct}.
\item \textbf{If $d'=S$ then} $\myop{choose}[l,s,r]=s$ so we immediately have that
$\infnorm{\realstep{\mathcal{M}}(\bar{c},\mu)-\idealrealstep{\mathcal{M}}(\realenc{c})}\leqslant
\infnorm{\bar{c}-\realenc{c}}$. Conclude using Lemma~\ref{lem:real_step_correct}.
\item \textbf{If $d'=R$ then} $\myop{choose}[l,s,r]=r$ and everything else is similar to
the case of $d'=L$.
\end{itemize}
Finally apply Lemma~\ref{lem:lagrange_interp}, Theorem
\ref{th:comp:round}, and Theorem 
\ref{th:gpac_comp_arith} and Theorem~\ref{th:gpac_comp_composition}
to get that $\realstep{\mathcal{M}}\in\gpc$.
\end{proof}

\subsection{Proof of Theorem~\ref{th:gpac_comp_iter}}
\label{app:th:gpac_comp_iter}

\subsubsection{Some facts}

We first state some facts.

\begin{lemma}[Round, \cite{TheseAmaury}]\label{lem:rnd}
There exists $\rnd\in\gpval$ such that:
For any $n\in\Z$, $\lambda\geqslant2$, $\mu\geqslant0$,
$|\rnd(x,\mu,\lambda)-n|\leqslant\frac{1}{2}$ for all $x\in\left[n-\frac{1}{2},n+\frac{1}{2}\right]$
and $|\rnd(x,\mu,\lambda)-n|\leqslant e^{-\mu}$ for all $x\in\left[n-\frac{1}{2}+\frac{1}{\lambda},n+\frac{1}{2}-\frac{1}{\lambda}\right]$.
\end{lemma}

\begin{lemma}[Clamped exponential]\label{prop:clamped_exp}
For any $a,b,c,d,x\in\R$ such that $a\leqslant b$ and $\ell\in\Rp$, define $h$ as follows. Then $h\in\gpc$:
\[h(a,b,c,d,x)=\max(a,\min(b,ce^x+d))\]
\end{lemma}

\subsubsection{Computing limits}

Intuitively, this model of computation already contains the notion of limit.
More precisely, if $f$ is computable and is such that $f(x,t)\rightarrow g(x)$ when
$t\rightarrow\infty$ then $g$ is computable. This is just a reformulation of equivalence
between computability and weak-computability.
The result below extends this result to the case where the limit is restricted to $t\in\N$.
The optimality of the assumptions is discussed in Remark~\ref{rem:gpac_comp_limit}.

The idea of the proof is to show that $g$ is weakly-computable and use the equivalence
with computability. Given $x$ and $\mu$, we want to run $f$ on $(x,\ceil{\omega})\in I\times J$
where $\omega=\mho(\infnorm{x},\mu)$.
Unfortunately we cannot compute the ceiling value in a continuous fashion.
The trick is to run two systems in parallels: one on $(x,(\rnd{\omega}))$ and one on $(x,\rnd(\omega+\frac{1}{2}))$.
This way one system will always have a correct input value but we must select which one.
If $\rnd$ is a good rounding function around $[n-\frac{1}{3},n+\frac{1}{3}]$, we
build the selecting function to pick the first system in $[n,n+\frac{1}{6}]$,
a barycenter of both in $[n+\frac{1}{6},n+\frac{1}{3}]$ and the second system in
$[n+\frac{1}{3},n+\frac{2}{3}]$ and so on. The crucial point is that in the region where we
mix both system, both have correct inputs so the mixing process doesn't create any error.
Furthermore, we can easily build such a continuous selecting function and the mixing
process has already been studied in a previous section.

\begin{theorem}[Closure by limit]\label{th:gpac_comp_limit}
Let $f:I\times J\subseteq\R^{n+1}\rightarrow\R^m$, $g:I\rightarrow\R^m$ and
$\mho:\Rp^2\rightarrow\Rp$ a polynomial.
Assume that $f\in\gpc$ and that $J\supseteq\N$.
Further assume that for all $(x,\tau)\in I\times J$ and $\mu\geqslant 0$, if $\tau\geqslant\mho(\infnorm{x},\mu)$
then $\infnorm{f(x,\tau)-g(x)}\leqslant e^{-\mu}$. Then $g\in\gpc$.
\end{theorem}

\begin{proof}
First note that $\frac{1}{2}-e^{-2}\geqslant\frac{1}{3}$ and define for $x\in I$ and $n\in\N$:
\[\begin{array}{ll}
f_0(x,\tau)=f(x,\crnd(\tau,2))&\qquad \tau\in\left[n-\frac{1}{3},n+\frac{1}{3}\right]\\
f_1(x,\tau)=f(x,\crnd(\tau+\tfrac{1}{2},2))&\qquad \tau\in\left[n+\frac{1}{6},n+\frac{5}{6}\right]
\end{array}\]
By Definition~\ref{def:comp:round} and hypothesis on $f$, both are well-defined because $\N\subseteq J$.
Also note that their domain of definition overlap on $[n+\tfrac{1}{6},n+\tfrac{1}{3}]$ and $[n+\tfrac{2}{3},n+\tfrac{5}{6}]$
for all $n\in\N$. Apply Theorems~\ref{th:comp:round} and~\ref{th:gpac_comp_composition} to get that $f_0,f_1\in\gpc$.
We also need to build the indicator function: this is where the choice of above values will prove convenient.
Define for any $\tau\in\Rp$:
\[i(x,\tau)=\tfrac{1}{2}-\cos(2\pi\tau)\]
It is now easy to check that:
\begin{align*}
\{(x,n)\thinspace|\thinspace i(x)<1\}&=\Rp\cap\cup_{n\in\N}\left]n-\tfrac{1}{3},n+\tfrac{1}{3}\right[\subseteq\dom{f_0}\\
\{(x,n)\thinspace|\thinspace i(x)>0\}&=\Rp\cap\cup_{n\in\N}\left]n+\tfrac{1}{6},n+\tfrac{5}{3}\right[\subseteq\dom{f_1}\\
\end{align*}
Define for any $x\in I$ and $\mu\in\Rp$:
\[f^*(x,\tau)=\mix{i}{f_0}{f_1}(x,\tau)\]
We can thus apply Theorem~\ref{th:comp:mix} to get that $f^*\in\gpc$. Note that $f^*$ is defined over $I\times\Rp$.
We now claim that for any $x\in I$ and $\mu\in\Rp$, if $t\geqslant1+\mho(\infnorm{x},\mu)$ then $\infnorm{f^*(x,\tau)-g(x)}\leqslant 2e^{-\mu}$.
There are three cases to consider:
\begin{itemize}
\item If $\tau\in[n-\tfrac{1}{6},n+\tfrac{1}{6}]$ for some $n\in\N$ then $i(x)\leqslant0$ so $\mix{i}{f_0}{f_1}(x,\tau)=f_0(x,\tau)=f(x,n)$
and since $n\geqslant\tau-\frac{1}{6}$ then $n\geqslant\mho(\infnorm{x},\mu)$ thus $\infnorm{f^*(x,\tau)-g(x)}\leqslant e^{-\mu}$.
\item If $\tau\in[n+\tfrac{1}{3},n+\tfrac{2}{3}]$ for some $n\in\N$ then $i(x)\geqslant1$ so $\mix{i}{f_0}{f_1}(x,\tau)=f_1(x,\tau)=f(x,n+1)$
and since $n\geqslant\tau-\frac{2}{3}$ then $n+1\geqslant\mho(\infnorm{x},\mu)$ thus $\infnorm{f^*(x,\tau)-g(x)}\leqslant e^{-\mu}$.
\item If $\tau\in[n+\tfrac{1}{6},n+\tfrac{1}{3}]\cup[n+\tfrac{2}{3},n+\tfrac{5}{6}]$ for some $n\in\N$ then $i(x)\in[0,1]$ so
$f^*(x,\tau)=(1-i(x,\tau))f_0(x,\tau)+i(x,\tau)f_1(x,\tau)=(1-i(x,\tau))f(x,\round{\tau})+i(x,\tau)f(x,\round{\tau+\tfrac{1}{2}})$. Since
$\round{\tau},\round{\tau+\tfrac{1}{2}}\geqslant\mho(\infnorm{x},\mu)$ then $\infnorm{f(x,\round{\tau})-g(x)}\leqslant e^{-\mu}$
and $\infnorm{f(x,\round{\tau+\tfrac{1}{2}})-g(x)}\leqslant e^{-\mu}$ thus $\infnorm{f^*(x,\tau)-g(x)}\leqslant2e^{-\mu}$
because $|i(x,\tau)|\leqslant1$.
\end{itemize}

It follows that $g$ is the limit of $f^*$ and thus $g\in\gpwc$ (see
Remark~\ref{rem:gpwc_gpc}) and one concludes using that $\gpwc=\gpc$.
\end{proof}

\begin{remark}[Optimality]\label{rem:gpac_comp_limit}
The condition that $\mho$ be a polynomial is essentially optimal. Intuitively,
if $f\in\gpc$ and satisfies that $\infnorm{f(x,\tau)-g(x)}\leqslant e^{-\mu}$
whenever $\tau\geqslant\mho(\infnorm{x},\mu)$ then $\mho$ is a modulus of continuity for $g$.
By Theorem~\ref{th:comp_implies_cont}, if $g\in\gpc$ then it admits a polynomial modulus of continuity
so $\mho$ must be a polynomial. For a formal proof of this intuition, see
examples~\ref{ex:gpac_limit_poly1} and~\ref{ex:gpac_limit_poly2}
\end{remark}

\begin{example}[$\mho$ must be polynomial in $x$]\label{ex:gpac_limit_poly1}
Let $f(x,\tau)=\min(e^x,\tau)$ and $g(x)=e^x$. Trivially $f(x,\cdot)$ converges to $g$
because $f(x,\tau)=g(x)$ for $\tau\geqslant e^x$. But $g\notin\gpc$ because it is not
polynomially bounded. In this case $\mho(x,\mu)= e^x$ which is exponential and $f\in\gpc$
by Lemma~\ref{prop:clamped_exp}.
\end{example}

\begin{example}[$\mho$ must be polynomial in $\mu$]\label{ex:gpac_limit_poly2}
Let $g(x)=\frac{-1}{\ln x}$ for $x\in\left[0,e\right]$ which is defined in $0$
by continuity. Observe that $g\notin\gpc$, indeed its modulus of continuity is
exponential around $0$ because $g(\cramped{e^{-e^\mu}})=e^{-\mu}$ for all $\mu\geqslant0$.
However note that $g^*\in\gpc$ where
$g^*(x)=g(\cramped{e^{-x}})=\frac{1}{x}$ for $x\in[1,+\infty[$.
Let $f(x,\tau)=g^*(\min(-\ln x,\tau))$ and check, using that $g$ is increasing
and non-negative, that:
$|f(x,\tau)-g(x)|=\left|g(\max(x,\cramped{e^{-\tau}}))-g(x)\right|
    \leqslant g(\max(x,\cramped{e^{-\tau}}))\leqslant \frac{1}{\tau}$.
Thus $\mho(\infnorm{x},\mu)=e^\mu$ which is exponential and $f\in\gpc$ because
$(x,\tau)\mapsto\min(-\ln x,\tau)\in\gpc$ by a proof similar to Lemma~\ref{prop:clamped_exp}.
\end{example}

\subsubsection{Proof of Theorem~\ref{th:gpac_comp_iter}}

We now go to the Proof of Theorem~\ref{th:gpac_comp_iter}.

We use three variables $y$, $z$ and $w$ and build a cycle to be repeated $n$ times. At all time, $y$ is an online system
computing $f(w)$. During the first stage of the cycle, $w$ stays still and $y$ converges to $f(w)$.
During the second stage of the cycle, $z$ copies $y$ while $w$ stays still. During
the last stage, $w$ copies $z$ thus effectively computing one iterate.

The crucial point is in the error estimation, which we informally develop here.
Denote the $k^{th}$ iterate of $x$ by $x^{[k]}$ and by $x^{(k)}$ the point computed
after $k$ cycles in the system. Because we are doing an approximation of $f$ at each
step step, the relationship between the two is that $x_0=x^{[0]}$ and
$\infnorm{x^{(k+1)}-f(x_k)}\leqslant e^{-\nu_{k+1}}$ where $\nu_{k+1}$ is the precision
of the approximation, that we control. Define $\mu_k$ the precision we need to achieve at step $k$:
$\infnorm{x^{(k)}-x^{[k]}}\leqslant e^{-\mu_k}$ and $\mu_n=\mu$. The triangle inequality
ensures that the following choice of parameters is safe:
\[\nu_k\geqslant\mu_k+\ln2\qquad
\mu_{k-1}\geqslant\mho\left(\infnorm{x^{[k-1]}}\right)+\mu_k+\ln2\]
This is ensured by taking $\mu_k\geqslant\sum_{i=k}^{n-1}\mho(\Pi(\infnorm{x},i))+\mu+(n-k)\ln2$
which is indeed polynomial in $k$, $\mu$ and $\infnorm{x}$. Finally a point
worth mentionning is that the entire reasoning makes sense because the assumption
ensures that $x^{(k)}\in I$ at each step.

Formally, apply Lemma~\ref{lemma:online} to get that
$f\in\guc{\Upsilon}{\Omega}{\Lambda}{\Theta}$
where $\Upsilon,\Lambda,\Theta,\Omega$ are polynomials.
Without loss of generability we assume that $\Upsilon,\Lambda,\Theta,\mho$ and $\Pi$
are increasing functions.
Apply Lemma~\ref{lem:gpuc_time_rescaling} to
get $\omega\geqslant1$ such that for all $\alpha\in\R,\mu\in\Rp$:
\[\Omega(\alpha,\mu)=\omega\geqslant1\]
Apply Definition~\ref{def:guc} to get $\delta,d$ and $g$. Define:
\[\tau=\omega+2\]
We will show that $f_0^*\in\gpwc=\gpc$:
let $n\in\N$, $x\in I_n$, $\mu\in\Rp$ and consider the following system:
\[
\left\{\begin{array}{@{}r@{}l}
\ell(0)&=\norm_{\infty,1}(x)\\
\mu(0)&=\mu\\
n(0)&=n\\
\end{array}\right.
\qquad
\left\{\begin{array}{@{}r@{}l}
\ell'(t)&=0\\
\mu'(t)&=0\\
n'(t)&=0
\end{array}\right.
\qquad
\left\{\begin{array}{@{}r@{}l}
y(0)&=0\\
z(0)&=x\\
w(0)&=x
\end{array}\right.
\]
\[
\left\{\begin{array}{@{}r@{}l}
y'(t)&=g(t,y(t),w(t),\nu(t))\\
z'(t)&=\sample_{[\omega,\omega+1],\tau}(t,\nu(t),z(t),y_{1..n}(t))\\
w'(t)&=\hxl_{[0,1]}(t-n\tau,\nu(t)+t,\sample_{[\omega+1,\omega+2],\tau}(t,\nu^*(t)+\ln(1+\omega),w(t),z(t)))\\
\end{array}\right.
\]
\[\ell^*=1+\Pi(\ell,n)\qquad
\nu=n\mho(\ell^*)+n\ln6+\mu+\ln3\qquad
\nu^*=\nu+\Lambda(\ell^*,\nu)\]

First notice that $\ell,\mu$ and $n$ are constant functions and we identify $\mu(t)$ with $\mu$
and $n(t)$ with $n$. Apply Lemma~\ref{lem:norm} to get that $\infnorm{x}\leqslant\ell\leqslant\infnorm{x}+1$,
so in particular $\ell^*,\nu$ and $\nu^*$ are polynomially bounded in $\infnorm{x}$ and $n$.
We will need a few notations:
for $i\in\intinterv{0}{n}$, define $x^{[i]}=\fiter{f}{i}(x)$ and $x^{(i)}=w(i\tau)$. Note
that $x^{[0]}=x^{(0)}=x$. We will show by induction for $i\in\intinterv{0}{n}$ that:
\[\infnorm{x^{(i)}-x^{[i]}}\leqslant e^{-(n-i)\mho(\ell^*)-(n-i)\ln6-\mu-\ln3}\]
Note that this is trivially true for $i=0$. Let $i\in\intinterv{0}{n-1}$
and assume that the result is true for $i$, we will show that it holds for $i+1$
by analyzing the behavior of the sytem during period $[i\tau,(i+1)\tau]$.
\begin{itemize}
\item\textbf{For $y$ and $w$, if $t\in[i\tau,i\tau+\omega+1]$ then}
  apply Lemma~\ref{lem:lxh_hxl} to
get that $\hxl\in[0,1]$ and Lemma~\ref{lem:sample} to get that $\infnorm{w'(t)}\leqslant e^{-\nu^*-\ln(1+\omega)}$.
Conclude that $\infnorm{w(i)-w(t)}\leqslant e^{-\nu^*}$, in other words
$\infnorm{w(t)-x^{(i)}}\leqslant e^{-\Lambda(\infnorm{x^{(i)}},\nu)}$ since
$\infnorm{x^{(i)}}\leqslant\infnorm{x^{[i]}}+1\leqslant1+\Pi(\infnorm{x},i)\leqslant\ell^*$.
Thus, by definition of \unaware{} computability, $\infnorm{f(x^{(i)})-y_{1..n}(u)}\leqslant e^{-\nu}$
if $u\in[i\tau+\omega,i\tau+\omega+1]$ because $\Omega\left(\infnorm{x^{(i)}},\nu\right)=\omega$.
\item\textbf{For $z$, if $t\in[i\tau+\omega,i\tau+\omega+1]$ then}
  apply Lemma~\ref{lem:sample}
to get that $\infnorm{f(x^{(i)})-z(i\tau+\omega+1)}\leqslant2e^{-\nu}$.
\item\textbf{For $z$ and $w$, if $t\in[i\tau+\omega+1,i\tau+\omega+2]$
    then} apply Lemma~\ref{lem:sample}
to get that $\infnorm{z'(t)}\leqslant e^{-\nu}$ thus $\infnorm{f(x^{(i)})-z(t)}\leqslant3e^{-\nu}$.
Apply Lemma~\ref{lem:lxh_hxl} to get that
$\infnorm{y'(t)-\sample_{[\omega+1,\omega+2],\tau}(t,\nu^*+\ln(1+\omega),w(t),z(t))}\leqslant e^{-\nu-t}$.
Apply Lemma~\ref{lem:sample} again to get that $\infnorm{f(x^{(i)})-w(i\tau+\omega+2)}\leqslant4e^{-\nu}+e^{-\nu^*}\leqslant5e^{-\nu}$.
\end{itemize}
Our analysis concluded that $\infnorm{f(x^{(i)})-z((i+1)\tau)}\leqslant5e^{-\nu}$.
Also, by hypothesis, $\infnorm{x^{(i)}-x^{[i]}}\leqslant e^{-(n-i)\mho(\ell^*)-(n-i)\ln6-\mu-\ln3}\leqslant
e^{-\mho\left(\infnorm{x^{[i]}}\right)-\mu^*}$
where $\mu^*=(n-i-1)\mho(\ell^*)+(n-i)\ln6+\mu+\ln3$ because $\infnorm{x^{[i]}}\leqslant\ell^*$.
Consequently, $\infnorm{f(x^{(i)})-x^{[i+1]}}\leqslant e^{-\mu^*}$ and thus:
\[\infnorm{x^{(i+1)}-x^{[i+1]}}\leqslant 5e^{-\nu}+e^{-\mu^*}\leqslant6e^{-\mu^*}
\leqslant e^{-(n-1-i)\mho(\ell^*)-(n-1-i)\ln6-\mu-\ln3}\]

From this induction we get that $\infnorm{x^{(n)}-x^{[n]}}\leqslant e^{-\mu-\ln3}$.
We still have to analyze the behavior after time $n\tau$.
\begin{itemize}
\item \textbf{If $t\in[n\tau,n\tau+1]$ then} apply Lemmas~\ref{lem:sample} and~\ref{lem:lxh_hxl}
to get that $\infnorm{w'(t)}\leqslant e^{-\nu^*-\ln(1+\omega)}$ thus
$\infnorm{w(t)-x^{(n)}}\leqslant e^{-\nu^*-\ln(1+\omega)}$.
\item \textbf{If $t\geqslant n\tau+1$ then} apply Lemma~\ref{lem:lxh_hxl}
to get that $\infnorm{w'(t)}\leqslant e^{-\nu-t}$ thus $\infnorm{w(t)-w(n\tau+1)}\leqslant e^{-\nu}$.
\end{itemize}
Putting everything together we get for $t\geqslant n\tau+1$ that:
\begin{align*}
\infnorm{w(t)-x^{[n]}}&\leqslant e^{-\mu-\ln3}+e^{-\nu^*-\ln(1+\omega)}+e^{-\nu}\\
    &\leqslant3e^{-\mu-\ln3}\leqslant e^{-\mu}
\end{align*}
We also have to show that the system does not grow to fast. The analysis during
the time interval $[0,n\tau+1]$ has already been done (although we did not write
all the details, it is an implicit consequence). For $t\geqslant n\tau+1$,
have $\infnorm{w(t)}\leqslant\infnorm{x^{[n]}}+1\leqslant\Pi(\infnorm{x},n)+1$
which is polynomially bounded. The bound on $y$ comes from Definition~\ref{def:guc}:
\[
\infnorm{y(t)}\leqslant\Upsilon\left(\pastsup{\delta}{\infnorm{w}}(t),\nu,0\right)
    \leqslant\Upsilon(\Pi(\infnorm{x},n),\nu,0)
    \leqslant\poly(\infnorm{x},n,\mu)
\]
And finally, apply Lemma~\ref{lem:sample} to get that:
\[\infnorm{z(t)}\leqslant2+\pastsup{\tau+1}{\infnorm{y_{1..n}}}(t)\leqslant\poly(\infnorm{x},n,\mu)
\]
This conclude the proof that $f_0^*\in\gpwc$.

We will now tackle the case of $\eta>0$. Let $\eta\in]0,\tfrac{1}{2}[$ and
define $g_\eta(x,\mu)=\rnd(x,\mu,\tfrac{1}{2}-\eta)$ for $x\in\Z+\left]-\eta,\eta\right[$.
Apply Lemma~\ref{lem:rnd} to get that $\rnd\in\gpval$ and Theorem~\ref{th:gpac_gen_implies_comp} to
get\footnote{Although this is a forward reference, the proof does not relies
on the iteration of functions} that $g_\eta\in\gpc$. By definition, $\infnorm{g_\eta(x,\mu)-n}\leqslant e^{-\mu}$
if $x\in[n-\eta,n+\eta]$ thus we can apply Theorem~\ref{th:gpac_comp_limit} to get that
$g_\eta^*(x)=\lim_{\mu\rightarrow\infty}g_\eta(x,\mu)$ belongs to $\gpc$
and $g_\eta^*(x)=n$ for any $x\in[n-\eta,n+\eta]$. Now define
$f_\eta^*(x,u)=f_0^*(x,g_\eta^*(u))$ and apply Theorem~\ref{th:gpac_comp_composition}
to conclude. As a final remark, note that $g_\eta^*$ is a pretty good rounding
function but we can do much better: see Theorem~\ref{th:comp:round} for more details.

\begin{remark}[Optimality of growth constraint]\label{rem:gpac_iter_opt_growth}
It is easy to see that without any condition, the iterates can produce an exponential
function. Pick $f(x)=2x$ then $f\in\gpc$ and $\fiter{f}{n}(x)=2^nx$ which is clearly not polynomial
in $x$ and $n$. More generally, by Lemma~\ref{prop:gp_growth}, it is necessary
that $f^*$ be polynomially bounded so clearly $f^{[n]}(x)$ must be polynomially
bounded in $\infnorm{x}$ and $n$.
\end{remark}

\begin{remark}[Optimality of modulus constraint]\label{rem:gpac_iter_opt_mod}
Without any constraint, it is easy to build an iterated function with exponential
modulus of continuity. Define $f(x)=\sqrt{x}$ then $f\in\gpc$ and $\fiter{f}{n}(x)=x^{\frac{1}{2^n}}$.
For any $\mu\in\R$, $\fiter{f}{n}(e^{-2^n\mu})-\fiter{f}{n}(0)=(e^{-2^n\mu})^{\frac{1}{2^n}}=e^{-\mu}$.
Thus $f^*$ has exponential modulus of continuity in $n$.
\end{remark}

\begin{remark}[Domain of definition]\label{rem:gpac_iter_opt_domain}
Intuitively we could have written the theorem differently, only requesting that $f(I)\subseteq I$,
however this has some problems. First if $I$ is discrete, the iterated
modulus of continuity becomes useless and the theorem is false. Indeed, define $f(x,k)=(\sqrt{x},k+1)$
and $I=\left\{(\sqrt[2^n]{e},n),n\in\N\right\}$: $\frestrict{f}{I}$ has polynomial
modulus of continuity $\mho$ because $I$ is discrete, yet $\frestrict{f^*}{I}\notin\gpc$
as we saw in Remark~\ref{rem:gpac_iter_opt_mod}. But in reality, the problem is more subtle than
that because if $I$ is open but the neighbourhood of each point is too small, a
polynomial system cannot take advantage of it. To illustrate this issue, define
$I_n=\left]0,\sqrt[2^n]{e}\right[\times\left]n-\tfrac{1}{4},n+\tfrac{1}{4}\right[$
and $I=\cup_{n\in\N}I_n$. Clearly $f(I_n)=I_{n+1}$ so $I$ is $f$-stable but
$\frestrict{f^*}{I}\notin\gpc$ for the same reason as before.
\end{remark}

\begin{remark}[Classical error bound]\label{rem:gpac_iter_classic_err}
The third condition in Theorem~\ref{th:gpac_comp_iter} is usually far more subtle than necessary.
In practice, is it useful to note this condition is satisfied in $f$ verifies for some
constants $\varepsilon,K>0$ that
\[\text{for all }x\in I_n,y\in\R^m,\text{ if }
    \infnorm{x-y}\leqslant \varepsilon\text{ then }
    y\in I\text{ and }\infnorm{f(x)-f(y)}\leqslant K\infnorm{x-y}\]
\end{remark}

\begin{remark}[Dependency of $\mho$ in $n$]\label{rem:gpac_iter_modulus_n}
In the statement of thereom, $\mho$ is only allowed to depend on $\infnorm{x}$ whereas
it might be useful to also make it depend on $n$. In fact the theorem is still true
if the last condition is modified to be $\infnorm{x-y}\leqslant e^{-\mho(\infnorm{x},n)-\mu}$.
The proof is straightfoward:
\end{remark}

\subsection{Proof of Theorem~\ref{th:p_gpac}}
\label{app:th:p_gpac}

\subsubsection{ $\FP$ iff emulable}

In this subsection, we fix an alphabet $\Gamma$ and all languages are considered over
$\Gamma$, so in particular $P\subset\Gamma^*$. It is common to take $\Gamma=\{0,1\}$
but the proofs work for any finite alphabet. We will assume that $\Gamma$ comes with an
injective mapping $\gamma:\Gamma\rightarrow\Ns$, in other words every letter has an
uniquely assigned positive number. By extension, $\gamma$ applies
letterwise over words.

We start by proving that functions of $\FP$ (i.e. computable in
polynomial time) are emulable and conversely. 

Before, we state that the following lemma can be proved:

\begin{lemma}[Size recovery, \cite{TheseAmaury}]\label{cor:size_recovery}
For any machine $\mathcal{M}$, there exists a function
$(\myop{tsize}_{\mathcal{M}}:\realenc{\machcfg{\mathcal{M}}}\times\N\rightarrow\N)\in\gpc$ such
that for any word $w\in\left(\Sigma\setminus\{b\}\right)^*$ and any
$n\geqslant|w|$, the size of the tape satisfies $\myop{tsize}_{\mathcal{M}}(0.w, n)=|w|$.
\end{lemma}

\begin{definition}[Discrete emulation]\label{def:fp_gpac_embed}
$f:\Gamma^*\rightarrow\Gamma^*$ is called \emph{emulable} if there exists $g\in\gpc$
and $k\geqslant1+\max(\gamma(\Gamma))$ such that for any word $w\in\Gamma^*$:
\[g(\psi(w))=\psi(f(w))\qquad\text{where}\quad
\psi(w)=\left(\sum_{i=1}^{|w|}\gamma(w_i)k^{-i},|w|\right)\]
We say that $g$ \emph{emulates} $f$ with $k$.
\end{definition}

\begin{theorem}[$\FP$ equivalence]\label{th:fp_gpac}
$f\in\FP$ if and only if $f$ is emulable  (with $k=2+\max(\gamma(\Gamma))$).
\end{theorem}

\begin{proof}
Let $f\in\FP$, then there exists a Turing machine
$\mathcal{M}=(Q,\Sigma,b,\delta,q_0,F)$ where $\Sigma=\intinterv{0}{k-2}$ and
$\gamma(\Gamma)\subset\Sigma\setminus\{b\}$,
and a polynomial $p_\mathcal{M}$ such that for any word $w\in\Gamma^*$,
$\mathcal{M}$ halts in at most $p_\mathcal{M}(|w|)$ steps, that is
$\fiter{\mathcal{M}}{p_\mathcal{M}(|w|)}(c_0(\gamma(w)))=c_\infty(\gamma(f(w)))$.
Note that we assume that $p_\mathcal{M}(\N)\subseteq\N$. Also note that $\psi(w)=(0.\gamma(w),|w|)$
for any word $w\in\Gamma^*$.

Let $\mu=\ln(4\cramped{k^2})$ and $h(c)=\mathcal{M}(c,\mu)$ for all $c\in\R^4$.
Define $I_\infty=\realenc{\machcfg{\mathcal{M}}}$ and
$I_n=I_\infty+\left[-\varepsilon_n,\varepsilon_n\right]^4$
where $\varepsilon_n=\tfrac{1}{4\cramped{k^{2+n}}}$ for all $n\in\N$. Note that
$\varepsilon_{n+1}\leqslant\tfrac{\varepsilon_n}{k}$ and that
that $\varepsilon_0\leqslant\tfrac{1}{2\cramped{k^2}}-e^{-\mu}$.
Apply Theorem~\ref{th:real_setp_robust} to get that $h\in\gpc$ and $h(I_{n+1})\subseteq I_n$.
In particular
$\infnorm{\fiter{h}{n}(\bar{c})-\fiter{h}{n}(c)}\leqslant k^n\infnorm{c-\bar{c}}$
for all $c\in I_\infty$ and $\bar{c}\in I_n$, for all $n\in\N$.
Let $\delta\in\left[0,\tfrac{1}{2}\right[$ and define
$J=\cup_{n\in\N}I_n\times[n-\delta,n+\delta]$. Apply Theorem~\ref{th:gpac_comp_iter} to
get $(h^*:J\rightarrow I_0)\in\gpc$ such that for all $c\in I_\infty$ and $n\in\N$
and $h^*(c,n)=\fiter{h}{n}(c)$.

Let $\pi_3$ denote the third projection, that is $\pi_3(a,b,c,d)=c$, then $\pi_3\in\gpc$.
Define $g(y,\ell)=\pi_3(h^*(0,b,y,q_0,p_\mathcal{M}(\ell)))$ for $y\in \psi(\Gamma^*)$ and $\ell\in\N$.
Note that $g\in\gpc$ and is well-defined. Indeed, if $\ell\in\N$ then $p_\mathcal{M}(\ell)\in\N$
and if $y=\psi(w)=0.w$ then $(0,b,y,q_0)=\realenc{(\emptyword,b,w,q_0)}=\realenc{c_0(w)}\in I_\infty$.
Furthermore, by construction, for any word $w\in\Gamma^*$ we have:
\begin{align*}
g(\psi(w),|w|)&=\pi_3\left(h^*(\realenc{c_0(w)},p_\mathcal{M}(|w|))\right)\\
    &=\pi_3\left(\fiter{h}{p_\mathcal{M}(|w|)}(c_0(w))\right)\\
    &=\pi_3\left(\realenc{\fiter{\machcfg{\mathcal{M}}}{p_\mathcal{M}(|w|)}(c_0(w))}\right)\\
    &=\pi_3\left(\realenc{c_\infty(\gamma(f(w)))}\right)\\
    &=0.\gamma(f(w))=\psi(f(w))
\end{align*}
Furthermore, the size of the tape cannot be greater than the initial size plus
the number of steps, thus
$|f(w)|\leqslant|w|+p_\mathcal{M}(|w|)$. Apply Lemma~\ref{cor:size_recovery}
to get that $\myop{tsize}_\mathcal{M}(g(\psi(w),|w|),|w|+p_\mathcal{M}(|w|))=|f(w)|$
since $f(w)$ does not contain any blank character (this is true because $\gamma(\Gamma)\subset\Sigma\setminus\{b\})$.
This proves that $f$ is emulable because $g\in\gpc$ and the tape size $\myop{tsize}_\mathcal{M}\in\gpc$.

Conversely, assume that $f$ is emulable and apply Definition~\ref{def:fp_gpac_embed} to get
$g\in\gc{\Upsilon}{\Omega}$ where $\Upsilon,\Omega$ are polynomials, and $k\in\N$.
Let $w\in\Gamma^*$: we will describe an $\FP$ algorithm
to compute $f(w)$. Apply Definition~\ref{def:gc} to $g$ to get $d,p,q$ and consider the following system:
\[y(0)=q(\psi(w))\qquad y'(t)=p(y(t))\]
Note that by construction, $y$ is defined over $\Rp$. Also note\footnote{and that is
absolutely crucial} that the coefficients of $p,q$ belong to $\Rpoly$ which means
that they are polynomial time computable. And since $\psi(w)$ is a pair of rational numbers
with polynomial size (with respect to $|w|$), then $q(\psi(w))\in
\Rpoly^d$
.

The algorithm works in two steps: first we compute a rough approximation of the output
to guess the size of the output. Then we rerun the system with enough precision to
get the full output.

Let $t_w=\Omega(|w|,2)$ for any $w\in\Sigma^*$, note that $t_w\in\Rpoly$ and that
it is polynomially bounded in $|w|$ because $\Omega$ is a
polynomial. Apply Theoorem~\ref{th:pivp_comp_analysis}
to compute $\tilde{y}$ such that $\infnorm{\tilde{y}-y(t_w)}\leqslant e^{-2}$: this takes a time polynomial
in $|w|$ because $t_w$ is polynomially bounded
and because $\glen{y}(0,t_w)\leqslant\poly(t_w,\sup_{[0,t_w]}\infnorm{y})$
and by construction, $\infnorm{y(t)}\leqslant\Upsilon(\infnorm{\psi(w)},t_w)$ for $t\in[0,t_w]$
where $\Upsilon$ is a polynomial. Furthermore, by definition $\infnorm{y(t_w)-g(\psi(w))}\leqslant e^{-2}$
thus $\infnorm{\tilde{y}-\psi(f(w))}\leqslant2e^{-2}\leqslant\frac{1}{3}$. But
since $\psi(f(w))=(0.\gamma(f(w)),|f(w)|)$, from $\tilde{y}_2$ we can find $|f(w)|$ by
rounding to the closest integer (which is unique because at distance at most $\frac{1}{3}$).
In other words, we can compute $|f(w)|$ in polynomial time in $|w|$. Note that this
implies that $|f(w)|$ is at most polynomial in $|w|$.

Let $t_w'=\Omega(|w|,2+|f(w)|\ln k)$ which is polynomial in $|w|$ because $\Omega$
is a polynomial and $|f(w)|$ is at most polynomial in $|w|$. We can use the
same reasoning and apply Theorem~\ref{th:pivp_comp_analysis} to get $\tilde{y}$ such
that $\infnorm{\tilde{y}-y(t_w')}\leqslant e^{-2-|f(w)|\ln k}$. Again this
takes a time polynomial in $|w|$. Furthermore,
$\infnorm{\tilde{y}_1-0.\gamma(f(w))}\leqslant2e^{-2-|f(w)|\ln k}\leqslant \frac{1}{3}k^{-|f(w)|}$.
We claim that this allows to recover $f(w)$ unambiguously in polynomial time in $|f(w)|$.
Indeed, it implies that $\infnorm{k^{|f(w)|}\tilde{y}_1-k^{|f(w)|}0.\gamma(f(w))}\leqslant\frac{1}{3}$.
Unfolding the definition shows that $k^{|f(w)|}0.\gamma(f(w))=\sum_{i=1}^{|f(w)|}\gamma(f(w)_i)k^{|f(w)|-i}\in\N$
thus by rounding $k^{|f(w)|}\tilde{y}_1$ to the nearest integer, we recover $\gamma(f(w))$,
and then $f(w)$. This is all done in polynomial time in $|f(w)|$, which proves that $f$
is polynomial time computable.
\end{proof}

An question arises when looking at this theorem: does the choice
of $k$ in Definition~\ref{def:fp_gpac_embed} matters, especially for the equivalence
with $\FP$ ? Fortunately not, as long as $k$ is large enough, as shown in the next lemma.

\begin{lemma}[Emulation reencoding, \cite{TheseAmaury}]\label{lem:fp_gpac_embed_reenc}
Assume that $g\in\gpc$ emulates $f$ with $k\in\N$. Then for
any $k'\geqslant k$, there exists $h\in\gpc$ that emulates $f$ with $k'$.
\end{lemma}

\subsubsection{ Proof of Theorem~\ref{th:p_gpac}}

We start by some technical lemma.

\begin{lemma}[``low-X-high'' and ``high-X-low'', \cite{TheseAmaury}]\label{lem:lxh_hxl}
There exists $\lxh_I,\hxl_I\in\gpval$ such that: 
Let $I=[a,b]$, $\mu\in\Rp$, then $\forall t,x\in\R$:
\begin{itemize}
\item $\exists\phi_1,\phi_2$ such that $\lxh_I(t,\mu,x)=\phi_1(t,\mu,x)x$ and $\hxl_I(t,\mu,x)=\phi_2(t,\mu,x)x$
\item if $t\leqslant a, |\lxh_I(t,\mu,x)|\leqslant e^{-\mu}$ and $|x-\hxl_I(t,\mu,x)|\leqslant e^{-\mu}$
\item if $t\geqslant b, |x-\lxh_I(t,\mu,x)|\leqslant e^{-\mu}$ and $|\hxl_I(t,\mu,x)|\leqslant e^{-\mu}$
\item in all cases, $|\lxh_I(t,\mu,x)|\leqslant|x|$ and $|\hxl_I(t,\mu,x)|\leqslant|x|$
\end{itemize}
\end{lemma}

The proof is based on the equivalence between $\gpc$ and $\gplc$, and the $\FP$
equivalence. Indeed, decidability can be seen as the computability of particular
functions with boolean output. The only technical point is to make sure that the decision of
the system is irreversible. To do that, we run the system from the $\FP$ equivalence
(which will output $0$ or $1$) for long enough so that the output is approximate
but good enough. Only then will another variable reach $-1$ or $1$. The fact that
the decision complexity is based on the length of the curve also makes the proof
slightly more complicated because the system we build essentially takes a decision
after a certain time (and not length).

Let $\mathcal{L}\in\PTIME$, then there exists $f\in\FP$ and two distinct symbols
$\bar{0},\bar{1}\in\Gamma$ such that for any $w\in\Gamma^*$,
$f(w)=\bar{1}$ if $w\in\mathcal{M}$ and $f(w)=\bar{0}$ otherwise.
Let $\myop{dec}$ be defined by $\myop{dec}(k^{-1}\gamma(\bar{0}))=-2$
and $\myop{dec}(k^{-1}\gamma(\bar{1}))=2$. Recall that
$\lagrange{\myop{dec}}\in\gpc$ by Lemma~\ref{lem:lagrange_interp}.
Apply Theorem~\ref{th:fp_gpac} to get $g$ and $k$ that emulate $f$. Note in particular that for any $w\in\Gamma^*$,
$f(w)\in\{\bar{0},\bar{1}\}$ so $\psi(f(w))=(\gamma(\bar{0})k^{-1},1)$ or $(\gamma(\bar{1})k^{-1},1)$.
Define $g^*(x)=\lagrange{\myop{dec}}(g_1(x))$ and check that $g^*\in\gpc$.
Furthermore, $g^*(\psi(w))=2$ if $w\in\mathcal{L}$ and $g^*(\psi(w))=-2$ otherwise,
by definition of the emulation and the interpolation.
Let $\Omega$ and $\Upsilon$ be polynomials such that $g^*\in\gc{\Upsilon}{\Omega}$
and assume, without loss of generality, that they are increasing functions.
Apply Definition~\ref{def:gc} to get $d,p,q$. Let $w\in\Gamma^*$ and consider the following system:
\[
\left\{\begin{array}{@{}r@{}l}
y(0)&=q(\psi(w))\\
v(0)&=\psi(w)\\
z(0)&=0\\
\tau(0)&=0\\
\end{array}\right.
\qquad
\left\{\begin{array}{@{}r@{}l}
y'(t)&=p(y(t))\\
v'(t)&=0\\
z'(t)&=\lxh_{[0,1]}(\tau(t)-\tau^*,1,y_1(t)-z(t))\\
\tau'(t)&=1
\end{array}\right.
\]
\[\tau^*=\Omega(v_2(t),\ln2)
\]
In this system, $y$ computes $g^*$ $f$, $v$ is a constant variable used to store
the input and in particular the input size ($v_2(t)=|w|$), $\tau(t)=t$
is used to keep the time and $z$ is the decision variable.
Let $t\in[0,\tau^*]$, then by Lemma~\ref{lem:lxh_hxl}, $\infnorm{z'(t)}\leqslant e^{-1-t}$
thus $\infnorm{z(t)}\leqslant e^{-1}<1$. In other words, at time $\tau^*$ the system
has still not decided if $w\in\mathcal{L}$ or not. Let $t\geqslant\tau^*$,
then by definition of $\Omega$ and since $v_2(t)=\psi_2(w)=|w|=\infnorm{\psi(w)}$,
$\infnorm{y_1(t)-g^*(\psi(w))}\leqslant e^{-\ln2}$.
Recall that $g^*(\psi(w))\in\{-2,2\}$ and let $\varepsilon\in\{-1,1\}$ such that $g^*(\psi(w))=\varepsilon2$.
Then $\infnorm{y_1(t)-\varepsilon2}\leqslant\frac{1}{2}$ which means that $y_1(t)=\varepsilon \lambda(t)$
where $\lambda(t)\geqslant\frac{3}{2}$.
Apply Lemma~\ref{lem:lxh_hxl} 
to conclude that $z$ satisfies for $t\geqslant\tau^*$:
\[z(\tau^*)\in[-e^{-1},e^{-1}]\qquad z'(t)=\phi(t)(\varepsilon\lambda(t)-z(t))\]
where $\phi(t)\geqslant0$ and $\phi(t)\geqslant1-e^{-1}$ for $t\geqslant\tau^*+1$.
Let $z_\varepsilon(t)=\varepsilon z(t)$ and check that $z_\varepsilon$ satisfies:
\[z_\varepsilon(\tau^*)\in[-e^{-1},e^{-1}]\qquad z_\varepsilon'(t)\geqslant\phi(t)(\tfrac{3}{2}-z_\varepsilon(t))\]
It follows that $z_\varepsilon$ is an increasing function and from a classical argument about differential inequalities that:
\[
z_\varepsilon(t)\geqslant\frac{3}{2}-\left(\frac{3}{2}-z_\varepsilon(\tau^*)\right)e^{-\int_{\tau^*}^t\phi(u)du}\]
In particular for $t^*=\tau^*+1+2\ln 4$
we have:
\[z_\varepsilon(t)\geqslant\frac{3}{2}-(\tfrac{3}{2}-z_\varepsilon(\tau^*))e^{-2\ln4(1-e^{-1})}
\geqslant\frac{3}{2}-2e^{-\ln4}\geqslant1\]
This proves that $|z(t)|$ is an increasing function, so in particular once it has
reached $1$, it stays greater than $1$. Furthermore, if $w\in\mathcal{L}$ then $z(t^*)\geqslant1$
and if $w\notin\mathcal{L}$ then $z(t^*)\leqslant1$. Also note that $\infnorm{(y,v,z,w)'(t)}\geqslant1$
for all $t\geqslant1$. Also note that $z$ is bounded by a constant, by a very similar reasoning.
This shows that if $Y=(y,v,z,\tau)$, then $\infnorm{Y(t)}\leqslant\poly(\infnorm{\psi(w)},t)$
because $\infnorm{y(t)}\leqslant\Upsilon(\infnorm{\psi(w)},t)$. Consequently, there is
a polynomial $\Upsilon^*$ such that $\infnorm{Y'(t)}\leqslant\Upsilon^*$ (this is immediate
from the expression of the system), and without loss of generality, we can assume that
$\Upsilon^*$ is an increasing function. And since $\infnorm{Y'(t)}\geqslant1$, we have that
$t\leqslant\glen{Y}(0,t)\leqslant t\sup_{u\in[0,t]}\infnorm{Y'(u)}\leqslant
t\Upsilon^*(\infnorm{\psi(w)},t)$. Define $\Omega^*(\alpha)=t^*\Upsilon^*(\alpha,t^*)$
which is a polynomial becase $t^*$ is polynomially bounded in $\infnorm{\psi(w)}=|w|$.
Let $t$ such that $\glen{Y}(0,t)\geqslant\Omega^*(|w|)$, then by the above reasoning,
$t\Upsilon^*(|w|,t)\geqslant\Omega^*(|w|)$ and thus $t\geqslant t^*$ so $|z(t)|\geqslant1$,
i.e. the system has decided.

\subsection{Proof of Theorem~\ref{main:th}}
\label{app:main:th}

\subsubsection{ $\FP$ iff emulable: extension to multiple inputs/outputs}

The equivalence between $\FP$ and the fact of beeing emulable has been
proved in Theorem~\ref{th:fp_gpac} for single input function, which is
sufficient in theory because we can always encode tuples of words
using a single word or give Turing machines several input/output
tapes. For what follows, it will be useful to have function with
multiple inputs/ouputs without going through an encoding.  We extend
the notion of discrete encoding in the natural way to handle this
case.

\begin{definition}[Discrete emulation]\label{def:fp_gpac_embed_multi}
$f:\left(\Gamma^*\right)^n\rightarrow\left(\Gamma^*\right)^m$ is called \emph{emulable}
if there exists $g\in\gpc$ and $k\in\N$ such that for any word $\vec{w}\in\left(\Gamma^*\right)^n$:
\[g(\psi(\vec{w}))=\psi(f(\vec{w}))\qquad\text{where}\quad
\psi(x_1,\ldots,x_\ell)=\left(\psi(x_1),\ldots,\psi(x_\ell)\right)\]
and $\psi$ is defined as in Definition~\ref{def:fp_gpac_embed}.
\end{definition}

\begin{remark}[Consistency]
It is trivial that Definition~\ref{def:fp_gpac_embed_multi} matches
Definition~\ref{def:fp_gpac_embed}
in the case of unidimensional functions, thus the two definitions are consistent
with each other.
\end{remark}

\begin{theorem}[Multidimensional $\FP$ equivalence]\label{th:fp_gpac_multi}
Let $f:\left(\Gamma^*\right)^n\rightarrow\left(\Gamma^*\right)^m$. Then $f\in\FP$ if and only if $f$ is emulable.
\end{theorem}

\begin{proof}
First note that we can always assume that $m=1$ by applying the result componentwise.
Similarly, we can always assume that $n=2$ by applying the result repeatedly. Since $\FP$
is robust to the exact encoding used for pairs, we choose a particular encoding to prove the result.
Let $\#$ be a fresh symbol not found in $\Gamma$ and define $\Gamma^\#=\Gamma\cup\{\#\}$.
We naturally extend $\gamma$ to $\gamma^\#$ which maps $\Gamma^\#$ to $\N^*$ injectively.
Let $h:{\Gamma^\#}^*\rightarrow\Gamma^*$ and define for any $w,w'\in\Gamma^*$:
\[h^\#(w,w')=h(w\#w')\]
It follows\footnote{This is folklore, but mostly because this particular encoding of pairs is polytime computable.} that
\[f\in\FP\text{ if and only if }\exists h\in\FP\text{ such that }h^\#=f\]

Assume that $f\in\FP$, then there exists $h\in\FP$ such that $h^\#=f$. Note that $h$ naturally
induces a function (still called) $h:{\Gamma^\#}^*\rightarrow{\Gamma^\#}^*$
so we can apply Theorem~\ref{th:fp_gpac} to get that $h$ is emulable over alphabet $\Gamma^\#$.
Apply Definition~\ref{def:fp_gpac_embed} to get $g\in\gpc$ and $k\in\N$ that
emulate $h$. In the remaining of the proof, $\psi$ denotes encoding of
Definition~\ref{def:fp_gpac_embed}
for this particular k, in other words:
\[\psi(w)=\left(\sum_{i=1}^{|w|}\gamma^\#(w_i)k^{-i},|w|\right)\]
Define for any $x,x'\in\R$ and $n,n'\in\N$:
\[\varphi(x,n,x',n)=\left(x+\left(\gamma^\#(\#)+x'\right)k^{-n-1},n+m+1\right)\]
We claim that $\varphi\in\gpc$ and that for any $w,w'\in\Gamma^*$, $\varphi(\psi(w),\psi(w'))=\psi(w\#w')$.
The fact that $\varphi\in\gpc$ is immediate using Theorem~\ref{th:gpac_comp_arith}
and the fact that $n\mapsto k^{-n-1}$ is analog-polytime-computable\footnote{Note that it works only because $n\geqslant0$.}. The second
fact is follows from a calculation:
\begin{align*}
\varphi(\psi(w),\psi(w'))&=\varphi\left(\sum_{i=1}^{|w|}\gamma^\#(w_i)k^{-i},|w|,\sum_{i=1}^{|w'|}\gamma^\#(w'_i)k^{-i},|w'|\right)\\
    &=\left(\sum_{i=1}^{|w|}\gamma^\#(w_i)k^{-i}+\left(\gamma^\#(\#)+\sum_{i=1}^{|w'|}\gamma^\#(w'_i)k^{-i}\right)k^{-|w|-1},|w|+|w'|+1\right)\\
    &=\left(\sum_{i=1}^{|w\#w'|}\gamma^\#((w\#w')_i)k^{-i},|w\#w'|\right)\\
    &=\psi(w\#w')
\end{align*}
Define $G=g\circ\varphi$, we claim that $G$ emulates $f$ with
$k$. First $G\in\gpc$ thanks to Theorem~\ref{th:gpac_comp_composition}.
Second, for any $w,w'\in\Gamma^*$, we have:
\begin{align*}
G(\psi(w,w'))&=g(\varphi(\psi(w),\psi(w')))\tag*{By definition of $G$ and $\psi$}\\
    &=g(\psi(w\#w'))\tag*{By the above equality}\\
    &=\psi(h(w\#w'))\tag*{Because $g$ emulates $h$}\\
    &=\psi(h^\#(w,w'))\tag*{By definition of $h^\#$}\\
    &=\psi(f(w,w'))\tag*{By the choice of $h$}
\end{align*}

Conversely, assume that $f$ is emulable.
Define $F:{\Gamma^\#}^*\rightarrow{\Gamma^\#}^*\times{\Gamma^\#}^*$ as follows for any ${w\in\Gamma^\#}^*$:
\[F(w)=\begin{cases}(w',w'')&\text{if }w=w'\#w''\text{ where }w',w''\in\Gamma^*\\(\emptyword,\emptyword)&\text{otherwise}\end{cases}\]
Clearly $F_1,F_2\in\FP$ so apply Theorem~\ref{th:fp_gpac} to get that they are emulable.
Thanks to Lemma~\ref{lem:fp_gpac_embed_reenc}, there exists $h,g_1,g_2$ that emulate
$f,F_1,f_2$ respectively with the same $k$. Define:
\[H=h\circ(g_1,g_2)\]
Clearly $H\in\gpc$ because $g_1,g_2,h\in\gpc$. Furthermore, $H$ emulates $f\circ F$ because for any $w\in{\Gamma^\#}^*$:
\begin{align*}
H(\psi(w))&=h(g_1(\psi(w)),g_2(\psi(w)))\\
    &=h(\psi(g_1(w)),\psi(g_2(w)))\tag*{Because $g_i$ emulates $F_i$}\\
    &=h(\psi(F(w)))\tag*{By definition of $\psi$}\\
    &=\psi(f(F(w)))\tag*{Because $h$ emulates $f$}
\end{align*}
Since $f\circ F:{\Gamma^\#}^*\rightarrow{\Gamma^\#}^*$ is emulable, we
can apply Theorem~\ref{th:fp_gpac} to
get that $f\circ F\in\FP$. It is now trivial so see that $f\in\FP$ because for any $w,w'\in\Gamma^*$:
\[f(w,w')=(f\circ F)(w\#w')\]
and $((w,w')\mapsto w\#w')\in\FP$
\end{proof}

\subsubsection{Some facts}

We need the following facts.

\begin{lemma}\label{prop:gp_growth}
Let $f\in\gpc$, there exists a polynomial $P$ such that $\infnorm{f(x)}\leqslant P(\infnorm{x})$
for all $x\in\dom{f}$.
\end{lemma}

\begin{proof}Assume that $f\in\gc{\Upsilon}{\Omega}$ and apply
  Definition~\ref{def:gc} to get $d,p,q$.
Let $x\in\dom{f}$ and let $y$ be the solution of $y(0)=q(x)$ and $y'=p(y)$.
Apply the definition to get that $\infnorm{f(x)-y_{1..m}(\Omega(\infnorm{x},0))}\leqslant1$
and $\infnorm{y(\Omega(\infnorm{x},0))}\leqslant\Upsilon(\infnorm{x},\Omega(\infnorm{x},0))\leqslant\poly(\infnorm{x})$
since $\Upsilon$ and $\Omega$ are polynomials.
\end{proof}

\begin{theorem}[Extraction, \cite{TheseAmaury}]\label{th:extract}
There exists $\myop{extract}\in\gpc$ such that for any $x\in\R$ and $n\in\N$:
\[\myop{extract}(x,n)=\cos(2\pi2^nx)\]
\end{theorem}

Notice that the proof of above theorem is obtained using an
iteration: See \cite{TheseAmaury}. 

\begin{definition}[Mixing function]
Let $f_0:\subseteq\R^n\rightarrow\R^d$, $f_1:\subseteq\R^n\rightarrow\R^d$ and $i:\subseteq\R^n\rightarrow\R$.
Assume that $\{x\thinspace|\thinspace i(x)<1\}\subseteq\dom{f_0}$ and
$\{x\thinspace|\thinspace i(x)>0\}\subseteq\dom{f_1}$, and define for $x\in\dom{i}$:
\[\mix{i}{f_0}{f_1}(x)=\begin{cases}
f_0(x)&\text{if }i(x)\leqslant0\\
(1-i(x))f_0(x)+i(x)f_1(x)&\text{if }0<i(x)<1\\
f_1(x)&\text{if }i(x)\geqslant1\end{cases}\]
\end{definition}

\begin{theorem}[Closure by mixing, \cite{TheseAmaury}]\label{th:comp:mix}
Let $f_0:\subseteq\R^n\rightarrow\R^d$, $f_1:\subseteq\R^n\rightarrow\R^d$ and $i:\subseteq\R^n\rightarrow\R$.
Assume that $f_0,f_1,i\in\gpc$, that $\{x\thinspace|\thinspace i(x)<1\}\subseteq\dom{f_0}$ and that
$\{x\thinspace|\thinspace i(x)>0\}\subseteq\dom{f_1}$. Then $\mix{i}{f_0}{f_1}\in\gpc$.
\end{theorem}

\begin{remark}[Limit computability]\label{rem:gpwc_gpc}
A careful look at Definition~\ref{def:gwc} shows that analog weak computability
is a form of limit computability. Formally, let $f:I\times\Rps\rightarrow\R^n$, $g:I\rightarrow\R^n$ and $\mho:\Rp^2\rightarrow\Rp$
a polynomial. Assume that $f\in\gpc$ and that for any $x\in I$ and $\tau\in\Rps$,
if $\tau\geqslant\mho(\infnorm{x},\mu)$ then $\infnorm{f(x,\tau)-f(x)}\leqslant e^{-\mu}$.
Then $g\in\gpwc$ because the analog system for $f$ satisfies all the items of the
definition.
\end{remark}

\begin{theorem}[Word decoding, \cite{TheseAmaury}]\label{th:decoding}
Let $k_1,k_2\in\N^*$ and $\kappa:\intinterv{0}{k_1-1}\rightarrow\intinterv{0}{k_2-1}$.
There exists a function $\left(\myop{decode}_\kappa:\subseteq\R\times\N\times\R\rightarrow\R\right)\in\gpc$
such that for any word $w\in\intinterv{0}{k_1-1}^*$ and $\mu,\varepsilon\geqslant0$:
\[\text{if }\varepsilon\leqslant k_1^{-|w|}(1-e^{-\mu})
\text{ then }\myop{decode}_\kappa\left(\sum_{i=1}^{|w|}w_ik_1^{-i}+\varepsilon,|w|,\mu\right)=
\left(\sum_{i=1}^{|w|}\kappa(w_i)k_2^{-i},\#\{i|w_i\neq0\}\right)\]
\end{theorem}

\begin{lemma}[Reencoding]\label{cor:reencoding}
Let $k_1,k_2\in\N^*$ and $\kappa:\intinterv{1}{k_1-2}\rightarrow\intinterv{0}{k_2-1}$.
There exists a function $\left(\myop{reenc}_\kappa:\subseteq\R\times\N\rightarrow\R\times\N\right)\in\gpc$
such that for any word $w\in\intinterv{1}{k_1-2}^*$
and $n\geqslant|w|$ we have:
\[\myop{reenc}_\kappa\left(\sum_{i=1}^{|w|}w_ik_1^{-i},n\right)=\left(\sum_{i=1}^{|w|}\kappa(w_i)k_2^{-i},|w|\right)\]
\end{lemma}

\begin{proof}
The proof is immediate: extend $\kappa$ with $\kappa(0)=0$ and define
\[\myop{reenc}_\kappa(x,n)=\myop{decode}_\kappa(x,n,0)\]
Since $n\geqslant|w|$, we can apply Theorem~\ref{th:decoding} with $\varepsilon=0$ to get the result.
Note that stricly speaking, we are not applying the theorem to $w$ but rather to $w$ padded with as
many $0$ symbols as necessary, ie $w0^{n-|w|}$. Since $w$ does not contain the symbol $0$ so its length is the same as the number of non-blank
symbols it contains.
\end{proof}

\subsubsection{Proof of Theorem~\ref{main:th}}

Assume that the theorem is true for functions in $C^0([0,1/2])$, then we claim
the theorem follows. Indeed, if $f\in C^0([a,b],\R)$ is polynomial time computable,
then there exists\footnote{To see that, observe that any polytime computable function is bounded by a polynomial.}
$m,M\in\Rpoly$ such that $m< f(x)< M$ for all $x\in[a,b]$. Define for $\alpha\in[0,1/2]$:
\[g(\alpha)=\frac{f(a+2\alpha(b-a))-m}{2(M-m)}\]
then clearly $g\in C^0([0,1/2])$ and is polytime computable because $a,b,m,M\in\Rpoly$ . It follows that $g\in\gpc$
and then $f\in\gpc$ by the closure properties of $\gpc$. Conversely, if $f\in C^0([0,1/2])$
belongs to $\gpc$ then there also exists\footnote{By Lemma~\ref{prop:gp_growth}, functions in $\gpc$ are
bouned by a polynomial.} $m,M\in\Rpoly$ as above and the reasoning is exactly the same.
In the remaining of the proof, we assume that $f\in C^0([0,1/2])$. This restriction is useful
to simplify the encoding used later in the proof.

Let $f\in C^0([0,1/2])$ be a polynomial time computable
function.  From classical recursive analysis arguments, there exists a computable (resp. polynomial time computable\footnote{The second argument of $g$ must be in unary.}) function
$g:(\Q\cap[a,b])\times\N\rightarrow\Q$ and a computable (resp. polynomial) function $m:\N\rightarrow\N$ such that:
\begin{itemize}
\item $m$ is a modulus of continuity for $f$
\item for any $n\in\N$ and $d\in[a,b]\cap\Q$, $|g(d,n)-f(d)|\leqslant2^{-n}$
\end{itemize}
Note that $g:\Q\cap[0,1/2]\times\N\rightarrow\Q\cap[0,1/2]$
has its second argument written in unary. In order to apply the $\FP$ characterization,
we need to discuss the encoding of rational numbers and unary integers.
Let us choose a binary alphabet $\Gamma=\{0,1\}$ with $\gamma(0)=1$ and $\gamma(1)=2$ and define for any $w,w'\in\Gamma^*$:
\[\psi_\N(w)=|w|\qquad\psi_\Q(w)=\sum_{i=1}^{|w|}w_i2^{-i}\]
Note that $\psi_\Q$ is a bijection from $\Gamma^*$ to $\Q\cap[0,1[$. Define for any $w,w'\in\Gamma^*$:
\[g_\Gamma(w,w')=\psi_\Q^{-1}(g(\psi_\Q(w),\psi_\N(w'))\]
Since $\psi_Q$ is a polytime computable encoding, then $g_\Gamma\in\FP$ because it has
running time polynomial in the size of $\psi_Q(w)$ and the (unary) value of $\psi_\N(w')$,
which are the size of $w$ and $w'$ respectively, by definition of $\psi_\Q$ and $\psi_\N$.
Apply Theorem~\ref{th:fp_gpac_multi} to get that $g_\Gamma$ is emulable. Thus there exits $h\in\gpc$ and $k\in\N$ such that
for all $w,w'\in\Gamma^*$:
\[h(\psi(w,w'))=\psi(g_\Gamma(w,w'))\]
where $\psi$ is defined as in Definition~\ref{def:fp_gpac_embed_multi}, for this specific value of $k$. 
Define $\kappa:\intinterv{0}{k-2}\rightarrow\{0,1\}$ by $\kappa(\gamma(0))=0$ and $\kappa(\gamma(1))=1$
and $\kappa(\alpha)=0$ otherwise, and define:
\[\psi_\Q^*(x,n)=\myop{reenc}_{\kappa,1}(x,n)\]
It follows from Lemma~\ref{cor:reencoding} that $\psi_\Q^*\in\gpc$ and:
\[
\psi_\Q^*(\psi(w))=\myop{reenc}_{\kappa,1}\left(\sum_{i=1}^{|w|}\gamma(w_i)k^{-i},|w|\right)
    =\sum_{i=1}^{|w|}\kappa(\gamma(w_i))2^{-i}
    =\sum_{i=1}^{|w|}w_i2^{-i}=\psi_\Q(w)
\]
We can now define:
\[g_\Gamma^*(x,n,x',n')=\psi_\Q^*(h(x,n,x',n')\]
and get that for any $w,w'\in\Gamma^*$:
\[
g_\Gamma^*(\psi(w,w'))=\psi_\Q^*(h(\psi(w,w')))
    =\psi_\Q^*(\psi(g_\Gamma(w,w'))
    =\psi_\Q(g_\Gamma(w,w'))
    =g(\psi_\Q(w),\psi_\N(w'))
\]
Let us summarize what we have done so far: we built $g_\Gamma^*\in\gpc$ that,
if provided with the encoding of $w,w'$, compute $g(\psi_\Q(w),\psi_\N(w'))$.
To use this function, we need to be able to compute, from the input $x\in[0,1]$
and the requested precision $\mu\geqslant0$, words $w,w'$ such that $|w'|\geqslant\mu$
and $|x-\psi_\Q(w)|\leqslant2^{-m(\psi_\N(w'))}$
so that we can run $g_\Gamma^*$ and get an approximation of $f(x)\pm2^{-\mu}$.
The problem is that for continuity reason, it is impossible to compute such $w,w'$ in
general. This is where mixing comes into play: given $x$ and $\mu$,
we will compute two pairs $w,w'$ and $u,u'$ such that at least one of them satisfies
the above criteria. We will then apply $g_\Gamma^*$ on both of them and mix the result.

Define\footnote{This is a technicality because $\myop{decode}_\iota$ will encode the output
in basis $k$ if $\iota:\Gamma\rightarrow\intinterv{0}{k-1}$.} $\iota:\Gamma\rightarrow\intinterv{0}{k-1}$ by $\iota=\gamma$.
Apply Theorems~\ref{th:decoding} and~\ref{th:extract} to
get $\myop{decode}_\iota,\myop{extract}\in\gpc$. Define for any $n\in\N$:
\[u(n)=\left(\tfrac{1-k^{-n}}{k-1},n\right)\]
Clearly $u\in\gpc$ and one checks that $u(n)=\psi(0^n)$ because:
\[\psi(0^n)=\left(\sum_{i=1}^n\gamma(0)k^{-i},n\right)=\left(\tfrac{1}{k}\tfrac{1-k^n}{1-k},n\right)=u(n)\]
Now define for any $n\in\N$ and relevant\footnote{We will discuss the domain of definition of $v$ right after.} $x\in[0,1]$:
\[v(x,n)=\left(\myop{decode}_{\iota,1}(x,n,2),n\right)\]
It follows from Theorem~\ref{th:decoding} and the fact that $1-e^{-2}\geqslant\tfrac{2}{3}$ that:
\[\text{if }x=\psi_\Q(w)+\varepsilon\text{ for some }w\in\Gamma^n\text{ and }\varepsilon\in\left[0,2^{-n}\tfrac{2}{3}\right]\text{ then }
v(x,n)=\psi(w)\]
Now define for any $n\in\N$ and relevant $x\in[0,1]$:
\begin{align*}
f_0(x,n)&=g_\Gamma^*(v(x,n),u(n))\\
f_1(x,n)&=g_\Gamma^*\left(v\left(x+2^{-n-1},n\right),u(n)\right)\\
i(x,n)&=\tfrac{1}{2}+\myop{extract}\left(x+2^{-n}\tfrac{1}{6},n\right)
\end{align*}
From the domain of definition of $v$, it follows that:
\[\smashoperator{\bigcup_{w\in\Gamma^{n}}}\left[\psi_\Q(w),\psi_\Q(w)+2^{-n}\tfrac{2}{3}\right]\subseteq\dom{f_0}\]
\[\smashoperator{\bigcup_{w\in\Gamma^{n}}}\left[\psi_\Q(w)-2^{-n-1},\psi_\Q(w)+2^{-n}\tfrac{1}{6}\right]\subseteq\dom{f_1}\]
First off, check that for any $n\in\N$:
\[\smashoperator{\bigcup_{w\in\times\Gamma^{n}}}\left[\psi_\Q(w),\psi_\Q(w)+2^{-n}\right[=\left[0,1\right[\]
Check that for any $n\in\N$, $\varepsilon\in[0,2^{-n}[$, $n\in\N$ and $w\in\Gamma^n$:
\[i(\psi_\Q(w)+\varepsilon,n)=\tfrac{1}{2}+\cos(2\pi2^n\varepsilon+\tfrac{\pi}{3})\]
It follows that any $n\in\N$, $\varepsilon\in[0,2^{-n}[$, $w\in\Gamma^{n}$ and $x=\psi_\Q(w)+\varepsilon$ we have:
\begin{align*}
\varepsilon\in\left[0,2^{-n}\tfrac{1}{6}\right[&\qquad\Rightarrow\qquad i(x,n)\in[0,1[\\
\varepsilon\in\left[2^{-n}\tfrac{1}{6},2^{-n}\tfrac{1}{2}\right]&\qquad\Rightarrow\qquad i(x,n)\leqslant0\\
\varepsilon\in\left]2^{-n}\tfrac{1}{2},2^{-n}\tfrac{2}{3}\right[&\qquad\Rightarrow\qquad i(x,n)\in]0,1[\\
\varepsilon\in\left[2^{-n}\tfrac{2}{3},2^{-n}\right[&\qquad\Rightarrow\qquad i(x,n)\geqslant1
\end{align*}
Thus:
\[\{(x,n)\thinspace|\thinspace i(x,n)<1\}\subseteq\dom{f_0}\qquad\qquad\{(x,n)\thinspace|\thinspace i(x,n)>0\}\subseteq\dom{f_1}\]
Define for any $x\in[0,1/2]$ and $n\in\N$:
\[g^*(x,n)=\mix{i}{f_0}{f_1}(x,n)\]
We can thus apply Theorem~\ref{th:comp:mix} to get that $g^*\in\gpc$. Note that $g^*$ is defined over $[0,1[\times\N$
which obviously contains $[0,1/2]\times\N$.
We will now see that $g^*$ approximates $f$ and conclude that $f\in\gpwc$. To do so, we will show
the following statement by a case analysis, for all $x\in[0,1/2]$ and $n\in\N$:
\[\exists y,z\in\Q\cap[0,1/2], \alpha\in[0,1],|x-y|,|x-z|\leqslant2^{-n}\text{ and }g^*(x,n)=\alpha g(y,n)+(1-\alpha)g(z,n)\]
To see that, first note that there exists\footnote{It may not be unique since we closed the interval on both sides
in order to get all of $[0,1/2]$ with words in $\{0\}\times\Gamma^n$. If we opened the interval on the right, we would only get
$[0,1/2[$ with such words.} $w\in\Gamma^n$ such that\footnote{The use of this assumption will become later on. Essentially,
it is there to ensure that the $y$ we construct belongs to $[0,1/2]$ so that we can apply the function $g$ to it.} $x\in[\psi_\Q(w),\psi_\Q(w)+2^{-n}]$.
Furthermore, since $x\in[0,1/2]$, we can always assume that $w\in\{0\}\times\Gamma^{n-1}$.
Write $\varepsilon=x-\psi_\Q(w)$, then there are four possible cases. It will be useful to keep in mind that $u(n)=\psi(0^n)$ as shown previously
and that $\psi_\N(0^n)=n$. Also remember that we showed that $g_\Gamma^*(\psi(w,w'))=g(\psi_\Q(w),\psi_\N(w'))$.
In almost all cases, we will define $y=\psi_\Q(w)$ and thus $|x-y|=\varepsilon\leqslant2^{-n}$.
\begin{itemize}
\item\textbf{If $\varepsilon\in\left[0,2^{-n}\tfrac{1}{6}\right[$ then} $i(x,n)\in[0,1[$ thus $g^*(x,n)=i(x,n)f_0(x,n)+(1-i(x,n))f_1(x,n)$.
By construction of $v$, $v(x,n)=\psi(w)$ thus $f_0(x,n)=g_\Gamma^*(\psi(w),\psi(0^n))=g_\Gamma^*(\psi(w,0^n))=g(\psi_\Q(w),\psi_\N(0^n))=g(y,n)$.
Since, $x+2^{-n-1}-\psi_\Q(w)\in[0,2^{-n}\tfrac{2}{3}]$ we similarly have $v(x+2^{-n-1},n)=\psi(w)$ and
thus $f_1(x,n)=f_0(x,n)=g(y,n)$. It follows that $g^*(x,n)=g(y,n)$. So in this case, $z=y\in[0,1/2]$ and $\alpha$ can be anything.
\item\textbf{If $\varepsilon\in\left[2^{-n}\tfrac{1}{6},2^{-n}\tfrac{1}{2}\right]$ then} $i(x,n)\leqslant0$ thus $g^*(x,n)=f_0(x,n)$.
By construction of $v$, $v(x,n)=\psi(w)$ thus $f_0(x,n)=g_\Gamma^*(\psi(w),\psi(0^n))=g_\Gamma^*(\psi(w,0^n))=g(\psi_\Q(w),\psi_\N(0^n))=g(y,n)$.
It follows that $g^*(x,n)=g(y,n)$. So in this case, $z=y\in[0,1/2]$ and $\alpha$ can be anything.
\item\textbf{If $\varepsilon\in\left]2^{-n}\tfrac{1}{2},2^{-n}\tfrac{2}{3}\right[$ then} $i(x,n)\in[0,1[$ thus $g^*(x,n)=i(x,n)f_0(x,n)+(1-i(x,n))f_1(x,n)$.
By construction of $v$, $v(x,n)=\psi(w)$ thus $f_0(x,n)=g_\Gamma^*(\psi(w),\psi(0^n))=g_\Gamma^*(\psi(w,0^n))=g(\psi_\Q(w),\psi_\N(0^n))=g(y,n)$.
However, $x+2^{-n-1}-\psi_\Q(w)\in[2^{-n},2^{-n}(1+\tfrac{1}{6})]$. Thus define $w'\in\Gamma^n$ such that\footnote{This
is always possible, formally if $w$ is seen as a number, written in binary, then $w'$ is $w+1$.}\saveFN\sftwplusone $\psi_\Q(w')=\psi_\Q(w)+2^{-n}$
and define $z=\psi_\Q(w')$. It follows that $x-\psi_\Q(w')\in[0,2^{-n}\tfrac{1}{6}]$ thus $v(x+2^{-n-1},n)=\psi(w')$ and
thus $f_1(x,n)=f_0(x,n)=g(z,n)$. It follows that $g^*(x,n)=\alpha g(y,n)+(1-\alpha)g(z,n)$ where $\alpha=i(x,n)\in[0,1]$.
Furthermore, $|z-x|\leqslant2^{-n}$ by construction of $w'$.
\item\textbf{If $\varepsilon\in\left]2^{-n}\tfrac{2}{3},2^{-n}\right]$ then} $i(x,n)\geqslant1$ thus $g^*(x,n)=f_1(x,n)$.
Define $w'\in\Gamma^n$ such that\useFN\sftwplusone
$\psi_\Q(w')=\psi_\Q(w)+2^{-n}$
and define $z=\psi_\Q(w')$. It follows that $x-\psi_\Q(w')\in[0,2^{-n}\tfrac{1}{2}]$ thus $v(x+2^{-n-1},n)=\psi(w')$ and
thus $f_1(x,n)=f_0(x,n)=g(z,n)$. So in this case, $y=z\in[0,1/2]$ and $\alpha$ can be anything.
\end{itemize}
We are now in position to conclude thanks to the modulus of continuity of $f$. Recall that
by definition, $m$ is a polynomial such that for any $x,y\in[0,1/2]$ and $k\in\N$, if $|x-y|\leqslant2^{-m(k)}$ then
$|f(x)-f(y)|\leqslant 2^{-k}$. Without loss of generality, we can assume\footnote{Do do so,
consider the same polynomial where each coefficient is the ceiling value of the absolute
value of the corresponding coefficient of $m$, and add the monomial $x\mapsto x$.} that $m(\N)\subseteq\N$ and $m(n)\geqslant n$. Now define for any $x\in[0,1/2]$ and $n\in\N$:
\[g^{**}(x,n)=g^*(x,m(n+1))\]
Clearly $g\in\gpc$ since $g\in\gpc$ and $m$ is a polynomial. Let $x\in\N$ and $n\in\N$. Then
we have shown that there exists $y,z\in\Q\cap[0,1/2]$ and $\alpha\in\N$ such that $|x-y|,|x-z|\leqslant2^{-m(n+1)}$
and $g^*(x,m(n+1))=\alpha g(y,m(n+1))+(1-\alpha)g(z,m(n+1))$. By definition of $g$,
$|g(y,m(n+1))-f(y)|\leqslant 2^{-m(n+1)}\leqslant 2^{-n-1}$ since $m(n+1)\geqslant2$. Similarly,
$|g(z,m(n+1))-f(z)|\leqslant 2^{-n-1}$. Furthermore, $|f(y)-f(x)|,|f(z)-f(x)|\leqslant2^{-n-1}$.
Thus:
\begin{align*}
|g^{**}(x,n)-f(x)|&\leqslant\alpha|g^*(y,m(n+1))-f(x)|+(1-\alpha)|g^*(z,m(n+1))-f(x)|\\
    &\leqslant\alpha(2^{-n-1}+|f(y)-f(x)|)+(1-\alpha)(2^{-n-1}+|f(z)-f(x)|)\\
    &\leqslant\alpha2^{-n}+(1-\alpha)2^{-n}\\
    &\leqslant2^{-n}
\end{align*}
Using Remark~\ref{rem:gpwc_gpc}, we have thus shown that $f\in\gpwc$ and
since $\gpwc=\gpc$, $f\in\gpc$.

\end{document}